\RequirePackage{fix-cm}
\documentclass[smallcondensed]{svjour3}     % onecolumn (ditto)
\smartqed  % flush right qed marks, e.g. at end of proo
\usepackage{graphicx}
\usepackage[numbers]{natbib}
\usepackage[hyphens]{url}
\usepackage{hyperref}
\usepackage{listings}
\usepackage{multirow}
\usepackage{longtable}
\usepackage{multirow}
\usepackage[table]{xcolor}
\usepackage{array}
\usepackage{subcaption}
\usepackage{wrapfig}

\usepackage[symbol*]{footmisc}
\usepackage{tcolorbox}
\usepackage{rotating}
\usepackage{booktabs}
\usepackage{adjustbox}
\usepackage{hhline}
\usepackage{wrapfig}
\usepackage{lineno}
\newcommand{\RRChange}[1]{{{#1}}}

\newcommand{\realsearchgoal}{The goal of this research is to assist managers and other decision-makers in making informed choices about the use of software vulnerability detection techniques through an empirical study of the  efficiency and effectiveness of four techniques on a Java-based web application.}

\newcommand{\rqEffective}{What is the effectiveness, in terms of number and type of vulnerabilities, for each technique?}
\newcommand{\rqEfficiency}{How does the reported efficiency in terms of vulnerabilities per hour differ across techniques?}
\newcommand{\rqQualitative}{What other factors should we consider when comparing techniques?}

\newcolumntype{S}{>{\centering\arraybackslash\columncolor[HTML]{EFEFEF}}  m}

\newcolumntype{A}{>{\centering\arraybackslash}  m}

\newcolumntype{I}{>{\centering\arraybackslash}  m}

\newcolumntype{T}{>{\centering\arraybackslash\columncolor[HTML]{CDCDCD}}  m}

\newcolumntype{H}{>{\centering\arraybackslash\columncolor[HTML]{E1E1E1}}  p}

\newcolumntype{B}{>{\raggedright\arraybackslash}  p}
\newcolumntype{C}{>{\centering\arraybackslash}  p}

\newcolumntype{R}[2]{%
  >{\begin{turn}{#1}\begin{minipage}{#2}\small\raggedright\hspace{0pt}}c%
  <{\end{minipage}\end{turn}} %
}

\newcounter{savefootnote}
\newcounter{tablefootnote}

\begin{document}
% \linenumbers
\setcounter{tablefootnote}{0}
\setlength\tabcolsep{1.5pt}
\renewcommand{\thefootnote}{\arabic{footnote}}

\title{PREPRINT: Do I really need all this work to find vulnerabilities?}
\subtitle{An empirical case study comparing vulnerability detection techniques on a Java application}

\author{Sarah~Elder        \and
        Nusrat~Zahan        \and
        Rui~Shu        \and
        Monica~Metro        \and
        Valeri~Kozarev       \and
        Tim~Menzies       \and
        Laurie~Williams
}

\authorrunning{Elder et al.} % if too long for running head

\institute{S. Elder \at
              North Carolina State University (NCSU)\\
              Department of Computer Science\\
              \email{seelder@ncsu.edu} 
           \and
        %   N. Zahan \at
        %       North Carolina State University (NCSU)\\
        %       Department of Computer Science\\
        %       \email{nzahan@ncsu.edu} 
        %   \and
        %   R. Shu \at
        %       North Carolina State University (NCSU)\\
        %       Department of Computer Science\\
        %       \email{rshu@ncsu.edu}
        %     \and
        %     M. Metro \at
        %       North Carolina State University (NCSU)\\
        %       Department of Computer Science\\
        %       \email{mgmetro@ncsu.edu}
        %     \and
        %     V. Kozarev \at
        %       North Carolina State University (NCSU)\\
        %       Department of Computer Science\\
        %       \email{vikozare@ncsu.edu}
        %     \and
        %     T. Menzies \at
        %       North Carolina State University (NCSU)\\
        %       Department of Computer Science\\
        %       \email{timm@ieee.org}
        %     \and
           L. Williams \at
              North Carolina State University (NCSU)\\
              Department of Computer Science\\
              College of Engineering\\
              890 Oval Drive\\
              Engineering Building II\\
              Raleigh, NC 27695\\
              \email{laurie\_williams@ncsu.edu}  
}

\date{Received: date / Accepted: date}

\maketitle

\begin{abstract}
\textit{\textbf{Context:}} Applying vulnerability detection techniques is one of many tasks using the limited resources of a software project.
\\\textit{\textbf{Objective:}} \textit{\realsearchgoal} 
\\\textit{\textbf{Method:}} We apply four different categories of vulnerability detection techniques \textendash~ systematic manual penetration testing (SMPT), exploratory manual penetration testing (EMPT), dynamic application security testing (DAST), and static application security testing (SAST) \textendash\ to an open-source medical records system.  
\\\textit{\textbf{Results:}}We found the most vulnerabilities using SAST. However, EMPT found more severe vulnerabilities. With each technique, we found unique vulnerabilities not found  using the other techniques. The efficiency of manual techniques (EMPT, SMPT) was comparable to or better than the efficiency of automated techniques (DAST, SAST) in terms of Vulnerabilities per Hour (VpH). 
\\\textit{\textbf{Conclusions:}} The vulnerability detection technique practitioners should select may vary based on the  goals and available resources of the project. If the goal of an organization is to find ``all'' vulnerabilities in a project, they need to use as many techniques as their resources allow. 

\end{abstract}

\section{Introduction}
\label{intro}
Detecting software vulnerabilities efficiently and effectively is necessary to reduce the risk that hackers will exploit vulnerabilities before developers can find and patch them. However, as noted by Alomar et al.~\cite{alomar2020you}, security teams often struggle to justify the costs of vulnerability detection and other vulnerability management activities. This need to improve vulnerability detection efforts while not expending unnecessary resources is highlighted in Section 7 of  U.S. Presidential Executive Order 14028, which begins ``\textit{The Federal Government shall employ all appropriate resources and authorities to maximize the early detection of cybersecurity vulnerabilities...}''~\cite{executiveorder2021}. The executive order also emphasizes the need for improved evaluation of security practices, including vulnerability detection.

\textit{\realsearchgoal} We perform a theoretical replication\footnote{A theoretical replication seeks to investigate  the  scope  of  the  underlying  theory, for example by redesigning the study for a different target population, or by testing a variant of the original hypothesis of the work~\cite{theoretical}} of work done by Austin et al.~\cite{austin2011onetechniquenotenough,austin2013comparison}. \RRChange{Since 2011 when the original Austin et al. work was published, the vulnerability detection landscape has changed from the types of applications being tested to the types of vulnerabilities found~\cite{OWASP2021TopTen,OWASP2017TopTen,OWASP2013TopTen,OWASP2010TopTen}. For example, the number of vulnerabilities in the United States National Vulnerability Database assigned to Cross Site Scripting (XSS) has increased faster than prevalence of other vulnerability types such as Code Injection~\cite{NVDCWEOverTime}. Our methodology and findings may also be useful to future evaluations of new vulnerability detection techniques being introduced.}

We examined the 4 vulnerability detection techniques from Austin et al.~\cite{austin2011onetechniquenotenough,austin2013comparison}.  
\begin{itemize}%[noitemsep,leftmargin=0.25in]
    \item \textbf{Systematic Manual Penetration Testing (SMPT)}: the analyst manually and systematically develops, documents, then executes test cases which verify the security objectives of the System Under Test (SUT)~\cite{smith2011systematizing,austin2011onetechniquenotenough,smith2012effective,austin2013comparison}
    \item \textbf{Exploratory Manual Penetration Testing (EMPT)}: the analyst ``spontaneously designs and executes tests based on the [analyst]'s existing relevant knowledge''~\cite{2013ISO29119-1}, searching for vulnerabilities.
    \item \textbf{Dynamic Application Security Testing (DAST)}: automatic tools generate and run tests based on security principles, without access to source code\cite{sei2018sectesttools}.   
    \item \textbf{Static Application Security Testing (SAST)}: automatic tools scan source code for patterns that indicate vulnerabilities~\cite{cruzes2017security,hafiz2016game,scandariato2013static}. 
\end{itemize} 

{\noindent}These four techniques can all be applied during and after software implementation. Applying vulnerability detection during and after the implementation phase of software development is more common in many industry settings~\cite{cruzes2017security} when compared to security testing techniques which focus on earlier phases of software development such as requirements or model-focused testing. \RRChange{ We used an industry standard, the Open Web Application Security Project's Application Security Verification Standard (OWASP ASVS), to systematically develop test cases for SMPT. The two DAST tools and two SAST tools are currently used in industry settings. Two of these tools, the OWASP Zed Attack Proxy (OWASP ZAP)\footnote{https://owasp.org/www-project-zap/} DAST tool and the Sonarqube\footnote{https://www.sonarqube.org/} SAST tool, are open-source. The other tools, which we will refer to as DAST-2 and SAST-2, are proprietary.}
 
 We applied each technique to OpenMRS (\url{https://openmrs.org/}), a large open source medical records system used both in medical research and clinical settings throughout the world. OpenMRS is a web application written in Java and JavaScript, containing 3,985,596 lines of code\footnote{ as measured by CLOC v1.74 (\url{https://github.com/AlDanial/cloc})}. We consider our work to be a case study, since we only examine a single System Under Test (SUT).

 The only previous work comparing as many different types of vulnerability detection techniques  we are aware of is by  Austin et al.~\cite{austin2011onetechniquenotenough,austin2013comparison} published in 2011 and 2013.  The systems in those studies were less than 500,000 lines of code. We know of no other study applying multiple different vulnerability detection techniques to a system this large. Collecting data for this study required a team of four graduate students a combined eleven months of full-time work and twenty months of part-time work; four months part-time work from an undergraduate student; and the results of assignments from a large graduate-level software security course. Our experiences in structuring the software security course have been reported previously in Elder et al.\cite{elder2021structuring}.

We answer the following research questions:
\begin{itemize}
    \item RQ1:\rqEffective
    \item RQ2: \rqEfficiency
\end{itemize}
As part of the software security course, students were asked to discuss and compare the four techniques. Two researchers performed qualitative analysis on the answers, addressing the following research question: 
\begin{itemize}
    \item RQ3: \rqQualitative
\end{itemize}

Our research makes the following contributions: 

 \begin{itemize}
    \item Analysis from our comparison of the efficiency and effectiveness of the four vulnerability detection techniques.
    \item \RRChange{A detailed description of the methodology and related findings, which may be useful for future comparisons of vulnerability detection techniques}
 \end{itemize}
  \RRChange{{\noindent}We are releasing our vulnerability dataset once the vulnerabilities are safely mitigated at \url{https://github.com/RealsearchGroup/vulnerability-detection-20}.}% We will work with OpenMRS to release the remaining data in the first quarter of 2022.}% \MyToDos{Find a way to re-include the dataset better.}

The rest of this paper is structured as follows.  In Section~\ref{sec:keyconcepts}, we provide explanations for key concepts used throughout this paper. In Section~\ref{sec:AustinStudy} we provide a brief overview of the previous work. We discuss other related work in Section~\ref{sec:relwork}. In Section~\ref{sec:VulnDetectionTechniques} we describe the vulnerability detection techniques used in this paper. In Section~\ref{sec:method-sut-openmrs} we discuss the SUT used in our Case Study, OpenMRS. In Section~\ref{sec:dataSources} we discuss the sources of data for the Case Study. In Sections~\ref{sec:method-effective}, \ref{sec:method-efficiency}, and~\ref{sec:method-qualitative} we outline our research methodology for RQ1, RQ2, and RQ3 respectively. Once the methodology is explained we discuss the equipment used in Section~\ref{sec:method-sut-equipment}. We report our results in Section~\ref{sec:results}.  We discuss our findings in Section~\ref{sec:discussion}. We discuss limitations of our study in Section~\ref{sec:limitations}. We discuss the findings in Section~\ref{sec:discussion} and conclude with Section~\ref{sec:conclusion}.  

\section{Key Concepts}\label{sec:keyconcepts}
In this section, we define key concepts used in this paper.

\begin{description}
\item [\textbf{Vulnerability:}] We use the definition of vulnerability from the U.S. National Vulnerability Database\footnote{\url{https://nvd.nist.gov/vuln}}. Specifically, a vulnerability is ``\textit{A weakness in the computational logic (e.g., code) found in software and hardware components that, when exploited, results in a negative impact to confidentiality, integrity, or availability.}'~\cite{NVDVulnDef}. 
\medskip
\item [\textbf{Common Weakness Enumeration (CWE):}] Per the CWE  website, ``\textit{CWE is a community-developed list of software and hardware weakness types.}''\cite{CWEmain}. \linebreak Many security tools, such as the OWASP Application Security Verification Standard (ASVS) and most vulnerability detection tools, use CWEs to identify the types of vulnerabilities relevant to a security requirement, test case, or tool alert. We use the CWE list in this paper to standardize and compare the vulnerability types found by different vulnerability detection techniques.
\medskip
\item [\RRChange{\textbf{OWASP Top Ten:}}] \RRChange{The OWASP Top Ten is a regularly updated list of ``the most critical security risks to web applications.''\cite{OWASPTopTen}. The OWASP Top Ten categories and ranking are is developed by security experts based on the incidence and severity of vulnerabilities associated with different CWEs.  A convenient mapping\cite{CWEOWASP} allows for vulnerabilities to be mapped from CWEs to OWASP Top Ten categories. We use the OWASP Top Ten in this paper to summarize the types vulnerabilities found, and to understand the relative severity of the vulnerabilities found. The latest (2021) Top Ten, which were used in our analysis, are: A01 - Broken Access Control, A02 - Cryptographic Failures, A03 - Injection, A04 - Insecure Design, A05 - Security Miscofiguration, A06 - Vulnerable and Outdated Components, A07 - Identification and Authentication Failures, A08 - Software and Data Integrity Failures, A09 - Security Logging and Monitoring Failures, and A10 - Server-Side Request Forgery (SSRF. Additional information on the OWASP Top Ten may be found at \url{https://owasp.org/Top10/}.}
\medskip
\item[\textbf{OWASP Application Security Verification Standard (ASVS):}] \hfill OWASP\linebreak ASVS is an open standard for performing web application security verification.  The ASVS provides a high-level set of ``\textit{requirements or tests that can be used by architects, developers, testers, security professionals, tool vendors, and consumers to define, build, test and verify secure applications}''~\cite{OWASP2019ASVS}. In the ASVS, each requirement or test is referred to as a ``control'' and must be adapted to a SUT. Each OWASP ASVS control is mapped to a CWE type. We used OWASP ASVS version 4.0.1 released in March 2019\footnote{https://github.com/OWASP/ASVS/tree/v4.0.1}, which was the current version when we began collecting data in Spring 2020. The OWASP ASVS has three levels of requirements. If a requirement falls within a level, it also falls within higher levels. ASVS describes Level 1 as ``\textit{the bare minimum that any application should strive for}''~\cite{OWASP2019ASVS}. 
\end{description}

\section{Previous Work by Austin et al.}\label{sec:AustinStudy}
Our study is a theoretical replication\footnote{A theoretical replication seeks  to  investigate  the  scope  of  the  underlying  theory,for example by redesigning the study for a different target population, or by testing a variant of the original hypothesis of the work~\cite{theoretical}} of previous work done by Austin et al.~\cite{austin2011onetechniquenotenough,austin2013comparison}. The goals of the previous work are ``\textit{to improve vulnerability detection by comparing the effectiveness of vulnerability discovery techniques and to provide specific recommendations to improve vulnerability discovery with these techniques}''\cite{austin2011onetechniquenotenough}. In their first study~\cite{austin2011onetechniquenotenough} Austin et al. applied SMPT, EMPT, DAST, and SAST to two electronic medical records systems, Tolven Electronic Clinician Health Record (eCHR), a Java-based application with 466,538 lines of code, and OpenEMR, a PHP-based application with 277,702 lines of code. The second paper by Austin et al~\cite{austin2013comparison} added a third SUT, PatientOS, a Java-based mobile application with 487,437 lines of code.  In both studies, the authors used one tool for each automated technique. The DAST tool used by Austin et al. was only applicable to web applications. Consequently, they only applied SMPT, EMPT, and SAST to PatientOS. 

Austin et al. calculated how long it took to apply each technique in terms of the number of hours. For each SUT, the authors compare the number and types of vulnerabilities found by each technique, as well as the rate of vulnerabilities per hour for each technique. In the current study, we examine these same metrics, referring to the number and types of vulnerabilities as ``effectiveness'' and the vulnerabilities per hour as ``efficiency''.

Austin et al. found that SAST identified the most vulnerabilities~\cite{austin2011onetechniquenotenough,austin2013comparison}. They emphasize that the vulnerabilities may not be as exploitable as vulnerabilities found using other techniques, and relying on SAST alone would be insufficient since each technique found vulnerabilities not found by other techniques. The authors also found automated tools to be faster in terms of vulnerabilities per hour, although automated techniques found fewer types of vulnerabilities.

\section{Related Work}\label{sec:relwork}
Several related studies have focused on a single category of techniques, such as comparisons of DAST tools or comparisons of SAST tools. In 2010, Doup{\'e} et al.~\cite{doupe2010johnny} compared eleven ``point and click'' DAST tools to each other. The authors found that while some types of vulnerabilities could be found reliably, other types of vulnerabilities could not be found by the tools examined in the study. More recently, Klees et al.~\cite{klees2018evaluating} performed a rigorous comparison of DAST tools, providing insights on the biases and limitations of DAST tool studies. They examined 32 papers on fuzz testing and performed an experiment comparing two tools against five benchmark applications. The U.S. National Institute of Standards and Technology (NIST) Software Assurance Metrics and Tool Evaluation (SAMATE) program has performed a series of Static Analysis Tool Expositions (SATE)~\cite{delaitre2018sate,okun2013report,okun2011report,okun2009static}. On a regular basis, the SAMATE program establishes a set of trials which they refer to as tests or test cases. For example, a test case may require running the tool to be evaluated against a benchmark. Organizations which the SATE report authors refer to as ``toolmakers''~\cite{delaitre2018sate} sign up to participate in the experiment and run their tools against the trials. The results were reviewed by SAMATE organizers. These studies, particularly the experiments run by Klees et al. and the SAMATE program, inform our methodology for SAST and DAST techniques but differ from our work in that they make comparisons between tools of a similar type and do not examine manual techniques. These are merely examples, as there are many more comparison studies within a single technique\cite{amankwah2020empirical,bau2012vulnerability}

Another common comparison between vulnerability detection techniques is between static techniques that primarily analyze source code, and techniques that are based on interactions with the running software system without access to the source code (Dynamic or DAST). Studies that compared static and dynamic analyses include a controlled experiment by Scandariato et al.~\cite{scandariato2013static}, which compared the use of SAST with the use of DAST. Scandariato et al. conducted an experiment in which nine participants performed vulnerability detection tasks. The authors examine the user experience of SAST and DAST tools, and analyze the efficiency of using SAST and DAST. Scandariato et al. found that although participants found DAST tools more ``fun'' to use, the participants were more efficient with SAST tools and considered SAST tools a better starting point for new security teams. Similarly, in 2009, prior to the previous work by Austin et al., Antunes and Viera~\cite{antunes2009comparing} performed a comparison between SAST and DAST tools. Similar to our study, Antunes and Viera found that different tools within the same technique found different vulnerabilities, and that SAST found more vulnerabilities than DAST. In contrast with Scandariato et al. and Antunes and Viera, we further subdivide Dynamic analyses into SMPT, EMPT, and DAST, exploring each of these techniques separately.

Similarly, many surveys and comparisons exist which focus on a single type of vulnerability. For example, Chaim et al. perform a survey of Buffer Overflow Detection techniques. They note that existing vulnerability detection techniques are either impractical or have a high false positive rate, but that emerging Hybrid techniques are ``promising''. Liu et al\cite{liu2019survey} perform a survey of automated, state-of-the-art techniques for finding and exploiting Cross-Site-Scripting (XSS) vulnerabilities, categorizing them as ``static'', ``dynamic'', or ``hybrid''. They note that the increasing size of web applications may be hindering the effectiveness of these automated techniques, but do not perform an empirical comparison. Fonseca et al. perform an empirical comparison of the effectiveness of different DAST tools~\cite{fonseca2007testing} for finding XSS vulnerabilities. While practitioners may prioritize some types of vulnerabilities over others and these studies may assist practitioners in understanding how vulnerability detection techniques compare against a single type of vulenrability. However, applications are rarely threatened only by a single type of vulnerability. Similar to the studies of a single type of detection technique; our study, which examines the effectiveness of techniques across a range of vulnerability types; gains insight from and provides additional insights into results from studies which focus on a single type of vulnerability.

An additional area of related work is the development and application of benchmarks for security testing tools, such as the 2010 work by Antunes and Viera~\cite{antunes2010benchmarking} on developing a benchmark for SAST and DAST tools. As noted in the SATE V report~\cite{delaitre2018sate}, benchmark studies have an important role in evaluating security testing techniques. The use of vulnerability detection techniques in benchmark studies may differ from how security vulnerability detection techniques would be applied in practice. The three web infrastructure performance benchmark systems used by Antunes and Viera\cite{antunes2010benchmarking} to develop security benchmarks contained a combined 2,654 lines of code which could be manually reviewed by security experts in a reasonable amount of time. The results of our study on OpenMRS, which has over 3,000,000 lines of code, may not generalize to smaller systems such as those examined in the Antunes and Viera study. However, our results suggest that for larger software systems; alerts, failing test cases, or other outputs of a vulnerability detection technique which appear to be ``false positives'' when compared against a benchmark should be examined with care to ensure that they are not true positives not found by the techniques used to create the benchmark. We encourage any future users of the set of vulnerabilities and test cases that were generated for this study to be aware that applying additional techniques and tools to the same SUT is likely to find additional vulnerabilities. 

\section{Vulnerability Detection Techniques} \label{sec:VulnDetectionTechniques}
 We begin this section by explaining analysis types which are frequently used to distinguish between vulnerability detection techniques. We then describe the specific types of techniques from our case study. We differentiate between analysis type and detection technique since many common names for vulnerability detection techniques are derived from the analysis types that are part of the technique. For example, Dynamic Application Security Testing (DAST) is a common name for a category of vulnerability detection techniques where analysts use \emph{automated} tools to perform \emph{dynamic} analysis. However, EMPT and SMPT also involve \emph{dynamic} analysis.

\subsection{Analysis Types} \label{sec:def-attributes}
In this section, we explain \emph{automated} and \emph{manual} analysis, \emph{static} and \emph{dynamic} analysis, \emph{source code} analysis, as well as \emph{systematic} and \emph{exploratory} analysis.

\textbf{Automated} vs \textbf{Manual} analysis: Some techniques are based on \emph{automated} analysis performed by a tool. Manual effort may be required to use vulnerability detection tools. However, for the purpose of this study we reserve the phrase \emph{manual analysis} to describe techniques where no automated tool is needed. 

\textbf{Static} vs \textbf{Dynamic} analysis: Static analysis is performed on artifacts such as source code or binaries, where the source code or binary is not executed~\cite{2013ISO29119-1}.  Dynamic analysis is any form of analysis that \emph{does} require code to be executed~\cite{2013ISO29119-1}. 

\textbf{Source code} analysis: Source code analysis is any form of analysis that requires access to source code. While source code analysis is sometimes used as a synonym of static analysis, static analysis and source code analysis are distinct concepts~\cite{mcgraw2006software,austin2013comparison}. For example, static analysis can include analyzing binaries and other artifacts that are not source code. Analysis which does not have access to source code is sometimes referred to as ``black box'' analysis.  

\textbf{Systematic} vs \textbf{Exploratory} analysis: \emph{Systematic} analysis is performed in a very prescriptive, methodical manner; in contrast with \emph{exploratory} analysis which is less formally planned. For example, ISO 29119~\cite{2013ISO29119-1} defines exploratory testing, a form of \emph{exploratory} analysis, as ``experience-based testing in which the [analyst] spontaneously designs and executes tests based on the [analyst]'s existing relevant knowledge, prior exploration of the test item (including the results of previous tests), and heuristic `rules of thumb' regarding common software behaviours and types of failure''. The concepts of \emph{systematic} and \emph{exploratory} analysis primarily apply to \emph{manual} analysis. Whether an automated tool has knowledge and experience is a philosophical discussion outside the scope of this paper.

\subsection{Case Study Techniques} \label{sec:CS-Techniques}
In this section, we provide greater detail on the four categories of vulnerability detection techniques we examine in our case study.  For automated techniques, we examine two tools for each category.

\subsubsection{Manual Techniques} \label{sec:Manual-CS-Techniques}
\begin{figure}[!htb]%[!t]
\centering \includegraphics[width=2.16in]{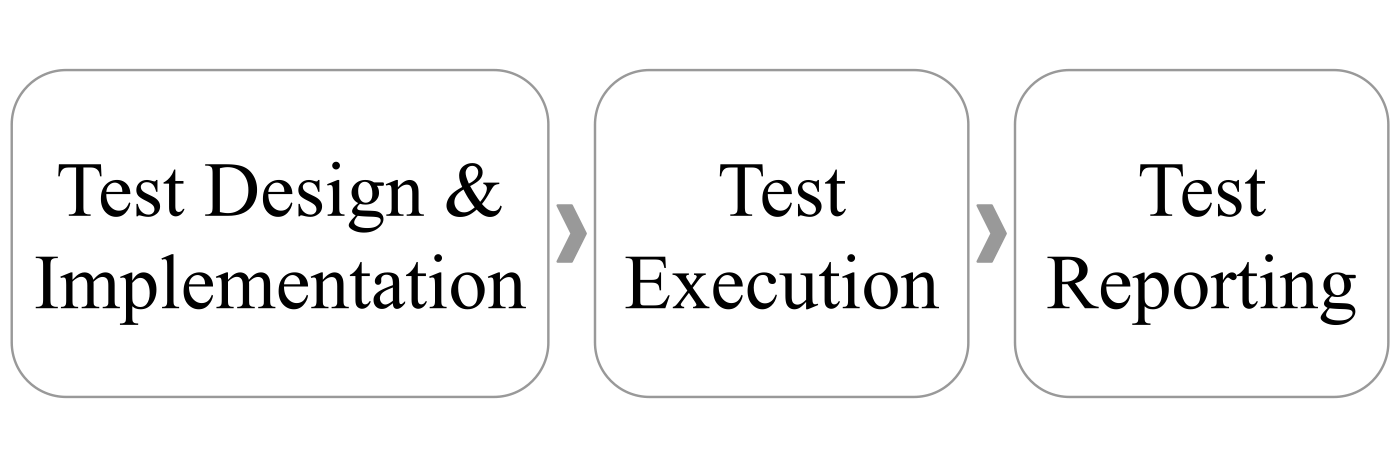} \caption{Applying Manual Test Techniques (based on ISO/IEC/IEEE 29119-1)}
% \centering \includegraphics[width=2.16in,natwidth=1400,natheight=466]{ManualTesting_ProcessChart_v3a.png} \caption{Applying Manual Test Techniques (based on ISO/IEC/IEEE 29119-1)}
\label{fig:ManualProcess} 
\end{figure}

Both \emph{manual} techniques examined in this study are \emph{dynamic} techniques that do not have access to source code. Manually examining the entire source code for system as large as OpenMRS is infeasible. A high-level overview of how  manual dynamic testing techniques, particularly systematic techniques, are applied is shown in Figure \ref{fig:ManualProcess}. This figure is based on the process for dynamic techniques presented in ISO/IEC/IEEE 29119-1~\cite{2013ISO29119-1}.

\paragraph{Systematic Manual Penetration Testing (SMPT)}
SMPT involves \emph{dynamic}, \emph{manual}, and \emph{systematic} analysis. Specifically, SMPT is a form of scripted testing defined by ISO 29119-1~\cite{2013ISO29119-1} as ``dynamic testing in which the [analyst]'s actions are prescribed by written instructions in a test case''. SMPT \emph{does not require access to source code}. In SMPT, the analyst begins  by writing a set of test cases and planning how the test suite will be run for a particular test execution in what 29119-1 refers to as the in the \textit{Test Design \& Implementation} stage, as shown in Figure \ref{fig:ManualProcess}. Many test cases are combined into a test plan. The number of test cases depends on the system and scope of the testing being performed. The tests are then executed. In the final stage, the test results are documented and reported.

 \RRChange{Figure~\ref{fig:TestCaseExample} shows an example SMPT test case from our case study. As we can see in the figure, the steps recorded in an SMPT test case are the actions a person would take when interacting with the system. As we will discuss further in our methodology in Section~\ref{RQ1-DC-Apply-SMPT} and indicated in the Figure, we used the ASVS, described in Section~\ref{sec:keyconcepts} as the basis for our test cases. The test case in Figure \ref{fig:TestCaseExample}, is based on ASVS Control 2.1.7 - \textit{Verify that passwords submitted during account registration, login, and password change are checked against a set of breached passwords either locally (such as the top 1,000 or 10,000 most common passwords which match the system's password policy) or using an external API}. In the test case, the administrator attempts to create a user with the common password ``Passw0rd''. }

\begin{figure}[!htb]
{
\begin{lstlisting}[breaklines=true]

Test Case ID: XXX
ASVS Control: 2.1.7 

Steps:
01) Open the OpenMRS web app to the login screen
02) Type ``admin'' as the username and ``Admin123'' as the password
03) Select ``Inpatient Ward'' as the location
04) Click ``login''
05) Select ``System Administration'', then select ``Manage Accounts''. 
06) Click ``Add New Account''
07) Enter the following information:
         Family Name: Potter 
         Given Name: Harry 
         Gender: Male
08) Select ``Add User Account?''
09) Enter the following information: 
         Username: Hedwig
         Privilege Level: Full
         Password: Passw0rd
         Confirm Password: Passw0rd
10) Leave all other defaults as they are
11) Click ``Save''

Expected Results: The password, ``Passw0rd'', should be rejected as it is on the list of the 10,000 most commonly-used passwords.
Actual Results:
\end{lstlisting}
\caption{Example SMPT Test Case}\label{fig:TestCaseExample}
}
\end{figure}

\paragraph{Exploratory Manual Penetration Testing (EMPT)}
Exploratory Manual Penetration Testing is a \emph{manual}, unscripted, \emph{exploratory}, {dynamic} technique that \emph{does not require access to source code}. Previous studies of functional exploratory testing have suggested that knowledge and experience may play a significant role in exploratory testing~\cite{itkonen2013role,pfahl2014exploratory}.

EMPT follows a similar process to diagram in Figure \ref{fig:ManualProcess}.  As found by Votipka et al.~\cite{votipka2018hackers}, security analysts who perform exploratory testing spend time learning about the system prior to beginning exploration. The analyst then moves on to activities such as the ``exploration'' and ``vulnerability recognition''~\cite{votipka2018hackers}.  The analyst also still documents and reports all vulnerabilities found.  However, the process is less formal and more iterative for EMPT as compared with SMPT.

\subsubsection{Automated Techniques} \label{sec:Automated-CS-Techniques}
\begin{figure}[!htb]
\centering \includegraphics[width=3in]{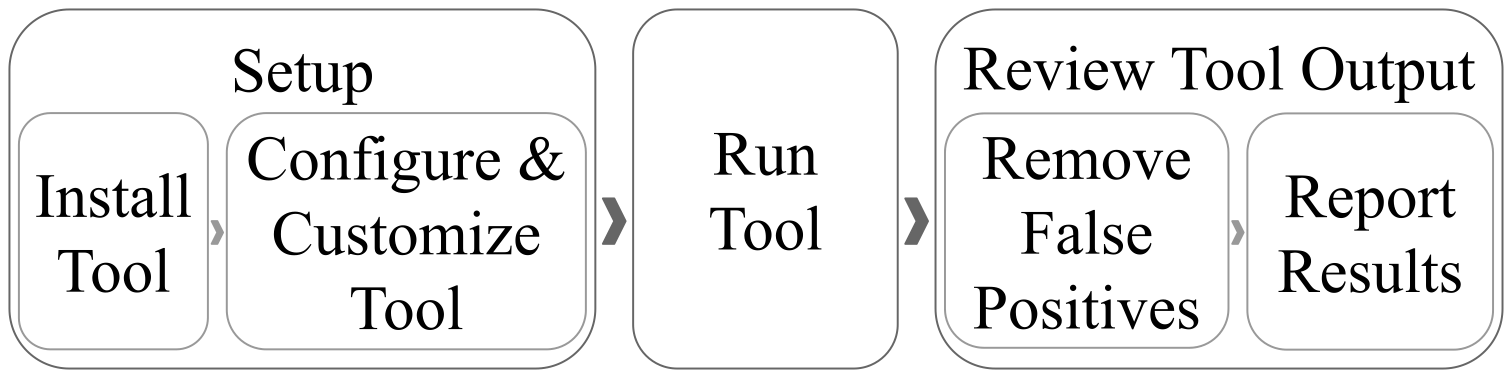} \caption{Applying Tool-Based Techniques}
% \centering \includegraphics[width=3in,natwidth=1510,natheight=377]{ToolBasedChart17.png} \caption{Applying Tool-Based Techniques}
\label{fig:ToolBasedProcess} 
\end{figure}

We examine two categories of \emph{automated} vulnerability detection techniques, Dynamic Application Security Testing (DAST) and Static Application Security Testing (SAST). Figure \ref{fig:ToolBasedProcess} provides an overview of how Tool-based techniques are applied. The tools must first be setup, which includes installing the tool as well as configuring and customizing the tool, if appropriate. The analyst then runs the tool. Once the automated portion of the analysis is complete, the analyst must review the tool output to remove false positives and prepare the report. 

\paragraph{Dynamic Application Security Testing (DAST)}\label{sec:DAST-Desc}
DAST uses \emph{automated} tools to perform \emph{dynamic} analysis. We only include techniques that \emph{do not have access to source code} in the DAST category.DAST is sometimes referred to as Automated Penetration Testing~\cite{antunes2010benchmarking,austin2011onetechniquenotenough,austin2013comparison}, Black-Box Web Vulnerability Scanning~\cite{doupe2010johnny}, Fuzzing~\cite{klees2018evaluating}, or Dynamic Analysis~\cite{cruzes2017security}.  \RRChange{As part of our study, we examined two general-purpose DAST tools. The first tool was the Open Web Application Security Project's Zed Attack Proxy version 2.8.1 (OWASP ZAP, further abbreviated as ZAP in tables)\footnote{\url{https://www.zaproxy.org/}}, is a free, open-source, dynamic analysis tool which describes itself as ``the world's most widely used web app scanner''\cite{OWASPZapMainPage}. The second DAST tool, which we will refer to as DAST-2 (further abbreviated as DA-2 in tables), is a proprietary tool.} 

DAST tools automatically generate a set of malformed inputs to the SUT based on sample inputs provided by the analyst. \RRChange{Inputs to a web application, including both sample inputs provided by the analyst and malformed inputs generated by the  DAST tool, are represented as HTTP messages. When a human interacts with the web application through a browser, as is done with SMPT or EMPT, HTTP messages are created by the web browser~\cite{HTTPmessages} to communicate with the application. Each HTTP message requests that some action be applied to a particular resource in the system~\cite{rfc7231}. Resources are identified through a Uniform Resource Identifier (URI). The action and URI are indicated in the header of an HTTP message. The HTTP request in Figure~\ref{fig:DASTinputexample} is a message from a sample input used in this study. In Figure~\ref{fig:DASTinputexample}, the browser is requesting that the information in the message be \texttt{POST}ed to the URI \texttt{http://127.0.0.1:8080/openmrs/login.htm}. The remainder of the HTTP message contains a \textit{representation}~\cite{rfc7231} of the resource requested.  The sample inputs provided by the analyst tell the DAST tool where the resources are, what the representations are, and provide an initial outline for the order in which messages should be sent. The sample message in Figure~\ref{fig:DASTinputexample} is just one message in a sequence. For example, if an analyst was executing the steps of the test case in Figure~\ref{fig:TestCaseExample}, the HTTP message in Figure~\ref{fig:DASTinputexample} would be generated as part of Step 04. GET requests to access the resources for the login page (Step 01-Step 03 of Figure~\ref{fig:TestCaseExample}) would be sent prior to the POST request in Figure~\ref{fig:DASTinputexample}. Additional requests would continue to be generated and sent based on the analyst's interactions with the application after the Request shown in Figure~\ref{fig:DASTinputexample}. Some DAST tools, such as OWASP ZAP, provide built-in ways to record HTTP messages. Other DAST tools require a standard file format such as HTTP Archive (.har)\footnote{e.g. \url{https://docs.rapid7.com/insightappsec/scan-scope/};{\newline}\url{https://www.netsparker.com/support/scanning-restful-api-web-service/};{\newline}\url{https://docs.gitlab.com/ee/user/application_security/api_fuzzing/create_har_files.html}} which can be generated by most web browsers. }

\RRChange{Some Web Application DAST tools, including OWASP ZAP but not DAST-2, also include a web crawler~\cite{ZAPspider,doupe2010johnny,scandariato2013static} sometimes referred to as a spider, to find additional resources that may not have been included in the original inputs. For example, the input may have not accessed a resource because the analyst creating the input did not know that the resource existed and was accessible. The fact that the resource is accessible may even be a vulnerability if the resource contains sensitive information.}

\begin{figure}[!htb]
{
\begin{lstlisting}[breaklines=true, basicstyle=\tiny]
POST http://127.0.0.1:8080/openmrs/login.htm HTTP/1.1
User-Agent: Mozilla/5.0 (Macintosh; Intel Mac OS X) Firefox/79.0
Accept: text/html,application/xhtml+xml,application/xml;q=0.9,image/webp,*/*;q=0.8
Accept-Language: en-US,en;q=0.5
Content-Type: application/x-www-form-urlencoded
Content-Length: 170
Origin: https://127.0.0.1:8800
Connection: keep-alive
Referer: https://127.0.0.1:8800/openmrs/login.htm
Cookie: JSESSIONID=owczzkvwskvc8n4580psmvbb
Upgrade-Insecure-Requests: 1
Host: 127.0.0.1:8800

username=admin&password=Admin123&sessionLocation=0&redirectUrl=/openmrs/referenceapplication/home.page
\end{lstlisting}
\caption{Message from a Sample Input Provided to a DAST Tool by the Analyst}\label{fig:DASTinputexample} 
}
\end{figure}

Once the sample inputs are provided, the DAST tool applies a set of security rules to create a new set of malformed inputs. Figure \ref{fig:DASTfuzzExample} shows an HTTP message that could be created by DAST tools as part of a malformed input, based on the sample input in Figure~\ref{fig:DASTinputexample}. In this example, the \texttt{sessionLocation} parameter is changed from a number, \texttt{0} as seen in Figure~\ref{fig:DASTinputexample} which is the identifier associated with the ``Inpatient Ward'' value on the login screen, to a script designed to find Cross-Site Scripting (XSS) vulnerabilities, \verb|<script>alert(1);</script>|. The same XSS-focused rule could be applied to the \texttt{username} parameter instead of the  \texttt{sessionLocation} parameter to generate a different malformed input. A different rule could ignore parameters entirely and search the http header for sensitive information, our could re-order the sequence of HTTP messages. \RRChange{The CWEs specifically covered by the rules from each tool are discussed in Appendix \ref{app:AutomatedCWEs}. ZAP rules were associated with 33 CWEs while DAST-2 covered 44 CWEs for a combined 68 CWEs covered. Similarly, ZAP covered six of the OWASP Top Ten while DAST-2 also covered 6 of the OWASP Top Ten, 5 of which overlapped between tools for a combined 7 of the Top Ten covered.}

\begin{figure}[!htb]
\begin{lstlisting}[breaklines=true, basicstyle=\tiny]
POST http://127.0.0.1:8080/openmrs/login.htm HTTP/1.1
User-Agent: Mozilla/5.0 (Macintosh; Intel Mac OS X) Firefox/79.0
Accept: text/html,application/xhtml+xml,application/xml;q=0.9,image/webp,*/*;q=0.8
Accept-Language: en-US,en;q=0.5
Content-Type: application/x-www-form-urlencoded
Content-Length: 170
Origin: https://127.0.0.1:8800
Connection: keep-alive
Referer: https://127.0.0.1:8800/openmrs/login.htm
Cookie: JSESSIONID=owczzkvwskvc8n4580psmvbb
Upgrade-Insecure-Requests: 1
Host: 127.0.0.1:8800

username=admin&password=Admin123&sessionLocation=<script>alert(1);</script>&redirectUrl=/openmrs/referenceapplication/home.page
\end{lstlisting}
\caption{Message from a Malformed Input (Test Case) Produced by a DAST Tool}\label{fig:DASTfuzzExample} 
\end{figure}

With many combinations of rules and ways to apply them, DAST tools can create and run hundreds or thousands of malformed inputs. The malformed inputs generated by the DAST tool are sometimes referred to as \emph{test cases}. \RRChange{DAST tools execute the ``test cases'' (malformed inputs) automatically, which suggests that DAST tools may be able to perform more testing in less time compared with manual techniques such as SMPT, since computer systems can send electrical impulses to other computer systems faster than an end-user can interact with the system via keyboard or mouse~\cite{ackerman2019Superhuman}. One of the motivations for studies such as this one, is to understand whether the promise of ``more'' inputs executed ``faster'' by DAST produces equivalent or better results.}

\paragraph{Static Application Security Testing (SAST)}
We use the term SAST to refer to techniques that use \emph{automated} tools to perform \emph{static} \emph{source-code} analysis. SAST tools are a common way to comprehensively apply source code analysis, as manual source code analysis can be tedious and time-consuming~\cite{mcgraw2006software,johnson2013don,scandariato2013static,smith2015questions,cruzes2017security}. 
In practice, SAST tools are less likely to be applied by security analysts~\cite{cruzes2017security,hafiz2016game,votipka2018hackers}. SAST and source code analysis is performed by the developers themselves~\cite{cruzes2017security}. \RRChange{In this study, we use three SAST tools commonly used in industry. One tool, Sonarqube (abbreviated as Sonar in tables) version 8.2, has an open-source platform available. We used the open-source community edition of Sonarqube version 8.2. The other tools examined, SAST-2 and SAST-3 (further abbreviated as SA-2 and SA-3 in tables) are proprietary tools and cannot be named due to license restrictions.  Sonarqube and SAST-2 were used to answer RQ1, while SAST-2 and SAST-3 were used to answer RQ2.}

All SAST tools used in this study perform static analysis by first parsing the source code to build a tree representation of the source code, known as a \emph{syntax tree}. The tool then applies a set of rules to the syntax tree, where each rule describes a pattern within the syntax tree that may indicate a vulnerability~\cite{mcgraw2006software}.  As with DAST, SAST tools have evolved to include additional features. For example, Sonarqube uses symbolic execution as well as a traditional rules engine to identify vulnerabilities\cite{Mallet2016SonarAnalyzer}. Similarly, both SAST tools used in this study employ taint analysis~\cite{Campbell2020TaintAnalysis}, although it is not clear whether this is available in the free / open source version of Sonarqube used in this study.

SAST tools can be setup according to different architectures. The SAST tools used in this study could be configured as client-server tools where the SUT code is scanned on the ``client'' machine, and information is sent to a ``server''. The analyst then reviews the results through the server. For some tools, the automated analysis and rules are applied on the client, while for other tools the automated analysis and rules are applied on the server. The SAST tools used in this study each included an optional plugin for Integrated Development Environments (IDEs) such as Eclipse\footnote{https://www.eclipse.org/ide/}. The plugin allows developers to initialize SAST analysis and in some cases view alerts from the tool within the IDE itself. Some tools can be run without a server using only IDE plugins. Other tools require a server. Similar to the previous work by Austin et al.~\cite{austin2011onetechniquenotenough,austin2013comparison}, we found that the server GUI was easier to use when aggregating and analyzing all system vulnerabilities for RQ1. Consequently, a client-server configuration was used with SAST-2 and Sonarqube to answer RQ1. SAST-2 and SAST-3 were more easily configured to use locally within an IDE using consistent rules, as was done in RQ2.

\section{System Under Test - OpenMRS}
\label{sec:method-sut-openmrs}
The SUT for our case study was OpenMRS, an open-source medical records system.  OpenMRS is a ``Java-based web application capable of running on laptops in small clinics or large servers for nation-wide use''\cite{OpenMRSDevManual}. 

\subsection{OpenMRS Overview}
OpenMRS contains 3,985,596 lines of code as measured by CLOC v1.74\footnote{\url{https://github.com/AlDanial/cloc}} including 476,139 coding lines, i.e. not comments, of Java  as well as 1,884,233 coding lines of Javascript.  The OpenMRS architecture is modular. The source code for each module is stored in a separate repository on github\footnote{\url{https://github.com/openmrs}}.  In this study, we examined the 43 modules that are in the basic reference application for OpenMRS Version 2.9.  We compiled and ran OpenMRS using Maven and Jetty as described in the Developer's Manual\cite{OpenMRSDevManual}.  

\subsection{Why OpenMRS?}\label{sec:why-openmrs}
\RRChange{We selected OpenMRS as the SUT first due to its domain. The three SUT examined by Austin et al~\cite{austin2011onetechniquenotenough,austin2013comparison} were  medical records systems. Hence the SUT for the current study should also come from the medical domain. Although OpenMRS was not examined by Austin et al., OpenMRS has also been used in other research on software testing and security analysis\cite{tondel2019collaborative,purkayastha2020continuous}. The security of medical records systems is, if anything, a more important issue in 2021 than in 2011, with healthcare systems being an increasingly popular target for hackers~\cite{NewZealandHealthcare2021,CISA2021Report,MarylandHealthDepartment2021,TexasHealthcareBreachDec21}.} 

\RRChange{Second, OpenMRS is a ``real'' system that is actively used and actively under development. The 2018 U.S. National Institute of Standards and Technology (NIST) Software Assurance Metrics and Tool Evaluation (SAMATE)  Static Analysis Tool Exposition (SATE) report~\cite{delaitre2018sate}, provides the following example criterion for ``real, existing software'': ``their development should follow industry practices. Their size should align with similar software. Their programming language should be widely used for their purpose.'' }As of July 2021\cite{OpenMRS2021Atlas}, OpenMRS is actively in use in many contexts from Non-Governmental Organizations (NGOs) operating clinics in Sierra Leone, Lesotho, Mexico, and Haiti to hospitals and health networks in Tanzania, Pakistan, and Bangladesh.\RRChange{OpenMRS follows common development practices for open-source systems, as discussed in their Developer Guide\cite{OpenMRSDevManual}. With over 3 million lines of code, OpenMRS is comparable to other modern medical records systems, such as the VistA system used by the US Department of Veteran Affairs\cite{VA2021} and Epic\cite{Epic2020MilkyWay}, which involve millions of lines of code. The languages and frameworks used by OpenMRS including Java, Javascript, Node.js, SQL, and CSS, consistently appear on lists of most commonly used software technologies such as the 2021 StackOverflow Developer Survey\cite{StackOverflow2021DevSurvey}.}

\RRChange{Furthermore, OpenMRS was being used by the graduate-level security course at North Carolina State University, which enabled us to collect sufficient efficiency data to perform a statistical comparison across techniques with little impact on the participants themselves, as outlined in our Institutional Review Board Protocol 20569. While studies such as SATE provide controlled comparisons of the effectiveness of a limited range of vulnerability detection techniques, we do not know of another study that provides perspective on the efficiency of these techniques in terms of Vulnerabilities per Hour (VpH).} 

\subsection{\textbf{Security Practices at OpenMRS}}\label{sec:openmrs-security}
\RRChange{OpenMRS is open-source software. The OpenMRS team has received vulnerability reports from both volunteers and independent researchers in the past, based on SAST and other vulnerability detection techniques. When we reached out to OpenMRS with our results, our understanding was that SAST and DAST tools were not being used at the organizational level. Since then, OpenMRS's security posture has continued to mature as they implement more vulnerability detection.}

% \RRChange{We have worked with OpenMRS in remediating the vulnerabilities found in this study. In Spring 2021 NCSU graduate students worked with OpenMRS in an effort to help fix vulnerabilities found in this study. Vulnerability remediation efforts have continued other other projects at OpenMRS.}

\section{Data Sources}\label{sec:dataSources}
The data that was analyzed came from two sources:  1) a team of five (5) researchers and 2) sixty-three (63) students from a graduate-level security course. In this section, we provide background on these sources necessary to understand how information from each source was used in our methodology

\subsection{Researcher Data}
\label{sec:method-authors}
Three PhD student researchers, one Master's student researcher, and one undergraduate student researcher applied SMPT, DAST, and SAST; and reviewed the outputs of all four techniques as part of data collection for RQ1. The researchers also reviewed all student information that was used in this study to remove incorrect answers. All graduate-level researchers had participated in the graduate-level software security course. The undergraduate student researcher had taken two security-related undergraduate courses. 

\subsection{Student Data}\label{sec:studentData}
Sixty-three of 70 students in a graduate-level software security course allowed their data to be used for this study by signing an informed consent form. Student data was collected following North Carolina State University (NCSU) Institutional Review Board Protocol 20569. Students worked in teams of 3-4 students, with a total of 19 teams in the class. Where data could only be aggregated at the team level, we use the data of the 13 teams in which all team members consented to the use of their data. Where data is available at the student level, we use data from all 63 students who consented to the use of their data. Student EMPT and SMPT data was used as part of the data collection process for RQ1. Additionally, researchers analyzed students' reported efficiency scores to answer RQ2. 

\subsubsection{Student Experience}
\begin{wrapfigure}[18]{r}{0.42\textwidth}
  \begin{center}
    \includegraphics[width=0.42\textwidth]{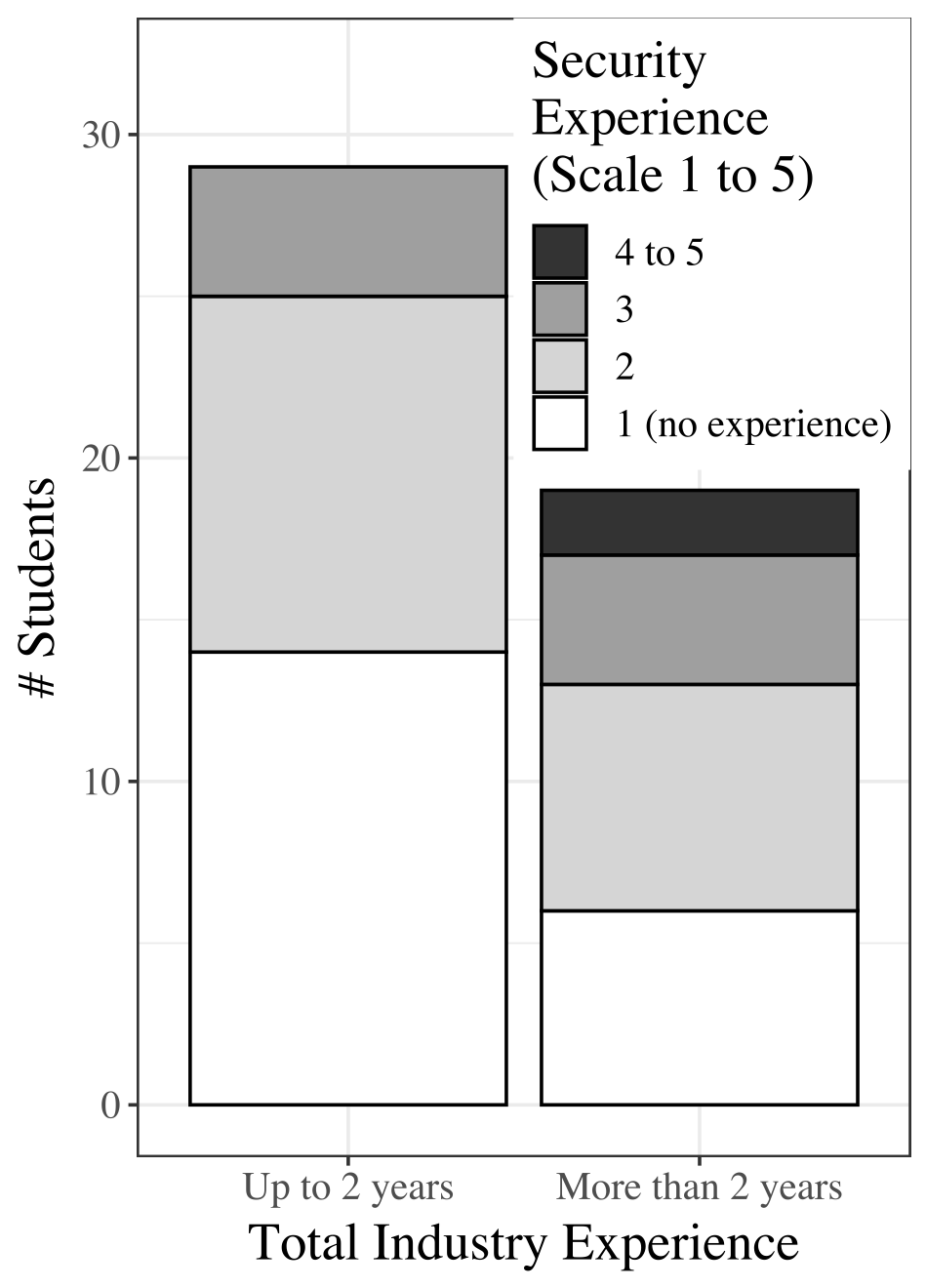}
  \end{center}
\caption{\centering{{Industry Security}\newline{Experience of Students}}}\label{fig:StudentExperience} 
\end{wrapfigure}

At the beginning of the course, students were asked to fill out a survey about their relevant experience. The full survey is available in Appendix \ref{app:ExperienceQuestions}. Fifty-five (55) of the 63 students whose data was used in this study provided valid survey responses.  Seven (7) of the 55 students had no industry experience. The median industry experience of students, including those with no experience, was 1 year. The average industry experience, including students with no experience, was 1 year 8 months.  Students were asked to note how much of their time in industry involved cybersecurity on a scale from 1 (none) to 5 (fully). The distribution of security experience for the 48 students with industry experience is shown in Figure \ref{fig:StudentExperience}. 
% \begin{wrapfigure}[23]{r}{0.42\textwidth}
The x-axis indicates students who had up to 2 years of industry experience as compared with more than 2 years of industry experience. The y-axis indicates the number of students. The shading within the bar chart indicates security experience on a scale from 1 to 5. Darker shades indicate higher security experience. Of the students with industry experience, the median value for security experience was 2, while the average was 1.85.  In addition to industry experience, 7 students had previously taken a course in security or privacy. Eight (8) students were currently taking a course in security or privacy in addition to the course from which the data was collected. Students had a diverse range of experience, but most students had little experience in cybersecurity at the start of the course.

\subsubsection{Course Assignments}\label{sec:studentAssignments}
Students were not required to perform any tasks that were not already part of their assignments for the course. The data used in this study that comes from student assignment responses is taken from the Course Project. The course project had four parts which were distributed over the duration of the semester. The verbatim text from the course project assignments is provided in Appendix \ref{app:Assignments}. A summary of the assignments that relate to our study is as follows:
\begin{itemize}
\item \textbf{SMPT Assignments:}
In the Project Part 1, students were required to write and execute a set of 15 systematic manual penetration test cases for the first assignment. Each test case mapped to at least one ASVS control. In Project Part 3, students were required to write and execute ten additional test cases for logging (ASVS V7 Levels 1 or 2), as well as to write and execute an additional five test cases to increase the ASVS coverage of their test suite. Correct, unique test cases and their results were used as part of the Data Collection for RQ1 (Effectiveness). The test cases were re-run by researchers and supplemented with additional test cases developed by researchers, as we will discuss further in Section~\ref{RQ1-DC-Apply-SMPT}. Student performance and experience with SMPT as part of these assignments also informed their response to the Comparison Assignment listed below, which was used to collect data for RQ2 (Efficiency) and RQ3 (Other Factors).
\item\textbf{EMPT Assignment:}
In Project Part 4, students spent three hours individually performing exploratory penetration testing. This activity occurred at the end of the course when students were familiar with the SUT and with many security concepts. Students produced a video recording of their three-hour session, speaking out loud about any vulnerabilities found; and created black box test cases to enable replication of each vulnerability found. The vulnerabilities found by students were used as part of the Data Collection for RQ1 (Effectiveness). Student performance and experience with EMPT informed their responses to the Comparison Assignment, which was used as part of the Data Collection for RQ2 (Efficiency) and RQ3 (Other Factors).
\item\textbf{DAST Assignment:} In Project Part 2 students used two DAST tools (OWASP ZAP and DAST-2), using 5 test cases from their SMPT assignments to provide the initial model for the DAST tool. Students reported the number of true positive vulnerabilities found and the amount of time required to review the output. Students' performance and experience with the DAST assignment contributed to their response to the Comparison Assignment, which in turn was used as part of the Data Collection for RQ2 (Efficiency) and RQ3 (Other Factors).
\item\textbf{SAST Assignment:} In Project Part 1, each student team ran two SAST tools (SAST-2 and SAST-3) on a subset of the SUT. Due to the length of many SAST reports, students were only required to review at least 10 of the reports from each tool to determine whether the alerts were true or false positives. Students had to identify at least 5 false positives even if it required reviewing more than 10 alerts to ensure that students were not incentivized to focus on trivial false positives. The students reported the number of true positive vulnerabilities found, as well as the amount of time required to review the alerts. The SAST assignment contributed to the students' response to the Comparison Assignment, which in turn was used as part of the Data Collection for RQ2 (Efficiency) and RQ3 (Other Factors).
\item\textbf{Comparison Assignment:}  At the end of the course in Part 3 and Part 4 of the project, each team created a table showing the number of vulnerabilities found by each activity, the amount of time it took to discover these vulnerabilities, and the resulting VpH. The students then reflected on their experience with the different vulnerability detection techniques in a free-response format. The numeric responses in the table were used to answer RQ2. Two researchers applied qualitative analysis to the students' free-response answers for RQ3.
\end{itemize}

\subsection{Overview of Data Sources Per RQ}
\RRChange{Table~\ref{tab:data-sources} provides an overview of where student data was used as part of Data Collection for each Research Question. Areas where student data was used are indicated with an S. Areas where researcher data was used are indicated with an R. All data analysis was performed by researchers, and is therefore not included in Table~\ref{tab:data-sources}. As can be seen in the table, RQ1 relied primarily on researcher efforts, although student data was used with SMPT and EMPT. Student data was used more extensively in RQ2, and was the source of the documents used in qualitative analysis for RQ3.The detailed methodology for each Research Question will be further explained in Sections \ref{sec:method-effective}, \ref{sec:method-efficiency}, and \ref{sec:method-qualitative}, respectively. }
\begingroup
\setcounter{savefootnote}{\value{footnote}}
\setcounter{footnote}{\value{tablefootnote}}
\renewcommand{\thefootnote}{\alph{footnote}}

% \begin{wrapfigure}[23]{r}{0.7\textwidth}

\begin{table}[htb] \renewcommand{\arraystretch}{1.3} \caption{Data Sources \newline {\tiny  R = Researchers; S = Students}} \label{tab:data-sources}
\centering
\begin{minipage}{\linewidth}\centering
{
\setcounter{footnote}{1}
\renewcommand{\thefootnote}{\alph{footnote}}
\renewcommand{\thempfootnote}{\alph{footnote}}

\begin{tabular}{ A{17pt}  A{78pt} || A{27pt} | A{27pt} | A{27pt} | A{27pt} |}
 &  &  \multicolumn{4}{ c|}{Technique}\\
\multicolumn{2}{ c||}{} & SMPT & EMPT & DAST & SAST \\\hline
\hline
\multirow{2}{=}{{RQ1}} & Applying Technique & {S\footnotemark[1] \& R} & {S\footnotemark[1]} & R & R\\\cline{2-6}
& Review Output & R & R & R & R \\\hline
\hline
\multirow{2}{=}{{RQ2}}& {Recording Efficiency} & {S} & {S} & {S\footnotemark[2]} & {S\footnotemark[2]}  \\\cline{2-6}
& {Data Cleaning} & {R} & {R} & {R} & {R}\\\hline
\hline
\multirow{2}{=}{{RQ3}} & {Document Source} & {S} & {S} & {S} & {S}\\\cline{2-6}
& Qualitative Coding & R & R & R & R \\\hline
\end{tabular}

\footnotetext[1]{\RRChange{Student results for RQ1 were reviewed by researchers for quality}}
\stepcounter{footnote}
\footnotetext[2]{\RRChange{For empirical comparison we base our answer on the recorded performance of Students. In Section \ref{sec:results-efficiency} we also include Researcher efficiency with automated tools for reference.}}
}
\end{minipage}
\end{table}

\setcounter{tablefootnote}{\value{footnote}}
\setcounter{footnote}{\value{savefootnote}}
\endgroup
\renewcommand{\thefootnote}{\arabic{footnote}}

\section{Methodology for RQ1 - Effectiveness}\label{sec:method-effective}
Our first research question is: \emph{\rqEffective} To answer RQ1, we need a comparable set of vulnerabilities found by each technique. Ensuring the vulnerability counts were comparable required an extensive Data Collection process described in Section~\ref{sec:RQ1-DataCollection} which is split into two phases. In the first phase of data collection in RQ1, \textit{Applying the Technique} we applied each vulnerability detection technique described in Section~\ref{sec:RQ1-1} to our SUT. The initial outputs of each technique are not comparable. For example, an analyst performing EMPT might document a single vulnerability where a malicious input script such as \verb|<script>alert(123)</script>| is saved in one part of the application due to an input validation vulnerability and executed by the application due to lack of output sanitization. A SAST tool, on the other hand may scan for input validation and output sanitization using different rules, resulting in two failing alerts for the same issue documented using EMPT. To reduce possible biases introduced by different vulnerability counting approaches or different vulnerability type classification approaches, we review the output of each technique in the second phase of Data Collection described in Section~\ref{sec:RQ1-2}. Once the data has been collected, we analyze the results as described in Section~\ref{sec:RQ1-3}.

\subsection{Data Collection}\label{sec:RQ1-DataCollection}
\RRChange{We begin this section with a set of guidelines which we will refer to throughout our description of the Data Collection process for RQ1. We then go into the details of how data collection was performed. As seen previously in Table~\ref{tab:data-sources}, we subdivide our Data Collection process for RQ1 into two phases. }

\RRChange{The first phase of Data Collection, \emph{Applying the Technique}, was accomplished through combination of student (S) and reviewer (R) work. We provide details of how each technique was applied in Section~\ref{sec:RQ1-1}. For DAST and SAST, a set of True / False Positive classification guidelines were required, which we highlight under General Guidelines in Section~\ref{sec:TFPositiveGuidelines}. }

\RRChange{The second phase of Data Collection, \emph{Reviewing the Technique Output}, was done entirely by researchers and is described for each technique in Section~\ref{sec:RQ1-2}. As part of this second phase, the researchers used the Counting, CWE Review, and Severity Guidelines  from Section~\ref{sec:rq1-guidelines}.}

\subsubsection{General Guidelines}\label{sec:rq1-guidelines}
\RRChange{This section provides key guidelines for the Data Collection process which we will refer to for the remainder of Section~\ref{sec:RQ1-DataCollection}. A common set of True / False Positive Classification guidelines is necessary for applying automated techniques (DAST and SAST), as described in Section~\ref{sec:VulnDetectionTechniques}. As part of Reviewing the Technique Output described in Section~\ref{sec:RQ1-2}, the output of each technique was reviewed to ensure that the vulnerability count, vulnerability type, and vulnerability severity are consistently evaluated using the guidelines provided in Sections~\ref{sec:methodCounting}, \ref{sec:cweReview}, and \ref{sec:Severity}}

\paragraph{True/False Positive Classification Guidelines}\label{sec:TFPositiveGuidelines}
A true positive alert or vulnerability is one that meets the the definition for a vulnerability from Section~\ref{sec:keyconcepts}.  We follow a conservative policy towards true- and false-positive classification based on the principle of Defense-in-Depth~\cite{NIST800-53R4}. We considered an alert or other finding to be a vulnerability if it could potentially lead to a security breach. For example, an alert is raised due to a particular malicious input. Upon review, we note that the input is stored in the database without encoding or other protection. We would classify the alert as a true positive even if we have not yet found another vulnerability where the malicious input is executed, e.g. by the application as part of an XSS attack. There may be vulnerabilities yet to be found that would result in the input being executed, and changes to the code could make the input validation vulnerability more exploitable in the future.
\paragraph{Counting Guidelines}\label{sec:methodCounting}
As mentioned in Section~\ref{sec:keyconcepts}, we used the CVE definition of a vulnerability. The CVE Program also provides a set of guidelines\cite{CVERules} to CVE Numbering Authorities (CNAs) for determining which vulnerability reports represent distinct true positive vulnerabilities. The guidelines help CNAs to identify and remove false positives, as well as to consolidate duplicate vulnerability reports. We based our counting process on the CVE Counting Rules\cite{CVERules} for determining the number of vulnerabilities identified using each technique\footnote{The CVE Counting rules have been updated since our original study. In future work, the authors may follow the updated rules: \url{https://cve.mitre.org/cve/cna/rules.html}}. The counting rules used in our analysis were:
\begin{itemize}
    \item  True/False Positive: The failure report must provide evidence of negative impact or that the security policy of the system is violated; and
    \item Independence: Each unique vulnerability must be independently fixable.
\end{itemize}

We applied these counting rules to the initial failures of each technique. For example, we applied the counting rules to the alerts produced by a tool. When one alert pointed to the same vulnerability as another alert,  we marked one of the alerts to be a ``duplicate'' of the other.

Where we were unsure of the independent fixability of different failures, we assumed that the failures represented independent vulnerabilities. For example, a vulnerability detection tool could raise two alerts for the same type of vulnerability, where each alert was triggered by a different checkbox in the same form. We counted each alert as a distinct vulnerability unless we knew that the checkboxes relied on the same server-side code. 

\RRChange{We consider an alert to be a ``True Positive'' even if the vulnerability found does not have the same CWE type as the original failure. CWE type is reviewed separately using the Vulnerability Type Guidelines in Section~\ref{sec:cweReview}.}

\paragraph{Vulnerability Type Guidelines}\label{sec:cweReview}
The vulnerability types assigned to each vulnerability are based on two systems - CWE and OWASP Top Ten - which are described in Section~\ref{sec:keyconcepts}. Each vulnerability was assigned a CWE based on the test case in SMPT, by the student who found the vulnerability in EMPT, or by the tool for DAST and SAST. The CWE values were reviewed as we discuss below for accuracy. The vulnerabilities were then mapped to the OWASP Top Ten (2021) categories based on the CWE values using the mapping provided by CWE\cite{CWEOWASP}.

Researchers reviewed the CWE type assigned to each vulnerability and corrected the CWE assignment when the assignment was missing, inaccurate, or inconsistent with the assignment for other vulnerabilities in our dataset.  For example, a DAST tool creates a malformed input (test case) designed to trigger XSS as described in Section~\ref{sec:DAST-Desc}. When the malformed input is executed against the SUT, it triggers an error message revealing sensitive information (CWE-209\footnote{\url{https://cwe.mitre.org/data/definitions/209.html}}). The error message is unexpected behavior which may result in the alert being flagged by the DAST tool. However, the alert will be assigned CWE-79 (XSS\footnote{\url{https://cwe.mitre.org/data/definitions/79.html}}) since that was the rule used to create the test case. In our study we would consider this alert to be ``true positive'' since the alert points to a vulnerability. However, the CWE type would need to be reclassified - i.e. we would consider the error message containing sensitive information to be a CWE-209 vulnerability even though the alert indicates CWE-79.  When a classification is incorrect, if there are already similar vulnerabilities in our dataset we reclassify the alert to the same CWE as the similar vulnerabilities. In the previous example of an alert reclassified as CWE-209, there were many CWE-209 vulnerabilities flagged by SMPT and EMPT prior to running the DAST tool. Otherwise, the analyst may need to perform a keyword search the CWE database~\cite{CWEmain} to find an appropriate CWE.

The CWE mapping to the OWASP Top Ten~\cite{CWEOWASP}, as well as general guidelines such as the list of ``Weaknesses for Simplified Mapping of Published Vulnerabilities''\cite{CWESimplifiedMapping} and relationships between CWEs provided by the CWE system~\cite{CWEMapping} were used to determine whether one CWE is more appropriate. For example, multiple students found issues where sensitive data was passed in the URI of a GET request, which results in the sensitive information being sent in cleartext that could be intercepted by a malicious actor. Some students associated these vulnerabilities with CWE-319 \textit{Cleartext Transmission of Sensitive Information}\footnote{\url{https://cwe.mitre.org/data/definitions/319.html}} while others used CWE-598 \textit{Use of GET Request Method With Sensitive Query Strings}\footnote{\url{https://cwe.mitre.org/data/definitions/598.html}}. CWE-319 maps to the OWASP Top Ten (2021) category \textit{A02 - Cryptographic Failures,} while CWE-598 maps to the OWASP Top Ten (2021)  category\textit{ A04:2021 - Insecure Design}. The vulnerability is not an issue with broken cryptography~\cite{OWASPGETUrl,PortSwiggerGETUrl}, and so only CWE-598 was assigned to the vulnerabilities where sensitive information was revealed in the URL of a GET request. On the other hand, when multiple CWEs were equally applicable, multiple CWEs could be assigned to the same vulnerability.

\RRChange{Fifty-seven (57) CWE types were found in our experiment. We mapped the vulnerabilities found to the OWASP Top Ten through their assigned CWE values. Mapping to the OWASP Top Ten provides a more readable summary of the types of vulnerabilities found, requiring 10 categories instead of 57.  The OWASP Top Ten, as described in Section~\ref{sec:keyconcepts}, categorizes and ranks vulnerability types based on how frequently they are seen in software systems and their severity, providing additional insight into the vulnerabilities found. }

\paragraph{\RRChange{Severity Guidelines}}\label{sec:Severity}
\RRChange{We use two different perspectives for examining the severity of the vulnerabilities found. Our first perspective on severity  is through the lens of the OWASP Top Ten.  As mentioned in Section~\ref{sec:keyconcepts} and reiterated in Section~\ref{sec:keyconcepts}, the categories of vulnerabilities included in the OWASP Top Ten are ranked based on how common the vulnerability is, and expert views on the severity of the vulnerability category. The OWASP Top Ten are ranked in a ``risk-based order'' suggesting that the first category of the OWASP Top Ten: A01 - Broken Access control, is considered a highest risk and is therefore more severe than vulnerabilities associated with lower-ranked categories.}

\RRChange{We also examine the severity based on the severity provided by tools, supplemented by analysis of high-frequency vulnerability types and discussions with OpenMRS. As discussed in Section \ref{sec:RQ1-1}, we excluded alerts that were labeled as insignificant or inconsequential by the tools themselves. In our results, we further split the vulnerabilities between those that are ``less severe'' and those that are ``more severe''. Vulnerabilities classified as ``less'' severe primarily include vulnerabilities where at least one tool indicated the vulnerability was of ``Low'' severity. Different tools have different labels for the different levels. We consider the lowest severity level for each tool to be ```Low''. Once vulnerabilities were detected, we reviewed ``more severe'' vulnerabilities but where more than 20 vulnerabilities associated with the same CWE were found by the same tool or technique. In our experience, if a tool or technique is flagging large amounts of vulnerabilities associated with the same vulnerability type, it is unlikely that those vulnerabilities are more severe. Additionally, large quantities of incorrectly classified vulnerabilities may skew the results. Additionally, we adjusted severity levels based on discussions with OpenMRS. Since the tool-based severity level was adjusted depending on the results themselves, we discuss the vulnerability types where severity was further reviewed in Section~\ref{sec:resultVulnSeverity}.}

\subsubsection{Applying the Technique}\label{sec:RQ1-1}
% \MyToDos{Improve Intro Paragraph}
Application of each technique is slightly different, as described in Section~\ref{sec:VulnDetectionTechniques}. We discuss how each technique was applied for our Case Study for SMPT in Section~\ref{RQ1-DC-Apply-SMPT}, for EMPT in Section~\ref{RQ1-DC-Apply-EMPT}, for DAST in Section~\ref{RQ1-DC-Apply-DAST}, and for SAST in Section~\ref{RQ1-DC-Apply-SAST}

\paragraph{SMPT}\label{RQ1-DC-Apply-SMPT}
As shown in Table \ref{tab:data-sources}, for applying SMPT as part of RQ1, we use data from both \emph{students} (S) and \emph{researchers} (R) who wrote and executed test cases based on the Open Web Application Security Project (OWASP) Application Security Verification Standard (ASVS)~\cite{OWASP2019ASVS} described in Section~\ref{sec:keyconcepts}.  While a healthcare application such as OpenMRS should strive for Level 2 or higher of the ASVS, as noted in the ASVS document itself ``\textit{Level 1 is the only level that is completely penetration testable using humans}''~\cite{OWASP2019ASVS}. In addition to the ASVS Level 1 controls, we used knowledge of the OpenMRS and documentation available on the OpenMRS wiki\footnote{https://wiki.openmrs.org/} to develop test cases specific to OpenMRS. We excluded 44 controls that were not applicable to the application. For example, the SUT did not include a mail server. ASVS control 5.2.3~\cite{OWASP2019ASVS} states ``\textit{Verify that the application sanitizes user input before passing to mail systems to protect against SMTP or IMAP injection}'' and hence is not applicable in our case study. 

\RRChange{In total, the test suite included 131 test cases which covered 63 of the remaining 87 controls. We used 86 test cases developed and executed by \emph{students} and 45 test cases developed and executed by \emph{researchers} to increase ASVS coverage. Students originally wrote over 390 test cases as part of their course assignments described in Section \ref{sec:studentAssignments}. However, many of the students' test cases were duplicates of each other since each of the 13 teams worked independently and generally wrote test cases for the easier security concepts. Additionally, test cases were removed due to quality concerns with the test case or the results recorded.  }
 
\RRChange{Each test case was executed by two independent analysts to reduce bias and inaccuracy due to human error and subjectivity.  For the 86 test cases used in this study that were developed by students, the first test case execution was performed by the students and the second execution was performed by researchers. For the 45 test cases developed by researchers, two different researchers each executed the test case one time. When the two executions of the test case produced different results, an additional researcher executed the test case and the result (pass or fail) given by two of the three test case executions was recorded as the actual result.}

\paragraph{EMPT}\label{RQ1-DC-Apply-EMPT}
\RRChange{For RQ1, EMPT was primarily applied by \emph{students} (S) as shown in Table \ref{tab:data-sources}, according to the assignment outlined in Section~\ref{sec:studentAssignments}. Students were required to spend three hours performing EMPT and record the results via video. Students documented their results as test cases to improve their replicability. An important factor in exploratory testing is ensuring that individuals have sufficient knowledge and experience~\cite{itkonen2013role,itkonen2014test,votipka2018hackers} since there is no formal process for the analysts to follow.  The students had limited security experience at the start of the course, as discussed in Section~\ref{sec:studentData}. However, our results suggest that the students had sufficient experience by the time of the EMPT assignment near the end of the course to be effective applying EMPT. One of the 63 students who agreed to participate in the study did not complete the EMPT assignment, and so data from 62 students was used as part of data collection for EMPT in RQ1. Since extensive review of the student results was needed, which we will discuss further in Section~\ref{method-rq1-dc-review-empt}, we distinguish between Student Reported Vulnerabilities (SRVs), and the final set of vulnerabilities found by EMPT used to answer RQ1.}

\paragraph{DAST}\label{RQ1-DC-Apply-DAST}
\RRChange{As shown in Table~\ref{tab:data-sources}, to collect DAST results for RQ1 researchers (R) setup and ran each tool using the default ruleset. \RRChange{The CWEs covered by each ruleset are shown in Table~\ref{tab:automated_cwe} of Appendix~\ref{app:AutomatedCWEs}.} As discussed in Section \ref{sec:DAST-Desc}, DAST tools require a set of application inputs, which are created by recording an analyst's interactions with a system. To better understand whether DAST tools would find the same vulnerabilities as SMPT and other techniques, the researchers recorded their interactions with the application as they performed actions based on six test cases from the SMPT suite. The test cases were selected based on their uniqueness, to maximize coverage of the SUT. Due to the complexity and resources required by the DAST-2 tool, we were unable to run additional test cases, a limitation discussed further in Section~\ref{sec:limitations}. Based on the interactions provided by analyst and default ruleset, each tool generated and tested a set of malformed inputs against the system.}

\RRChange{For OWASP ZAP, we manually explored the system through the ZAP proxy to record our interactions. We then ran the spider before running the security scan. Since OWASP ZAP was run after DAST-2, we limited the OWASP ZAP input to the six SMPT test cases used for DAST-2. With OWASP ZAP we were able to cover all six test cases in a single run of the tool, which took less than two hours to execute. }

\RRChange{DAST-2 was more resource-intensive and required significant configuration to run on our systems. For DAST-2, we recorded the interactions for each of the 6 SMPT test cases in an HTTP Archive file and uploaded it to the tool. Dozens or in some cases hundreds of http messages could be exchanged during the execution of a single SMPT test case. Unfortunately, the DAST-2 tool itself would eventually crash due to memory or processing constraints if more than approximately six HTTP messages were included in the initial model. Hence each test case had to be recorded separately, and used in a separate ``run'' of the DAST tool.  We removed all HTTP messages from the input sequence other than the messages that were key to the test case. To apply DAST-2 based on the interactions for the example test case in Figure \ref{fig:TestCaseExample},  test case we would include the HTTP messages for the GET request that loaded the initial OpenMRS login page in Step 01, the POST request which performed the login at step 04 containing the login details from steps 02-03, the GET request for the ``Add New Account'' page in Step 06, and the Post request for saving the user at step 11, containing the account information from steps 07-10. We would remove all other HTTP messages from the sequence, such as GET requests for CSS files for each page as well as intermediate pages such as the ``System Administration'' and ``Manage Accounts'' pages in step 05 which were not the target of the test case. For longer test cases such as the one shown in Figure \ref{fig:TestCaseExample}, we would start with a sequence of hundreds of requests, and remove all but 2-6 requests. We then used DAST-2 to generate malformed inputs based on the condensed set of requests and default rules. DAST-2 did not recommend using the entire set of malformed inputs that were generated, and enabled the analyst to run a randomized subset of the test cases. }

\RRChange{Two researchers performed multiple trials of the same tool in different configurations to determine how different choices would impact the alerts produced. For example, we applied the DAST-2 tool to the POST request from the login page with both a full set of test cases and a randomize subset to better understand the ramifications of using only a randomized subset of the test cases. We noted that due to the repetitive nature of the test cases generated by DAST-2, the alerts produced by the random subset were similar to the alerts produced by the full set, and the alerts produced by the full set did not seem to point to any additional vulnerabilities. The observation that vulnerability count may be stable is congruous with the results from Klees et al.~\cite{klees2018evaluating}, where the authors found that the unique bugs identified may be more consistent than the failures and crashes produced.  We then performed a final run for each test case recorded where we had all unnecessary HTTP messages from the recording and used a randomized subset of test cases to ensure that the DAST tool run for each test case was performed consistently. Seven separate models were setup and run for DAST-2 to cover the key http messages of six test cases selected from SMPT. The final run required between 2 hours and 3 days for each model.}

\RRChange{We reviewed the output taken from the last run for each tool for true and false positives using the guidelines in Section \ref{sec:TFPositiveGuidelines}. As noted by Klees et al.~\cite{klees2018evaluating}, fuzzing-based tools, such as DAST, can produce thousands of results. \RRChange{We excluded output where the tool indicated the alert is likely insignificant or inconsequential, for example with OWASP ZAP we exclude alerts where the Risk level noted by the alert was ``Informational''\cite{ZAPAlerts}. Unless otherwise noted, when we refer to the alerts from a DAST or SAST tool, we are excluding alerts marked as insignificant or inconsequential by the tools themselves.} For OWASP ZAP, two reviewers independently examined all alerts as was done with the SAST tools. It became apparent once DAST-2 produced over one thousand alerts, excluding insignificant alerts, that two researchers reviewing all alerts would be inefficient if coding could be done consistently. Consequently, researchers independently reviewed a random subset of the alerts.  For each of the seven models for DAST-2, two researchers reviewed at least 40 alerts individually to compare their agreement and determine whether continued review by two independent researchers was necessary for consistency. For models which produced less than 40 alerts, two researchers each reviewed every alert. }

\RRChange{For both tools, the researchers calculated their inter-rater reliability using Cohen's Kappa both manually and using R to determine if they were consistently classifying the results as true or false positive. For DAST-2 if the inter-rater reliability was at least 0.70\cite{lombard2002content,votipka2018hackers}, the reviewers then split the remaining alerts such that each reviewer only examined half of the remaining alerts. The inter-rater reliability for the classification of DAST alerts as true or false positive and the precision of the tools based on the final set of True/False Positives is reported in Section~\ref{sec:ToolPerformance}. }

\paragraph{SAST}\label{RQ1-DC-Apply-SAST}
Similar to applying DAST, as seen in Table~\ref{tab:data-sources}, applying SAST began with researchers (R) running the SAST tools on the SUT using the default security rules. \RRChange{The CWEs covered by each ruleset are shown in Table~\ref{tab:automated_cwe} of Appendix~\ref{app:AutomatedCWEs}.} SAST tools are deterministic, producing the same alerts for each run. \RRChange{Similar to DAST results, we excluded SAST alerts that the tool marked as insignificant or inconsequential such as Sonarqube's ``Security Hotspots'' which are used for ``security protections that have no direct impact on the overall application's security''\cite{SonarqubeSecurityRules}. Unless otherwise noted, when we refer to the alerts from a SAST or DAST tool, we are excluding alerts marked as insignificant or inconsequential by the tools themselves.} Since none of the SAST tools produced over 1000 results, the remaining alerts were reviewed by two researchers to identify and remove false positives using the guidelines in Section \ref{sec:TFPositiveGuidelines}. We computed Cohen's Kappa\cite{cohen1960coefficient} to determine whether the classification process was consistent and reproducible. A third researcher resolved disagreements.

\subsubsection{Reviewing the Technique Output}\label{sec:RQ1-2}
Each technique produces different outputs. For example, systematic, dynamic techniques such as SMPT and DAST produce a set of failing test cases. SAST, on the other hand, finds specific weaknesses in the codebase that should be changed to improve the security of the system. Multiple failing test cases may be due to the same weakness in the codebase, or a single failing test case may be due to multiple weaknesses in the codebase. Consequently, the raw count of the output may be higher or lower for one technique even though it is no more effective than another technique. To resolve these potential counting differences, we take the output of each technique, which we refer to as a list of ``failures'', and apply the Counting Rules described in Section~\ref{sec:methodCounting} to determine the number of vulnerabilities found by each technique. While most of the failures are already assigned a CWE type by the tool or by the ASVS control, the CWE type may not be correctly assigned. We also review the CWE assignments as part of reviewing the technique output. As shown in Table~\ref{tab:data-sources}, researchers (R) reviewed the output of all techniques.

\paragraph{SMPT}
\RRChange{Two researchers (R) independently reviewed all failing test cases from SMPT to determine how many vulnerabilities were found using the counting rules outlined in Section~\ref{sec:methodCounting}. The researchers discussed their differences with a third researcher, as needed to determine the final vulnerability count. The researchers also reviewed the CWE assigned to the vulnerabilities, as described in Section~\ref{sec:cweReview}. Each test case was linked to an ASVS control, which was associated with a CWE. However, a test case failure may have been due to a violation of a different security principle than the original CWE associated with the test case. Finally, after discussing the final set of vulnerabilities with OpenMRS, we separated ``less severe'' vulnerabilities from more critical vulnerabilities as discussed in  Section~\ref{sec:Severity}.}

\paragraph{EMPT}\label{method-rq1-dc-review-empt}
\RRChange{For EMPT, one researcher (R) reviewed each Student Reported Vulnerability (SRV) while a second researcher audited 100 randomly sampled SRV as well as 2 additional SRV at the request of the first reviewer. A third researcher performed additional auditing. The first reviewer examined each of the 484 SRV provided by students who participated in the study to determine if the SRV was reproducible. The researcher removed SRV if the researcher could not understand the students' documentation or if the researcher was unable to observe the result reported. The researcher determined if the SRV was a true positive using the True/False Positive guidelines described in Section~\ref{sec:method-effective}. While reviewing the student responses for reproducibility and removing false positives, researchers used the counting rules specified in Section~\ref{sec:methodCounting} to remove SRV that had already been reported by other students and to split SRV into multiple vulnerabilities when students did not follow the counting rules and reported multiple vulnerabilities as a single SRV. The researchers reviewed the CWE values assigned to each vulnerability, as described in Section~\ref{sec:cweReview}, removing inaccuracies due to typos and other errors. Finally, after discussing the final set of vulnerabilities with OpenMRS, we distinguished less severe vulnerabilities from more severe vulnerabilities following the guidelines in  Section~\ref{sec:Severity}.}

\RRChange{After reporting our results to OpenMRS, feedback from the OpenMRS team resulted in the removal of five additional EMPT vulnerabilities that were determined to be not reproducible. A team of Master's and Undergraduate students at NCSU working with OpenMRS to assist in fixing the vulnerabilities also provided feedback,  which resulted in consolidating three EMPT vulnerabilities. The three vulnerabilities consolidated had been found on three different pages of the application, in a shared search function.} 

\paragraph{DAST},
\RRChange{Once the true and false positive alerts were determined for each DAST tool,  two researchers (R) determined how many unique vulnerabilities were indicated by the alerts using the counting rules from Section~\ref{sec:methodCounting}. Researchers marked alerts that were triggered by the same vulnerability as ``duplicates'' of each other. The researchers assumed DAST alerts were distinct unless the malformed inputs triggering the alert shared similar characteristics such as the CWE type, URL, and targeted parameter. Research and experience were used to determine whether the alerts were unique or duplicate. If the researchers could not determine whether alerts were duplicates, they were assumed to be unique unless the alerts shared all three characteristics (CWE type, URL, and targeted parameter), in which case they were assumed to be duplicate. It is unlikely, for example, that the ``sessionLocation'' parameter for the ``/openmrs/login.htm'' URL shown in Figure~\ref{fig:DASTfuzzExample} would contain two distinct XSS vulnerabilities. Discussions of duplication and de-duplication continued in subsequent steps of the review.}

\RRChange{The CWE value of each vulnerability found by DAST was based on the CWE value of the alerts associated with that vulnerability. The CWE value of the alerts was assigned by the DAST tool(s). Researchers also reviewed the CWE values using the guidelines in Section~\ref{sec:cweReview}. The severity measures provided by the DAST tools were also used to distinguish less severe vulnerabilities from more severe vulnerabilities as described in Section~\ref{sec:Severity}.}

\paragraph{SAST}
\RRChange{Researchers (R) determined the number of distinct vulnerabilities indicated by the SAST alerts using the counting rules from Section~\ref{sec:methodCounting}. The researchers then reviewed the vulnerability CWE assignments provided by the SAST tools to ensure their accuracy following the guidelines from Section~\ref{sec:cweReview}. Additionally, severity measures provided by the SAST tools were used to distinguish less severe vulnerabilities from more severe vulnerabilities as described in Section~\ref{sec:Severity}. }

\subsection{Data Analysis}\label{sec:RQ1-3}
\RRChange{Once we had a comparable set of vulnerabilities, we calculate the number of vulnerabilities found by each technique for each type of vulnerability, using the CWE number and associated OWASP Top Ten as the type. Vulnerability count is commonly used in both academia and industry as a measure of security risk~\cite{morrison2018mapping}.  We use vulnerability counts and types to answer RQ1. }

\section{Methodology for RQ2 - Efficiency}\label{sec:method-efficiency}
\RRChange{For RQ2, we address the question \emph{\rqEfficiency}. To reduce the bias that could be introduced by a high-performing or low-performing participant, we cannot rely on results from a single individual or team~\cite{kirk2013experimentaldesign}. Using data from a graduate level security course worked well for three reasons. First, we have a wide participant pool. Second, the students are all required to perform exactly the same tasks, reducing the amount of external factors that could influence our results. Third, graduate students can be assumed to have some existing knowledge in computer science.}

\subsection{Data Collection}\label{sec:RQ2-1}
\RRChange{We collected efficiency scores recorded by students (S) as shown in Table~\ref{tab:data-sources} which we discuss in Section~\ref{sec:RQ2-DC-Recording}. We then needed to perform data cleaning, as we discuss in Section~\ref{sec:RQ2-DC-Clean} before the data could be analyzed.}

\subsubsection{Recording Efficiency}\label{sec:RQ2-DC-Recording}
\RRChange{To quantify efficiency, we started with information provided by the students (S) as shown in Table~\ref{tab:data-sources}. As discussed in Section~\ref{sec:studentData}, the students worked in Teams of 3-4 to apply SMPT, EMPT, DAST, and SAST to OpenMRS as part of their course project.  The students were given the assignments described in Section~\ref{sec:studentAssignments} which appear verbatim in Appendix~\ref{app:Assignments}. We do not know VpH at the student level for SMPT, DAST, and SAST since these assignments were only reported at the team level. However, students were allowed to work independently for EMPT and while some teams appear to have coordinated efforts within the team, e.g. team members seem to have looked at different parts of the system and only reported unique vulnerabilities, others did not, e.g. team members examined the same parts of the system and multiple team members reported the same vulnerability. Hence we use the average VpH across all participating members of each team for EMPT. For RQ2 we exclude data from students whose team members did not participate in the study as  as discussed in Section~\ref{sec:studentData}.}

\subsubsection{Data Cleaning}\label{sec:RQ2-DC-Clean}
\RRChange{Once we have collected the data, as shown in Table~\ref{tab:data-sources} the researchers (R) performed data cleaning as needed.} We used trimming\cite{wilcox2003modern,kitchenham2017robust} to remove outliers. We formally identified outliers for each technique using the median absolute deviation and median (MADN)\cite{wilcox2003modern,kitchenham2017robust}, applying MADN to the VpH scores for each technique. We removed outliers where the MADN was higher than 2.24, the threshold recommended in the literature\cite{wilcox2003modern,kitchenham2015evidence}. \RRChange{This data cleaning was needed to systematically identify and remove cases where students did not correctly follow the assignment to the extent that it impacted our analysis. For example, one team reported spending only 32 minutes on the SAST assignment described in Section~\ref{sec:studentAssignments} in contrast with the second-fastest team who spent 7.5 hours on the assignment. We detected and removed only four outliers, one in each technique.}

\subsection{Data Analysis}\label{sec:RQ2-2}
\RRChange{With outliers removed, we retained 12 efficiency scores for each technique.} We performed a statistical comparison to determine if the average efficiency across the groups for each technique was higher or lower than the average efficiency for other techniques. We first applied the Shapiro-Wilk test~\cite{razali2011power} to the data for each technique to assess normality. Based on the output of the Shapiro-Wilk test, we used Bartlett's test for homogeneity of variance~\cite{bartlett1937properties}. For our case study, the trimmed data was normal but the variance differed across techniques. We then apply the Games-Howell test~\cite{games1976pairwise,kirk2013experimentaldesign} to perform pairwise comparison across the different vulnerability detection techniques and determine which techniques were different. The Games-Howell test adjusts the p-value for multiple comparisons. 

\section{Methodology for RQ3 - Other Factors}\label{sec:method-qualitative}
\RRChange{Once the students had experience with the four vulnerability detection techniques}, the students (S) were asked to reflect on their experiences with each technique and to compare the techniques as part of the ``comparison assignment''  described in Section~\ref{sec:studentAssignments}. Students were instructed to discuss tradeoffs between the techniques, and `` Based upon your experience with these techniques, compare their ability to efficiently and effectively detect a wide range of types of exploitable vulnerabilities.'' as shown in Appendix~\ref{app:Assignments}. The student responses to the comparison assignment were the source documents for RQ3 as shown in Table~\ref{tab:data-sources}. The comparison assignment was answered at the team-level, and so the data used for RQ3 excludes students whose teammates did not agree to participate in the study. Two researchers (R) performed qualitative analysis on the student responses to understand what other factors may distinguish the different techniques. One researcher segmented the text by sentence, but left the sentences in order, to retain key contextual information. Both researchers independently coded each segment using ``open coding''\cite{corbin2008basics}. The researchers found that more than one code could apply to the same sentence. The researchers then compared and discussed their results. One researcher further standardized the codes and determined which codes were mentioned by more than one response. The resulting information was primarily used to understand the results of RQ1 and RQ2, but may also be informative for future work.

\section{Equipment}\label{sec:method-sut-equipment}
We faced several equipment constraints. OpenMRS could be run with relatively low resources such as CPU, memory, and disk space. However, the tools used for SAST and DAST were more resource-intensive. Additionally, for the course from which we collected student data, all 70 students needed independent access to the SUT as part of their coursework, such that students would not interfere with each others' work as they attempted to hack into the SUT. The equipment used to apply each technique to answer RQ2 and RQ3 was the same for all techniques. We used the Virtual Computing Lab\footnote{\url{https://vcl.apache.org}} (VCL) at our university, North Carolina State University (NCSU)\footnote{\url{https://vcl.ncsu.edu}}. VCL provided virtual machine (VM) instances. Researchers created a system image including the SUT (OpenMRS) as well as SAST and DAST tools. An instance of the image could be checked out by students or researchers and accessed remotely. Additional resources were needed when answering RQ1 to improve system coverage, including larger VCL instances, Virtualbox VMs based on the VCL images, and a large desktop machine with 24 CPUs, 32G RAM, and 500G disk space. Additional information on the systems is in Appendix~\ref{app:Equipment}.

\section{Results}
\label{sec:results}
 In this section, we describe our results. Table \ref{tab:summary} provides a high-level summary of our numeric results for RQ1 - \emph{\rqEffective} and RQ2 - \emph{\rqEfficiency}. Detailed results for RQ1 and RQ2 are provided in Sections \ref{sec:results-effectiveness} and \ref{sec:results-efficiency} respectively. We provide our qualitative results for RQ3 - \emph{\rqQualitative} in Section~\ref{sec:results-qualitative}.

\begingroup
\setcounter{savefootnote}{\value{footnote}}
\setcounter{footnote}{\value{tablefootnote}}
\renewcommand{\thefootnote}{\alph{footnote}}

\begin{table}[htb] \renewcommand{\arraystretch}{1.3} \caption{Results Summary} \label{tab:summary}
\centering
\begin{minipage}{\linewidth}\centering
\setcounter{footnote}{3}
\renewcommand{\thefootnote}{\alph{footnote}}
\renewcommand{\thempfootnote}{\alph{footnote}}
\begin{tabular}{ m{110pt} ||| A{40pt} | A{40pt} |  A{40pt} | A{40pt}  }
 & SMPT & EMPT & DAST & SAST\\\hline
\hline\hline
{\textit{Effectiveness}:\newline Vulnerabilities found} & 37 & 185 & 23 & 823 \\\hline
{\textit{Effectiveness}:\newline \# OWASP Top Ten Covered\footnotemark[3]} & {9}  & 7 & {7} & {7} \\\hline
{\textit{Efficiency}:\newline Average VpH} & 0.69\footnotemark[4] & 2.22 & 0.55\footnotemark[4] &1.17  \\
\end{tabular}
\footnotetext[3]{One category within the OWASP Top Ten is outside the scope of the tools in this study. Maximum possible coverage is 9 out of 10.}
\stepcounter{footnote}
\footnotetext[4]{The difference in efficiency between SMPT and DAST is not statistically significant.}

\end{minipage}
\end{table}

\setcounter{tablefootnote}{\value{footnote}}
\setcounter{footnote}{\value{savefootnote}}
\endgroup
\renewcommand{\thefootnote}{\arabic{footnote}}

\subsection{RQ1 - Technique Effectiveness}\label{sec:results-effectiveness}
In this section, we discuss the results for our question \textit{\rqEffective} First, we include information specific to tool-based techniques: the agreement of researchers when reviewing the output of vulnerability detection tools, and the tools' false positive rates. We then provide the number of vulnerabilities discovered by each technique, as well as the types of vulnerabilities identified by each technique as classified using the CWE.

\subsubsection{\RRChange{True and False Positive Tool Alerts}}\label{sec:ToolPerformance}
We examined two tool-based techniques in this study, SAST and DAST, for which we can calculate the precision of the tools. As noted in Section~\ref{sec:RQ1-1}, for tool-based techniques, two individuals classified each alert produced by the tool as true or false positive. We calculated the inter-rater reliability and present the results in Section~\ref{sec:agreement}. The reviewers discussed the results with a third reviewer, who assisted in resolving disagreements, to create a set of true and false positive determinations which could be used to determine the number of false positives and tool precision presented in Section~\ref{sec:precision} and shown in Table~\ref{tab:AllAlerts}. 

\paragraph{Reviewer Agreement}\label{sec:agreement}
 We calculate the inter-rater reliability of the reviewers for SAST and DAST using Cohen's Kappa~\cite{cohen1960coefficient}. Inter-rater reliability measures how much two reviewers agree. Statistics such as Cohen's Kappa also account for agreement due to chance, measuring the extent to which reviewers agree beyond whatever agreement would be expected due to chance. In a classification of two ratings such as true and false positive, if one of the ratings applies to an extremely high percentage of cases (e.g. 98\%) while the other rating applies to an extremely small percentage of cases (e.g. 2\%), the probability of agreement due to chance is estimated to be very high. The high estimated probability of agreement can lead to a paradox where reviewers who have high observed agreement, in other words - they apply the same rating to most of the objects being rated, but have low inter-rater reliability~\cite{feinstein1990high,cicchetti1990high,feng2013factors}. 
 
  We observe the paradox of high observed agreement but low inter-rater reliability for Sonarqube and SAST-2. Of the 698 alerts for Sonarqube, we calculate Cohen's Kappa on 693 alerts that were independently reviewed. One of the two reviewers found 12 of the 693 reports to be false positives, while the other reviewer did not consider any of the reports to be false positives. A third researcher reviewed the results and resolved disagreements for a final set of 4 false positives and 694 true positives out of the original 698 alerts. Based on the true/false positive classifications of the first two reviewers, the expected agreement, as estimated when calculating Cohen's Kappa, is 98.3\% which is identical to the observed agreement of 98.3\% with a resulting Cohen's Kappa of 0 (95\% confidence interval $\pm 0$). Similarly, the Cohen's Kappa for SAST-2 is 0.22 (95\% confidence interval $\pm 0.40$), in spite of a high observed agreement  of 93.1\%.  For SAST-2, the two reviewers met to discuss and resolve disagreements, while the third researcher participated in the discussion with a final false positive count of 16 out of 264 total alerts. These Kappa scores are low. However, given that our final false positive count for Sonarqube was only 4 of 698 total alerts and our final false positive count for SAST-2 was 16 of 264 alerts, even with dozens of reviewers, we may not be able to increase the inter-rater reliability statistics. Observed agreement between reviewers may never be above the expected agreement, considering that the expected agreement for Sonarqube was 98.3\%.
  
 Another way to examine the paradox is to consider that the reviewers agreed with the tool as frequently as they agreed with each other. While the reviewers had low inter-rater agreement as analyzed using the Cohen's Kappa statistic, they had high agreement in terms of the percentage of alerts on which they agreed upon the classification, with 98.3\% agreement for Sonarqube and 93.1\% for SAST-2. In both cases, the reviewer's observed agreement with the tool was as high with their agreement with each other, with observed agreement for each of the Sonarqube reviewers and the tool at 98.3\% and 100\%. Observed agreement for each of the SAST-2 reviewers and the tool was  at 94.3\% and 96.6\%. 
 
  The agreement for DAST tools was much less complicated. Both researchers reviewed all alerts from OWASP ZAP, since there were fewer alerts than for DAST-2. The inter-rater reliability for OWASP ZAP alerts was 0.97 (95\% confidence interval $\pm 0.28$). The output of OWASP ZAP was reviewed after the output of DAST-2, and the researchers may have been more familiar with the process. The inter-rater reliability for the 288 DAST-2 alerts reviewed by two individuals was 0.78 (95\% confidence interval $\pm 0.16$), which was above the recommended minimum cutoff of 0.70~\cite{lombard2002content,votipka2018hackers}. Only one researcher reviewed the remaining alerts, as described in Section~\ref{sec:RQ1-1}.

\paragraph{True / False Positives and Precision}\label{sec:precision}
 Table \ref{tab:AllAlerts} shows the Total Alerts (Tot. Alrt.), False Positives (FP), and Precision (Prec.) for all tools in the current study, where the SUT was OpenMRS (M). Subtable \ref{tab:AllAlerts}.a provides the Tot. Alrt., FP, and Prec. for DAST, while Subtable \ref{tab:AllAlerts}.b provides the Tot. Alrt., FP, and Prec. for SAST. In each subtable, the Total (M) column for Tot. Alrt. and for FP is the sum of the alerts and false positives from both tools applied to OpenMRS in this study. The Prec. row of the Total (M) column is the precision calculated based on all alerts from the two tools applied to OpenMRS for each technique. Table \ref{tab:AllAlerts} also shows the Total Alerts (Tot. Alrt.), False Positives (FP), and Precision (Prec.) from the work by Austin et al.~\cite{austin2011onetechniquenotenough,austin2013comparison}, as we will describe Section~\ref{sec:ToolAustinComp}.
 
The precision of most SAST and DAST vulnerability detection tools in the current study was relatively high. The precision of Sonarqube and SAST-2 was 0.99 and 0.94, respectively, and the precision across the combined alerts for Sonarqube and SAST-2 at 0.98.  The precision of OWASP ZAP was also high at 0.95. However, the precision of DAST-2 was 0.09, resulting in  0.23  precision across all DAST alerts.

 \begin{table}[!htb]
    \caption{\centering{Total Alerts (Tot. Alrt.), False Positives (FP), and Precision (Prec) for the Current Study Compared with Austin et al.}\newline{\tiny M indicates OpenMRS, E indicates OpenEMR, T indicates Tolven, and P indicates PatientOS}}\label{tab:AllAlerts}
        \begin{subtable}{.475\linewidth}\renewcommand{\arraystretch}{1.3}\label{tab:AADAST}
      %\centering
      \raggedright
        \caption{DAST}
        \begin{tabular}{ A{19pt} |||   S{21pt} |  A{21pt} | A{21pt}  ||  S{21.5pt} | S{21.5pt} |}
            &   \multicolumn{3}{ c||}{Current Study} & \multicolumn{2}{ c |}{Austin et al.}\\
             &{Total (M)} & ZAP (M) & DA-2 (M) & {Total (E)} & {Total (T)}\\\hline
             \hline
             {Tot. Alrt.} &  3414 & 550 & 2862 & 735 & 37 \\\hline
             {FP} & 2612 & 28 & 2597 & 25&15 \\\hline
            \hline
            {Prec.} & 0.23 & 0.95 & 0.09 & 0.97 & 0.59 \\\hline
        \end{tabular}
    \end{subtable} 
    \begin{subtable}{.52\linewidth}\renewcommand{\arraystretch}{1.3}\label{tab:AASAST}
      %\centering
      \raggedleft
        \caption{SAST}
        \begin{tabular}{ A{19pt} |||   S{21pt} |  A{21pt} | A{21pt}  ||  S{21pt} | S{21pt} | S{21pt} |}
            %\cline{2-7}
            &   \multicolumn{3}{ c||}{Current Study} & \multicolumn{3}{c|}{Austin et al.}\\
            &{Total (M)} & Sonar (M) & SA-2 (M) & {Total (E)} & {Total (T)}& {Total (P)}\\\hline
            \hline
            {Tot. Alrt.} & 962 & 698 & 264 & 5036& 2315 & 1789\\\hline
            {FP} & 20 & 4 & 16  & 3715 & 2265 &1644\\\hline
            \hline
            {Prec.} & 0.98 & 0.99 & 0.94  & 0.26& 0.02 & 0.08 \\\hline
        \end{tabular}
    \end{subtable}%
\end{table}

We examined possible reasons for the low precision of DAST-2. Table \ref{tab:DASTAlerts} shows the DAST alert counts for each CWE type originally assigned by the tool based on the test case or check that triggered the alert. In Table \ref{tab:DASTAlerts} we use the abbreviation Tot. Alrt. for total alerts, TP Alrt. for true positive alerts, FP Alrt. for false positive alerts, and \# Vuln. for number of vulnerabilities. 

More than one alert can point to the same vulnerability as discussed in Section~\ref{sec:RQ1-2}. For example, as can be seen in Table\ref{tab:DASTAlerts},  two (2) true positive alerts (TP Alrt.) found by DAST-2 were assigned by the tool to CWE-89 \textit{SQL Injection}\footnote{\url{https://cwe.mitre.org/data/definitions/89.html}}. One of the TP alerts assigned to CWE-89 was a ``duplicate'' of the other, pointing to the same vulnerability; resulting in a vulnerability count of 1 in the \# Vuln column for alerts with Tool-Assigned CWE-89.

When an alert correctly provided an indicator of a unique vulnerability, such as the malformed input from a DAST tool used to trigger the alert, the CWE provided by the tool did not always match the type of vulnerability found. If the CWE type assigned by the tool was not the same as the type of vulnerability found, we classified the alert as True Positive and reassigned the CWE type as described in Section \ref{sec:cweReview}. The Tool-Assigned CWE, shown in the first column of Table \ref{tab:DASTAlerts} is the CWE value provided by the tool. The final CWE types of the vulnerabilities found, reviewed and reassigned if necessary, are listed in the ``Final CWE'' column. Of the alerts originally assigned to CWE-89 \textit{SQL Injection} by DAST-2, the researchers determined that the alerts revealed an http message where sensitive patient information was visible in the URL. However, the researchers could not perform SQL injection with  the test case corresponding to the alert. Consequently, the vulnerability was reassigned  CWE-598 \textit{Use of GET Request Method With Sensitive Query Strings} 

An alert could also be a duplicate of another alert with a different tool-assigned CWE. For example, one of the Tool-Assigned CWE-352 alerts was a duplicate of an alert whose Tool-Assigned CWE was 77, but where the researchers agreed that CWE-20 \textit{Improper Input Validation}\footnote{\url{https://cwe.mitre.org/data/definitions/20.html}} was more accurate. The second TP tool-assigned CWE-352 alert was a duplicate of a CWE-79 \textit{Cross Site Scripting (XSS)} alert with Final CWE-79. Since both alerts were duplicates of alerts with other Tool-Assigned CWEs,  the \# Vuln column in Table \ref{tab:DASTAlerts} for DAST-2 Tool-Assigned CWE-352 alerts is 0, even though there were 2 TP alerts. For rows where there were no TP alerts and therefore no vulnerabilities, we leave the \# Vuln column blank, consistent with other tables.

The Tool-Assigned CWE was linked to the rules used to create the malformed input or test case that triggered each alert. Of the Tool-Assigned CWEs in Table \ref{tab:DASTAlerts}, only CWE-16 and CWE-79 were associated with more than one rule or check by OWASP ZAP. CWE-16 was associated with 3 rules by OWASP ZAP, while CWE-79 was associated with two distinct rules. As we can see in Table \ref{tab:DASTAlerts}, for OWASP ZAP, the alerts that had been assigned a given CWE were either all true positive or all false positive, except for CWE-79. Within CWE-79, the 4 true positive alerts were all from one of the two rules, while the 3 False Positive alerts were all associated with the other rule. For every DAST-2 rule, either all alerts were true positive or all alerts were false positive.

All Tool-Assigned CWEs in Table \ref{tab:DASTAlerts} were associated with only one rule or check by DAST-2. We can see that the most frequently occurring CWE type, and therefore rules used to create the malformed inputs, is CWE-35 \textit{Path Traversal} which accounted for 2537 of all DAST-2 alerts. 2484 of the Path Traversal alerts are false positives, and the remaining 53 alerts were all reclassified as other CWE types. If we exclude alerts for CWE-35, the precision of DAST-2 goes from 0.09 to 0.69. While still lower than the precision for the other tools, the improved precision without Tool-Assigned CWE-35 alerts suggests that further customization such as updating or removing rules that do not accurately model the SUT may be able to improve the performance of DAST-2.

\begin{table}[!htb] \renewcommand{\arraystretch}{1.2} \caption{DAST Alerts} \label{tab:DASTAlerts}
\centering
\begin{tabular}{ >{\raggedright\arraybackslash} m{95pt} ||| S{19.5pt}  |  A{17.5pt} |  A{17.5pt} ||  A{21pt} ||  A{19.5pt} ||| S{19.5pt}|  A{17.5pt}|  A{17.5pt} ||  A{25pt} ||  A{19.5pt}}
 &   \multicolumn{5}{ c|||}{OWASP ZAP} & \multicolumn{5}{c}{DAST-2}\\%\cline{2-7} 
{Tool-Assigned CWE}& {Tot. Alrt.} & TP Alrt. &  FP Alrt. & Final CWE &\# Vuln. & {Tot. Alrt.} & TP Alrt. & FP Alrt. & Final CWE &\# Vuln. \\\hline\hline
 Total & 550 & 522 & 28 & N/A & 12 & 2862 & 265 & 2597 & N/A & 14 \\\hline
 \hline
16 - Configuration & 262 & 262 &  & 16 & 3 &  & & &  \\\hline
35 - Path Traversal&  &  &  &  &  & 2537 & 53 & 2484 &\tiny{79, 209, 613}& 1\\\hline
77 - Command Injection&  &  &  &  &  & 2 & 1 & 1 & {20} & 1 \\\hline
79 - Cross-site Scripting & 7 & 4 & 3 & 79 & 3 & 307 & 207 & 100 & \tiny{79, 20, 209, 598} & 10\\\hline
89 - SQL Injection &  &  & &  &  & 8 & 2 & 6 & {598} & 1 \\\hline
120 - Buffer Overflow & 21 &  & 21 &  &  & &  &  &  \\\hline
134 - Use of Externally- Ctrl. Format String & 1 &  & 1 &  &  & & &   &  & \\\hline
200 - Exposure of Sensitive Info. to an Unauth. Actor & 68 & 68 & & {7, 548} & 1 &  &  &  & & \\\hline
326 - Inadequate Encryption Strength & 14 & 14 &  & 326 & 1 &  &  &  &  & \\\hline
345 - Insuf. Verification of Data Authenticity & 11 & 11 &  & 345 &  1 &  &  &  & & \\\hline
352 - Cross-Site Req. Forgery (CSRF) & 51 & 51 &  & 352 & 1  & 8 & 2 & 6 &  {20, 79} &  0 \\\hline
472 - External Ctrl. of Assumed-Immutable Web Param. & 3 &  & 3 & & &  &  &  & \\\hline
548 - Exposure of Info. Through Dir. Listing & 1 & 1 &  & 548 & 1 &  &  &  & & \\\hline
933 - Security Misconfiguration & 111 & 111 &  & 933 & 1 &  &  &  &  & \\
\end{tabular}
\end{table}

\paragraph{Tool \RRChange{False Positives and Precision} Comparison with Austin et al.}\label{sec:ToolAustinComp}
Table \ref{tab:AllAlerts} also shows the tool \RRChange{precision} reported by Austin et al. for comparison with the current study. As described in Section~\ref{sec:AustinStudy}, the SUT examined by Austin et al. were OpenEMR (E), Tolven eCHR (T), and PatientOS (P). Austin et al. used a single tool for each technique. PatientOS is not included in the DAST results since the DAST tool used by Austin et al. was not applicable to PatientOS. 

As seen in Table \ref{tab:AllAlerts}, the SAST tool applied to the SUT in the previous study had much lower precision than the tools and SUT examined in the current study. The highest precision in the previous study for SAST was 0.26, as compared with the lowest precision of 0.94 in the current study. The high precision we observed may be part of greater trends in SAST tools as seen in the recent NIST SAMATE project's regular Static Analysis Tool Expositions (SATE)~\cite{delaitre2018sate} where the precision and recall of SAST tools for Java was far higher than the precision and recall reported in previous work~\cite{austin2011onetechniquenotenough,austin2013comparison}.  

Austin et al. had similar results to our current study using DAST. In Austin et al.'s work, the DAST tool when applied on OpenEMR had a precision of 0.97. When applied to Tolven eCHR, the DAST tool used by Austin et al. only had a precision of 0.59. Austin et al.~\cite{austin2011onetechniquenotenough,austin2013comparison} also found that entire categories of alerts could be labeled true or false positive, similar to the results shown in Table \ref{tab:DASTAlerts}. The impact of not customizing tool rules on performance measures such as precision may be more apparent as tools become more advanced and precise. 

In the current section (\ref{sec:ToolAustinComp}) we have compared our work to Austin et al. on tool-based measures. Comparison with Austin et al. on effectiveness measures applicable to all four techniques may be found in Section~\ref{sec:effect-austin}. Comparison with Austin et al. on efficiency measures may be found in Section~\ref{sec:resultsEfficiencyAustin}.

\subsubsection{Number of Vulnerabilities}\label{sec:resultVulnCount}
Overall, SAST found the most vulnerabilities, at 824 vulnerabilities. EMPT found the second most vulnerabilities, with 229 vulnerabilities. We provide further information on the number of vulnerabilities found using each technique and tool in Table \ref{tab:VulnCounting}. The main results for each technique are shaded gray, while white-shaded columns indicate the results for each of the DAST and SAST tools independently. 

In the first row of Table \ref{tab:VulnCounting}, we provide the number of ``True Positive (TP) Failures''. For SMPT, these are failing test cases; for DAST and SAST the ``True Positive Failures'' are true positive alerts.   We do not have a true positive failure count for EMPT comparable to the failing test cases from SMPT or true positive alerts from DAST and SAST. Unlike SMPT where failing test cases could be assumed to be true positive since poorly written test cases had been removed, EMPT results required additional quality review. We mark the number of true positive failures for EMPT to be Not Applicable (N/A) in Table \ref{tab:VulnCounting}. The ``Total'' column for DAST and SAST ``True Positive Failures'' is the sum of all true positive alerts for the technique.

The second row of Table \ref{tab:VulnCounting} shows the total number of vulnerabilities indicated by the failures, while the third and fourth rows show how many of these vulnerabilities were more severe and less severe, respectively. The vulnerability counts are determined by applying our counting rules described in Section~\ref{sec:methodCounting}, with severity assigned according to the guidelines in Section \ref{sec:Severity}. The same vulnerability could be found by both SAST tools or both DAST tools. Hence the ``Total'' \textit{column} for SAST and DAST vulnerabilities is the total vulnerabilities found by the technique and \textit{not} simply the sum of the vulnerability counts for each tool. However, a vulnerability cannot be both ``more severe'' and ``less severe'' so the ``Total Vulnerabilities'' row is the sum of the more severe and less severe vulnerabilities. We found the most vulnerabilities using SAST with 824 total vulnerabilities, followed by EMPT with 229 total vulnerabilities. We found 38 vulnerabilities using SMPT, and 25 vulnerabilities using DAST.

The fifth row of Table \ref{tab:VulnCounting} is the ratio between the number of TP Failures and the total number of vulnerabilities (row 2). The ratio of TP Failures to Vulnerabilities varies across techniques, but is higher for DAST (32.08) than for SMPT (1.58) and SAST (1.12). We discuss the implications of this ratio in Section~\ref{sec:discussion}.

The sixth row of of Table \ref{tab:VulnCounting}, labeled ``Vuln. Unique to Tech./Tool'', shows the number of vulnerabilities that were only found by each technique or tool. It may be helpful to consider the ``Vuln. Unique to Technique/Tool'' as ``the number of vulnerabilities we would have missed if we had not used this technique or tool''. Similar to the Total Vulnerabilities for SAST and DAST in row 2, the Total columns for SAST and DAST  in row 3 indicate the count of  vulnerabilities found only by the technique. One (1) vulnerability was found by all techniques, including both DAST tools and one of the two SAST tools. Specifically, the fact that our instance of OpenMRS was configured such that the default server errors, e.g. 500 errors, revealed sensitive information about the system, which was associated with CWE-7 \textit{J2EE Misconfiguration: Missing Custom Error Page}\footnote{\url{https://cwe.mitre.org/data/definitions/7.html}}. Ten (10)  vulnerabilities were found by both of the SAST tools, but by no other technique. The 10 vulnerabilities found by both tools are included in the 822 vulnerabilities that were only found by SAST, but not in the vulnerabilities only found by Sonarqube and SAST-2 independently. Similarly, 1 vulnerability was found by both DAST tools but not by other techniques. The vulnerability found only by both DAST tools is included in the Total Vuln. Unique to DAST, but not the Vuln Unique to OWASP ZAP or DAST-2 independently.  Each technique or tool found vulnerabilities that were not found using other techniques and tools.% 

\begin{table}[htb]  \caption{Vulnerability Counts} \label{tab:VulnCounting} \renewcommand{\arraystretch}{1.6}
%\vspace*{2mm}
\centering
\begin{tabular}{ >{\raggedleft\arraybackslash} A{65pt}| A{50pt} ||| S{25.5pt} || S{25.5pt} || S{26.5pt} | A{20pt} | A{20pt} || S{26.5pt} |  A{21pt} | A{20pt}  }
\multicolumn{2}{c |||}{ } & SMPT & EMPT & DAST (Total) & ZAP & DA-2 & SAST (Total) & Sonar & SA-2  \\
\hline\hline
\multicolumn{2}{c |||}{True Positive (TP) Failures} & 60 & N/A & 787 & 522 & 265 & 948 & 694 & 254\\\hline
% \hline
\multicolumn{2}{c |||}{Total Vulnerabilities} & {37} & {185} & {23} & {12} & {13} & {823} & {598} & {235}\\\hline
% & {Total} & {37} & {185} & {23} & {12} & {13} & {823} & {598} & {235}\\\hhline{~|-|||-||-||-|-|-||-|-|-}
% & \RRChange{\textit{More Severe}} & \RRChange{\textit{32}} & \RRChange{\textit{165}} & \RRChange{\textit{16}} & \RRChange{\textit{6}} & \RRChange{\textit{11}} &  \RRChange{\textit{375}} & \RRChange{\textit{239}} & \RRChange{\textit{143}}\\\hhline{~|-|||-||-||-|-|-||-|-|-}
% \multirow{-3}{=}{\# Vulnerabilities} & \RRChange{\textit{Less Severe}} & \RRChange{\textit{5}} & \RRChange{\textit{20}} & \RRChange{\textit{7}} & \RRChange{\textit{6}} & \RRChange{\textit{2}} & \RRChange{\textit{448}} & \RRChange{\textit{359}} & \RRChange{\textit{92}} \\\hline
% \hline
\multicolumn{2}{c |||}{Ratio: $\frac{TP~Failures}{Total Vulnerabilities}$} & $1.58$ & N/A & {$34.22$} & $43.50$ & $20.38$ & $1.12$ & $1.16$ & $1.05$\\\hline
% \hline
\multicolumn{2}{c |||}{Vuln. Unique to Tech./Tool} & 11 & 157 & 13 & 8 & 4 & 822 & 588 & 225\\
\end{tabular}
\end{table}

\subsubsection{Vulnerability Severity}\label{sec:resultVulnSeverity}

As discussed in Section~\ref{sec:Severity}, we reviewed more severe vulnerabilities where the same tool or technique found greater than 20 vulnerabilities associated with the same CWE. First were 233 vulnerabilities found using SAST associated with CWE-52 \textit{Cross-Site Request Forgery}\footnote{\url{https://cwe.mitre.org/data/definitions/52.html}} labeled as at least ``High'' severity by the SAST tools.  The vulnerabilities found by SAST were all functions in source code which mapped to HTTP Requests where input parameters were not sufficiently restricted. For example, 220 of these functions used an @RequestParameter mapping but did not specify which methods (Post, Get, etc) could be used to call the function. Not specifying which types of requests can be used can result in additional access being granted where it otherwise would not be, and the lack of a method parameter can be particularly problematic if the application is using CSRF protection mechanism\footnote{\url{https://docs.spring.io/spring-security/site/docs/5.0.x/reference/html/csrf.html}}. The base application does not employ CSRF protection, but the OpenMRS team is working to employ better CSRF protection. However, the ``High'' or higher severity assigned by SAST tools contrasts with the single vulnerability associated by the DAST tools with CWE-352 \textit{Cross-Site Request Forgery}\footnote{\url{https://cwe.mitre.org/data/definitions/352.html}}, was labeled as ``Low'' severity by the DAST tool. Specifically, the DAST tool noted the lack of CSRF protection mechanisms in the version of the application we were using. The risks posed by each individual vulnerability are low. Hence, we classified the CSRF vulnerabilities found by the SAST tools as ``less severe''.

The second-largest group of more severe vulnerabilities found by a single tool or technique were 100 vulnerabilities found using EMPT associated with CWE-79 \textit{Cross-Site Scripting}\footnote{\url{https://cwe.mitre.org/data/definitions/79.html}}. An example XSS vulnerability would be if a field in a patient intake form accepts and saves the value \verb|<script>alert(1);</script>|, if the script is then executed when the user returns to the patient information page where the information from the intake form would otherwise be displayed this is considered Cross-Site Scripting. The XSS vulnerabilities found via EMPT were all found within a short period of time by students. We consider the risk of exploitability to be high and leave the classification of the vulnerabilities associated with CWE-79  as ``more severe''.

Third, SAST tools, specifically SAST-2, found 56 vulnerabilities CWE-404 \textit{Improper Resource Shutdown or Release}\footnote{\url{https://cwe.mitre.org/data/definitions/404.html}}. Of these 56 vulnerabilities, 39 were considered more severe while 17 were considered less severe. All 56 vulnerabilities were instances where a database connection or other resource could potentially be left open for certain executions of the code. The 17 less severe issues were considered to be on an ``exceptional'' execution path that the tool considered less likely to be executed, e.g. if an secondary failure happened on an unusual path within nested try-catch-finally blocks. The 39 more severe vulnerabilities were more likely to be executed system, e.g., if a connection was not inside a try-catch block at all in a function where an error is explicitly thrown under certain conditions. %These are the only high-more severe vulnerabilities 

Furthermore, after discussion with OpenMRS, we determined that some types of vulnerabilities were low priority for their organization in the context of the application. Specifically, a number of vulnerabilities involved errors which revealed potentially sensitive information about application source code. Since the tool is open-source, the threat posed by these vulnerabilities is minimal. Vulnerabilities associated with error messages that reveal too much information about the system are also classified as ``less severe''.

\subsubsection{Vulnerability Type (OWASP Top Ten)}\label{sec:VulnTypeResults}
\RRChange{Table \ref{tab:OWASPTopTen2} shows the distribution of the vulnerabilities found by each technique according to the OWASP Top Ten 2021 Categories. The Top Ten Category assignments are based on the CWEs assigned to the vulnerabilities, using the mapping to the OWASP Top Ten provided by CWE~\cite{CWEOWASP}, as discussed in Section~\ref{sec:cweReview}. The vulnerability counts for the specific CWE types within each Top Ten category is available in Appendix~\ref{app:CWEs}. The leftmost column of Table \ref{tab:OWASPTopTen2} indicates the OWASP Top Ten Category. Columns two through five indicate the vulnerabilities that were found for each technique. Column six of Table \ref{tab:OWASPTopTen2} shows the total vulnerabilities found within each Top Ten category across all techniques. Within each cell, the first value indicates the number of more severe vulnerabilities that were found in the OWASP Top Ten Category when using each technique. The second value (in parentheses) indicates the total vulnerabilities, including both more severe and less severe vulnerabilities, found in the corresponding OWASP Top Ten Category when using each technique.}

\begingroup
\setcounter{savefootnote}{\value{footnote}}
\setcounter{footnote}{\value{tablefootnote}}

\begin{table*}[htb] \renewcommand{\arraystretch}{1.3} \caption{\RRChange{\centering{More Severe Vulnerability Count based on OWASP Top Ten (2021)}\newline{(Total count, including both more and less severe vulnerabilities)}}} \label{tab:OWASPTopTen2}

\begin{minipage}{\linewidth}\centering
\setcounter{footnote}{5}

\renewcommand{\thefootnote}{\alph{footnote}}
\renewcommand{\thempfootnote}{\alph{footnote}}
{\begin{tabular}{  >{\raggedright\arraybackslash} m{135pt} ||| A{25pt} || A{25pt} || A{25pt}  ||  A{25pt} |||  A{25pt} }
OWASP Top Ten (2021) Category & SMPT & EMPT & DAST & SAST & Total Found\\
\hline\hline
\hline%\hline
{A01:2021 - Broken Access Control} & ~~2\footnotemark[5]{\newline}(2) & 15 (15) & {~ \newline (1) } & {28 (261)} & 58\footnotemark[5] (292) \\\hline
{A02:2021 - Cryptographic Failures} & ~~1{\newline}(1) & ~~1{\newline}(1) & {~~1{\newline}(1)} &  ~~2{\newline}(4) & ~~3{\newline}(6)\\\hline
{A03:2021 - Injection} & ~~5{\newline}(5) & 119 (119) & 11 (11) & 24 (58) &  150 (184)\\\hline
{A04:2021 - Insecure Design} & ~~5\footnotemark[5]{\newline}(7) & {~~8{\newline}(26)} & {~~1{\newline}(2)} & ~~8{\newline}(36) & 27\footnotemark[5] (73)\\\hline
{A05:2021 - Security Misconfiguration} & ~~2{\newline}(5) & ~~2{\newline}(4) & {~~2{\newline}(6)} & 14\footnotemark[7] (15) & 19\footnotemark[7] (23) \\\hline
{A06:2021 - Vulnerable and Outdated Components} & & & & & 0 \\\hline
{A07:2021 - Identification and Authentication Failures} & 13 (13) & 10 (10) & {~~1{\newline}(1)} & {~~2{\newline}(2)} & 17 (17)\\\hline
{A08:2021 - Software and Data Integrity Failures} & ~~1\footnotemark[6]{\newline}(1) & & {~ \newline (1)}& 10 (11) & 11\footnotemark[6] (13)\\\hline
{A09:2021 - Security Logging and Monitoring Failures} & ~~3{\newline}(3) & {~~9{\newline}(9)} & & & 12 (12)\\\hline
{A10:2021 - Server-Side Request Forgery (SSRF)} & ~~1\footnotemark[6]{\newline}(1) & & & & ~~1\footnotemark[6]{\newline}(1)\\\hline
\hline
{No Mapping to OWASP Top Ten} & ~~1{\newline}(1) & ~~1{\newline}(1) & & 54\footnotemark[7] (436) & 56\footnotemark[7] (438)\\\hline
\hline
{Total for Technique} & 32 (37) & 165 (185) & 17 (23) & 142 (823) & 329 (1033)\\
\end{tabular}}
\footnotetext[5]{\RRChange{One more severe vulnerability found using SMPT was mapped to both A01 and A04 through two different CWEs.}}
\stepcounter{footnote}
\footnotetext[6]{\RRChange{One more severe vulnerability found using SMPT was mapped to both A08 and A10 through two different CWEs.}}
\stepcounter{footnote}
\footnotetext[7]{\RRChange{14 more severe vulnerabilities found using SAST were associated with two CWEs, one of which mapped to A05 while the other CWE was not mapped to the OWASP Top Ten. We only include these vulnerabilities under A05 since they cannot be described as ``No Mapping''}}
\end{minipage}
\end{table*}

\setcounter{tablefootnote}{8}
\setcounter{footnote}{\value{savefootnote}}
\endgroup
\renewcommand{\thefootnote}{\arabic{footnote}}

\RRChange{This study is an evaluation of techniques used to find vulnerabilities that developers may have inserted themselves. Tools for finding vulnerabilities in third-party code, such as Software Composition Analysis (SCA) tools were excluded from our analysis. As can be seen in Table \ref{tab:OWASPTopTen2} of the tools examined found vulnerabilities in the OWASP Top Ten Category for Vulnerable and Outdated Components (A06), further suggesting that different techniques and categories of techniques are useful for finding different types of vulnerabilities.}

\RRChange{\textbf{SMPT} was most effective at finding vulnerabilities associated with Identification and Authentication Failures, finding more Identification and Authentication (A07) failures than any other vulnerability type and finding as many Identification and Authentication failures as EMPT.  While SMPT found fewer vulnerabilities than EMPT or SAST, most of the vulnerabilities found were more severe. Additionally, SMPT identified at least one vulnerability in every Top Ten category within scope of the tools in this study, providing better coverage of the Top Ten as compared with the other techniques.}

\RRChange{\textbf{EMPT} was one of the most effective techniques at finding severe vulnerabilities for the Broken Access Control (A01), Injection (A03), Insecure Design (A04), Identification and Authentication Failures (A07), and Security Logging and Monitoring Failures (A09) categories. EMPT was notably effective at finding Injection vulnerabilities, finding 119 of the 150 more severe Injection vulnerabilities in OpenMRS. }

\RRChange{\textbf{DAST} was most effective at finding Injection (A03) vulnerabilities relative to other categories of vulnerability. However, DAST found fewer injection vulnerabilities than EMPT and SAST, having found the least number of vulnerabilities overall in our study. }

\RRChange{\textbf{SAST} was the most effective at finding vulnerabilities associated with Security Misconfiguration (A05). However, the Security Misconfiguration vulnerabilities found by SAST were not the same as the Security Misconfiguration vulnerabilities found by other techniques. Overall, only one vulnerability found by other techniques was found by SAST in entire dataset. Hence SAST should not be seen as something that can substitute for other techniques, or be substituted for by other techniques. While many of the SAST vulnerabilities were marked as ``less severe'',  all of the less severe SAST vulnerabilities except the CSRF vulnerabilities classified as described in Section \ref{sec:Severity} were marked as low severity by the tools themselves. Between the two SAST tools there were also differences in the types vulnerabilities found. Although there were at least 184 total Injection vulnerabilities in OpenMRS, as can be seen in Tables~\ref{tab:long_CWE} and~\ref{tab:long_CWE_low} of Appendix~\ref{app:CWEs}, the 58 Injection vulnerabilities found using SAST, 24 of which were more severe, were found by SAST-2. Sonarqube did not find any Injection vulnerabilities. }

\subsubsection{Effectiveness Comparison with Austin et. al.}\label{sec:effect-austin}
A comparison with the previous study by Austin et al.~\cite{austin2011onetechniquenotenough,austin2013comparison} of vulnerability counts for each vulnerability type is shown in Table \ref{tab:PreviousWork}.  The first column of Table \ref{tab:PreviousWork} indicates the technique (Tech.), the second column of Table \ref{tab:PreviousWork} indicates whether the data is from the current study or Austin et al. The third column of Table \ref{tab:PreviousWork} indicates the SUT. As with previous tables, M indicates OpenMRS, the SUT from the current study. E indicates OpenEMR, T indicates Tolven, and P indicates PatientOS, the three SUT from Austin et al. The total vulnerability count calculated for each row is provided in the final \textit{TOTAL (TOT)} column. \RRChange{The remaining columns indicate the vulnerability counts for each of the OWASP Top Ten categories.} The total row indicates  the total number of vulnerabilities found in a particular study, either the  current study or Austin et al.\cite{austin2011onetechniquenotenough,austin2013comparison}. For the current study, the total is equivalent to the results from OpenMRS as indicated in the table. For Austin et al., the total is the sum of the vulnerabilities found across all three SUT. In Table \ref{tab:PreviousWork}, the row for the current study is shaded in the darkest gray, the total row from the Austin et al. study is in the medium gray color, and the rows for the individual SUT from Austin et al. are the lightest gray color. Austin et al. did not specify how severity was evaluated in their study, and vulnerabilities such as error messages containing sensitive information about the system (CWE-209) which would have been classified as ``less severe'' in our current study were included in their vulnerability counts. We assume that the Austin et al. counts reported\cite{austin2011onetechniquenotenough,austin2013comparison} include less severe vulnerabilities. Consequently, the vulnerability counts from the current study in Table~\ref{tab:PreviousWork} also include those that are less severe. We highlight insights from Table~\ref{tab:PreviousWork} that we find most interesting below.

\begingroup
\setcounter{savefootnote}{\value{footnote}}
\setcounter{footnote}{\value{tablefootnote}}
\renewcommand{\thefootnote}{\alph{footnote}}

\newcommand*{\PillarWidth}{155pt}
\newcommand*{\TechWidth}{8pt}

\begin{table*}[htb] \renewcommand{\arraystretch}{1.2} \caption{\RRChange{\centering{{Vulnerability Type Comparison with Austin Study}\newline{\tiny M indicates OpenMRS, E indicates OpenEMR, T indicates Tolven, and P indicates PatientOS}}}} \label{tab:PreviousWork}

\begin{minipage}{\linewidth}\centering
\setcounter{footnote}{6}

\renewcommand{\thefootnote}{\alph{footnote}}
\renewcommand{\thempfootnote}{\alph{footnote}}
{\begin{tabular}{  c | A{48pt} | c || c | c | c | c | c | c | c | c | c | c || c || c |} 
\multicolumn{3}{r||}{OWASP Top Ten Descr.} & \multicolumn{1}{R{90}{\PillarWidth} | }{Broken Access Control} & \multicolumn{1}{R{90}{\PillarWidth} | }{Cryptographic Failures} & \multicolumn{1}{R{90}{\PillarWidth} | }{Injection} & \multicolumn{1}{R{90}{\PillarWidth} | }{Insecure Design} & \multicolumn{1}{R{90}{\PillarWidth} | }{Security Misconfiguration} & \multicolumn{1}{R{90}{\PillarWidth} | }{Vulnerable and Outdated Components} & \multicolumn{1}{R{90}{\PillarWidth} | }{ Identification and Authentication Failures} & \multicolumn{1}{R{90}{\PillarWidth} | }{Software and Data Integrity Failures} & \multicolumn{1}{R{90}{\PillarWidth} | }{Security Logging and Monitoring Failures} & \multicolumn{1}{R{90}{\PillarWidth} || }{Server-Side Request Forgery (SSRF)} & \multicolumn{1}{R{90}{\PillarWidth} || }{Not Mapped to Top Ten} & \multicolumn{1}{R{90}{\PillarWidth} | }{TOTAL}\\\hline

 \multicolumn{3}{r ||}{OWASP Top Ten \#} & A01 & A02 & A03 & A04 & A05 & A06 & A07 & A08 & A09 & A10 & NM & TOT\\\hline
  \hline
  Tech. & Study & SUT &  \multicolumn{12}{r |}{ }\\\hline
\hline
\rowcolor[HTML]{CDCDCD} \cellcolor[HTML]{FFFFFF}  & \cellcolor[HTML]{FFFFFF} {Curr. Study} & M (Total) & 2 & 1 & 5 & 8 & 5 & & 13 & 1 & 3 & 1 & 1 & 37 \\\hhline{~|- |-||-|-|-|-|-|-|-|-|-|-||-||-}

\rowcolor[HTML]{E1E1E1} \cellcolor[HTML]{FFFFFF} &\cellcolor[HTML]{FFFFFF}  & Total & 4 & 3 & 17 & 4 &  & & 9 & & 99 & & & 136\\\hhline{~|~|-||-|-|-|-|-|-|-|-|-|-||-||-}
\rowcolor[HTML]{F1F1F1} \cellcolor[HTML]{FFFFFF} & \cellcolor[HTML]{FFFFFF} & E & 3 & 2 & 16 & 2 &  & & 3 & & 37 & & & 63\\\hhline{~|~|-||-|-|-|-|-|-|-|-|-|-||-||-}
\rowcolor[HTML]{F1F1F1} \cellcolor[HTML]{FFFFFF} & \cellcolor[HTML]{FFFFFF} & T & 1 & & & 2 & & & 4 & & 29 & & & 36\\\hhline{~|~|-||-|-|-|-|-|-|-|-|-|-||-||-}
\rowcolor[HTML]{F1F1F1}\parbox[c]{\TechWidth}{\multirow{-5}{*}{\rotatebox[origin=c]{90}{\cellcolor[HTML]{FFFFFF} SMPT}}}  & \multirow{-4}{*}{\cellcolor[HTML]{FFFFFF}{Austin et al.}} & P & & 1 & 1 & & & & 2 & & 33 & & & 37\\\hline
 \hline
\rowcolor[HTML]{CDCDCD} \cellcolor[HTML]{FFFFFF}  & \cellcolor[HTML]{FFFFFF} {Curr. Study} & M (Total) & 15 & 1 & 119 & 26 & 4 & & 10 & & 9 & & 2 & 185 \\\hhline{~|- |-||-|-|-|-|-|-|-|-|-|-||-||-}
\rowcolor[HTML]{E1E1E1} \cellcolor[HTML]{FFFFFF} &\cellcolor[HTML]{FFFFFF} & Total & & 1 & 8 & 1 & & & 1 & & & & 2 & 13\\\hhline{~|~|-||-|-|-|-|-|-|-|-|-|-||-||-}
\rowcolor[HTML]{F1F1F1} \cellcolor[HTML]{FFFFFF} & \cellcolor[HTML]{FFFFFF} & E & & & 8 & 1 & & & 1 & & & & 2 & 12\\\hhline{~|~|-||-|-|-|-|-|-|-|-|-|-||-||-}
\rowcolor[HTML]{F1F1F1} \cellcolor[HTML]{FFFFFF} & \cellcolor[HTML]{FFFFFF}  & T & &  & & & & & & & & & & \\\hhline{~|~|-||-|-|-|-|-|-|-|-|-|-||-||-}
\rowcolor[HTML]{F1F1F1} \parbox[c]{\TechWidth}{\multirow{-5}{*}{\rotatebox[origin=c]{90}{\cellcolor[HTML]{FFFFFF} EMPT }}}  & \multirow{-4}{*}{\cellcolor[HTML]{FFFFFF}{Austin et al.}}  & P & & 1 &  & & & & & & & & & 1\\\hline
 \hline
\rowcolor[HTML]{CDCDCD} \cellcolor[HTML]{FFFFFF}  & \cellcolor[HTML]{FFFFFF} {Curr. Study} & M (Total) & 1 & 2 & 11 & 2 & 6 & & 1 & 1 & & & & 23\\\hhline{~|- |-||-|-|-|-|-|-|-|-|-|-||-||-}
\rowcolor[HTML]{E1E1E1} \cellcolor[HTML]{FFFFFF} &\cellcolor[HTML]{FFFFFF} & Total & 502 & & 221 & & 9 & & & & & & & 732\\\hhline{~|~|-||-|-|-|-|-|-|-|-|-|-||-||-}
\rowcolor[HTML]{F1F1F1} \cellcolor[HTML]{FFFFFF} &\cellcolor[HTML]{FFFFFF} & E & 485 & & 221 & & 4 & & & & & & & 710\\\hhline{~|~|-||-|-|-|-|-|-|-|-|-|-||-||-}
\rowcolor[HTML]{F1F1F1} \parbox[c]{\TechWidth}{\multirow{-4}{*}{\rotatebox[origin=c]{90}{\cellcolor[HTML]{FFFFFF} DAST }}} & \multirow{-3}{*}{\cellcolor[HTML]{FFFFFF}{Austin et al.}} & T & 17 & & & &  5 & && & & & & 22\\\hline
\hline
\rowcolor[HTML]{CDCDCD} \cellcolor[HTML]{FFFFFF}  & \cellcolor[HTML]{FFFFFF} {Curr. Study} & M (Total) & 261 & 4 & 58 & 36 & 15 & & 2 & 11 & & & 436 & 822 \\\hhline{~|-|-||-|-|-|-|-|-|-|-|-|-||-||-}
\rowcolor[HTML]{E1E1E1} \cellcolor[HTML]{FFFFFF} &\cellcolor[HTML]{FFFFFF} & Total & 93 & & 1190 & 124 & 1 & & & & 22 & & 86 & 1516 \\\hhline{~~|-||-|-|-|-|-|-|-|-|-|-||-||-}
\rowcolor[HTML]{F1F1F1} \cellcolor[HTML]{FFFFFF} & \cellcolor[HTML]{FFFFFF} & E & 36 & & 1155 & 122 & 1 & & & & & & 7 & 1321\\\hhline{~~|-||-|-|-|-|-|-|-|-|-|-||-||-}
\rowcolor[HTML]{F1F1F1} \cellcolor[HTML]{FFFFFF} & \cellcolor[HTML]{FFFFFF}  & T & 13 & & 35 & 2 & & & & & & & & 50\\\hhline{~~|-||-|-|-|-|-|-|-|-|-|-||-||-}
\rowcolor[HTML]{F1F1F1} \parbox[c]{\TechWidth}{\multirow{-5}{*}{\rotatebox[origin=c]{90}{\cellcolor[HTML]{FFFFFF} SAST }}} & \multirow{-4}{*}{\cellcolor[HTML]{FFFFFF}{Austin et al.}} & P & 44 & & & & & & & & 22 & & 79 & 145 \\\hline
\end{tabular}}
\end{minipage}
\end{table*}

\setcounter{tablefootnote}{6}
\setcounter{footnote}{\value{savefootnote}}
\endgroup
\renewcommand{\thefootnote}{\arabic{footnote}}

\RRChange{SMPT in the current study was similar in effectiveness to the study by Austin et al.\cite{austin2011onetechniquenotenough,austin2013comparison}. Austin et al. found more vulnerabilities in OpenEMR using SMPT as compared with the current study, but a similar number of vulnerabilities in Tolven and PatientOS. The distribution of the vulnerabilities across the OWASP Top Ten categories differs between studies. One possible explanation is differences in the test suite. While the current study covered more of the OWASP Top Ten categories, we are using the the 2021 Top Ten and the first study by Austin et al. was published in 2011. Some vulnerability types were less prevalent, and some vulnerability types may have been considered less severe at the time of the Austin et al. work, and vulnerability detection techniques of the time may not have ensured coverage of any less prevalent or severe vulneraiblities. For example, Security Misconfiguration (A05) which had no vulnerabilities found by Austin et al., but five vulnerabilities found by the current study, was not included in the OWASP Top Ten until 2013. Similarly, in the Austin et al. test suite which was developed as part of a previous work\cite{smith2011systematizing}, 58 of the 137 test cases (i.e. 42\% of the test suite) were targeted towards logging and auditing security controls. The ASVS standard around which our test suite was built only has 2 level 1 controls relating to logging, and only 5 test cases out of 131 (i.e. 4\% of the test suite) were related to auditing and logging. The higher number of logging related test cases used by Austin et al. may help explain why Austin et al. were more effective at finding vulnerabilities associated with Security Logging and Monitoring Failures (A09).}

\RRChange{Comparing our results against Austin et al.\cite{austin2013comparison,austin2011onetechniquenotenough} for EMPT is more complicated for methodological reasons. For DAST and SAST our methodology was comparable to Austin et al.~\cite{austin2013comparison,austin2011onetechniquenotenough} since the analysis for RQ1was done by a small team of researchers in both cases. For SMPT, the procedure was also comparable as indicated by the size of the test suite: 131 test cases in the current study, compared with 137 per SUT for Austin et al.\cite{austin2013comparison,austin2011onetechniquenotenough}. The procedure for EMPT differed between the studies in order to take full advantage of the data generated by students to better understand the strengths and weaknesses of EMPT. As we note in Section~\ref{sec:RQ1-DataCollection}, the 229 vulnerabilities in Table~\ref{tab:VulnCounting} for EMPT are the result of efforts by 62 students as well as researcher review. Large numbers of individuals involved in EMPT is not uncommon, for example with bug bounty programs~\cite{finifter2013rewards}. However, the use of smaller, internal teams such the 6-person team used to apply EMPT to OpenEMR in Austin et al.~\cite{austin2013comparison,austin2011onetechniquenotenough}, or even individual hackers working alone on EMPT as was done for Tolven and PatientOS are also not uncommon~\cite{alomar2020you,votipka2018hackers}. The high number of students how applied EMPT for RQ1 in the current study should not impact the distribution of vulnerabilities across types. However, more participants may increased the number of vulnerabilities found and to enable comparison between the studies we must further evaluate individual effectiveness for EMPT. }

\RRChange{Our results suggest that even at the individual level, the average individual applying EMPT found more vulnerabilities in the current study as compared with Austin et al.~\cite{austin2013comparison,austin2011onetechniquenotenough}. In Austin et al. for OpenEMR a team of 6 individuals spent a combined 30 hours performing EMPT. The team found 8 vulnerabilities in total, as shown in Table~\ref{tab:PreviousWork}, with a per-person average of 1.33 vulnerabilities per student. For both Tolven and PatientOS, a single individual applied EMPT for 15 and 14 hours, respectively. Austin et al. found no vulnerabilities using EMPT against Tolven and only 1 vulnerability using EMPT against PatientOS, as shown in Table~\ref{tab:PreviousWork}. In the current study, EMPT was applied by 62 students, and reviewed by 3 researchers for a total of 65 people involved in collecting EMPT data. We found 185 unique vulnerabilities of which 165 were more severe, resulting in a per-student average of 2.94 for all vulnerabilities and 2.62 for more severe vulnerabilities; the overall vulnerabilities per-person was 2.85 for all vulnerabilities and 2.54 for the more-severe vulnerabilities.}

\RRChange{Austin et al. were more effective with DAST, particularly against OpenEMR, when compared against the current study. Austin et al. found 710 vulnerabilities using DAST against OpenEMR and 22 vulnerabilities in Tolven, as compared with 23 vulnerabilities found in OpenMRS in the current study. We suspect that this difference may be due to differences in how counting rules are applied. The number of true positive alerts appears to be the same or close to the total number of vulnerabilities reported in the Austin et al. study\cite{austin2013comparison,austin2011onetechniquenotenough} and the terms ``alert'' and ``vulnerability'' appear to be used interchangeably.  While SAST would also be impacted by any differences in counting rules, as reported in Table~\ref{tab:VulnCounting}, in the current study the ratio of alerts to vulnerabilities for DAST tools was 34.26 to 1 ratio. In contrast, the ratio of alerts to vulnerabilities for SAST tools in the current study is 1.12 to 1. The lower ratio for SAST may help explain why the effectiveness of SAST in Austin et al.'s work is more similar to the effectiveness of SAST in the current study, as compared with the DAST results from each study. In OpenEMR, the system where Austin et al. found the most vulnerabilities overall\cite{austin2011onetechniquenotenough,austin2013comparison}, Austin et al. were more effective with SAST as compared to the current study. For Tolven and PatientOS, Austin et al were less effective with SAST as compared to the current study.  Austin et al. do not provide their counting rules, and so our hypotheses that counting rules may contribute to the differences between the studies cannot be confirmed.  We provide our current counting rules as well as references to how they were derived to assist in future evaluations of vulnerability detection techniques.}

\bigskip
\begin{tcolorbox}[colback=white,colframe=black,title=RQ1 - \rqEffective ]
\textbf{Answer:} SAST found the largest number of vulnerabilities overall. However, over half of the vulnerabilities identified by SAST were of low severity. EMPT found the highest number of ``more severe'' vulnerabilities of any technique. Furthermore, if any particular tool or technique had been excluded from the analysis, at least 4 and as many as 588 vulnerabilities would not have been found.
\end{tcolorbox}

\subsection{RQ2 - Efficiency}\label{sec:results-efficiency}
In this section, we discuss the results for our question \emph{\rqEfficiency} The data was collected from students, as described in Section~\ref{sec:method-efficiency}.

\subsubsection{Data Cleaning}\label{sec:results-rq2-cleaning}
In analyzing the student efficiency, we noticed significant outliers. The most extreme outlier occurred for the SAST technique, where the outlier was 309\% greater than the second highest efficiency with any technique. As described in Section~\ref{sec:RQ2-DC-Clean}, we use trimming~\cite{wilcox2003modern,kitchenham2017robust} to identify and remove outliers based on MADN. Each category of vulnerability detection technique had one outlier. We removed the four outliers from the dataset. Consequently, for each technique we analyzed 12 efficiency reports.

\subsubsection{Data Analysis}\label{sec:results-rq2-analysis}
\begin{figure}[!htb]
\centering \includegraphics[width=3.5in]{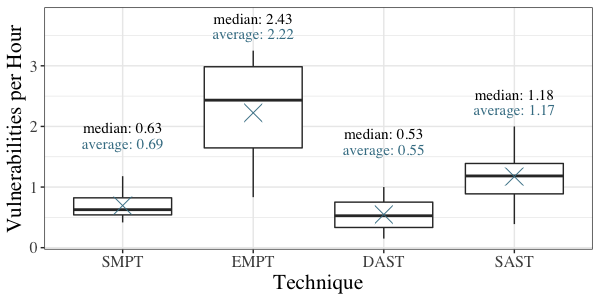} \caption{Vulnerability Detection Technique Efficiency}
\label{fig:EfficiencyBoxPlot} 
\end{figure}

Boxplots for each technique's efficiency as well as the values for the median and average are shown in Figure \ref{fig:EfficiencyBoxPlot}.  EMPT had the highest efficiency (Median 2.4 VpH, Average 2.22 VpH). SAST had the second highest efficiency (Median 1.18 VpH, Average 1.17 VpH). SMPT (Median 0.63 VpH, Average 0.69 VpH) and DAST (Median 0.53 VpH, Average 0.55 VpH) were least efficient for students. 
% \centering \includegraphics[width=3.5in,natwidth=600,natheight=300]{Boxplot_v4.png} \caption{Vulnerability Detection Technique Efficiency}

Table \ref{tab:EfficiencyGH} shows the Games-Howell test results for comparing each pair of techniques. As can be seen in the table, EMPT is statistically significantly more efficient than every other technique ($p<0.05$ for all comparisons). SAST is the second-most-efficient technique ($p<0.05$ when compared against both SMPT and DAST). We observed no statistically significant difference in the reported efficiency of SMPT when compared to the reported efficiency of DAST. In Table~\ref{tab:EfficiencyGH}, we round the p-values to the thousandths position. The p-value for the comparison between DAST and EMPT was less than one thousandth.

\begin{table*}[!htb] \renewcommand{\arraystretch}{1.3} \caption{Games-Howell Efficiency t-test Results} \label{tab:EfficiencyGH}
\vspace*{5mm}
\centering
\begin{tabular}{ p{75pt} || p{55pt} | p{30pt}  }
Techniques & t-value & $p$-value \\
\hline\hline
DAST - SMPT & $\ \ 0.15\ (\pm 0.29)$ & $0.499$ \\\hline
DAST - SAST & $\ \ 0.63\ (\pm 0.46)$ & $0.006$\\\hline
DAST - EMPT & $\ \ 1.68\ (\pm 0.76)$ & $<0.001$\\\hline
SAST - SMPT & $-0.48\ (\pm 0.45)$ & $0.034$ \\\hline
EMPT - SAST & $-1.05\ (\pm 0.81)$ & $0.009$ \\\hline
EMPT - SMPT & $-1.53\ (\pm 0.75)$ & $<0.001$ \\\hline
\end{tabular}
\end{table*}

\subsubsection{Efficiency Comparison with Austin et al.}\label{sec:resultsEfficiencyAustin}
Table \ref{tab:VPHComp} shows the average and median VpH of this study, compared with the VpH reported by Austin et al.  For our study, we report the average and median across the groups performing the task, as reported in Section~\ref{sec:results-rq2-analysis}. Austin et al. calculated the amount of time it took, on average, to find each vulnerability.
\begin{table*}[!htb]
\renewcommand{\arraystretch}{1.3} \caption{VpH Compared to Austin et al.} \label{tab:VPHComp}
\vspace*{5mm}
\centering
\begin{tabular}{ A{30pt}| A{40pt} ||| A{55pt} |  A{55pt} | A{55pt}| A{55pt}}
Study & SUT & SMPT & EMPT & DAST & SAST\\\hline\hline\hline
Current & OpenMRS & {0.69 (avg)\newline 0.63 (median)} & {2.22 (avg)\newline 2.43 (median)} & {0.55 (avg)\newline 0.53 (median)} &{1.17 (avg)\newline 1.18 (median)} \\\hline\hline
\multirow{3}{=}{Austin et al.}& OpenEMR & 0.55 & 0.40 & 71.00 & 32.40 \\\cline{2-6}
& Tolven & 0.94 & 0.00 & 22.00 & 2.78\\\cline{2-6}
& PatientOS & 0.55 & 0.07 & N/A & 11.15\\
\end{tabular}
\end{table*}

The differences in efficiency are not only due to differences between student performance and researcher performance. When researchers performed SAST and DAST to answer RQ1, they also reported lower efficiency than Austin et al. When applying SAST for RQ1, researchers' efficiency was estimated at 18-32 VpH with SAST, comparable to Austin et al. With DAST researchers were not much more efficient than the students with an estimated VpH of 0.34-1.8.  DAST-2 had a lower efficiency for the researchers, which may have impacted the results. However, efficiency for ZAP was still less than reported in Austin et al. One possible cause of the discrepancy, particularly with DAST efficiency, could be our focus in unique vulnerabilities as defined by our counting rules in Section~\ref{sec:methodCounting} compared with alerts or true positive failures. As shown previously in Table~\ref{tab:VulnCounting}, DAST had the highest ratio of true positive failures.

\bigskip
\begin{tcolorbox}[colback=white,colframe=black,title=RQ2 - \rqEfficiency]
\textbf{Answer:} EMPT was the most efficient (2.22 VpH), followed by SAST (1.17 VpH). SMPT (0.69 VpH) and DAST (0.55 VpH) were least efficient.
\end{tcolorbox}

\subsection{RQ3 - Other Factors to Consider when Comparing Tools}\label{sec:results-qualitative}
We discuss the results of a qualitative analysis of student responses for RQ3. As mentioned in Section~\ref{sec:studentData}, we were able to collect responses from 13 teams whose members all consented to the use of their data in this study. Of the free-form text responses used to answer RQ3, we discarded the response from one (1) team since both reviewers considered the text to be confusing and self-contradictory. Responses from 12 teams were analyzed. We use ``at least'' to describe how many teams discussed certain ideas because there may be instances that were missed due to human error, or where the author's intent was unclear. Some teams discussed ``automated'' and ``manual'' categories of techniques rather than the further subdivision used in the rest of this paper (i.e., EMPT and SMPT are ``manual'' techniques, while SAST and DAST are ``automated'' techniques). We discuss our results for RQ3 in terms of automated and manual techniques when those terms are used by one or more teams.

We focus on concepts that were not already covered in previous sections. Most teams included some discussion of efficiency and effectiveness, as well as the types of vulnerabilities found.  We summarize the three other concepts that were most frequently mentioned by students.

\subsubsection{Effort}
\noindent\textit{\textbf{Summary:} People do not like to do any more work than necessary. Effort was one of the most discussed topics by students. While effort was exclusively seen as a disadvantage with manual techniques, discussions of effort for automated techniques had more mixed views.}

Every response in some way mentioned human effort beyond VpH. Automation itself is seen as an advantage for at least two teams, one of which explicitly stated ``\emph{Dynamic application security testing is better than manual blackbox testing because you can automate the tests}''. Effort was perceived as a disadvantage for SMPT and EMPT more than SAST and DAST. Eight (8) teams mentioned effort as a disadvantage of one or both of the manual techniques, while 0 teams mentioned advantages relating to effort for either of the manual techniques. In contrast, for one or both automated techniques, 4 teams mentioned effort as a disadvantage, 3 teams mentioned effort as an advantage, and 2 teams mentioned both advantages and disadvantages.  The students' focus on effort and the frequency of comments that effort was a disadvantage of manual techniques with less focus on automated techniques contrasts with the numeric data they reported. The numeric efficiency of tool-based techniques in terms of VpH of effort used to answer RQ2 was reported by the students themselves in the same assignment, generally on the same page of their responses. The recorded numeric efficiency for DAST and SAST was comparable to the numeric efficiency of SMPT and EMPT.

\subsubsection{Time}\label{sec:results-rq3-time}

\textit{\textbf{Summary:} \RRChange{Although the amount of time spent on an activity was a component of our metric for efficiency (VpH), Time was discussed distinctly from efficiency. Manual techniques were seen as requiring more time, while for automated techniques some teams considered time an advantage while others considered time a disadvantage. Additionally, students conjectured that the efficiency of each technique relative to the other techniques may change if the techniques were applied over a longer timeframe. }}

\RRChange{Similar to Effort, Time was frequently seen as a disadvantage for manual techniques, particularly SMPT. At least 10 of the 12 teams mentioned time in some way. Eight (8) teams explicitly mentioned time spent on manual tasks, while 5 of those teams also explicitly discussed the amount of time it takes for tools to run, even though tool running time is not active time for the individual applying the technique. Additionally, 8 teams mentioned time as a disadvantage for manual techniques, with 4 teams specifically mentioning time as a disadvantage for SMPT and 3 teams mentioning time as a disadvantage for EMPT. One of the teams who considered time a disadvantage for SMPT considered time an advantage for EMPT. No other team noted time as an advantage for any manual technique. Responses for manual techniques were more mixed, with time seen as an advantage for SAST by 4 teams and for DAST by 1 of the 4. However, 2 teams considered time to be a disadvantage for SAST and 3 teams considered time to be a disadvantage for DAST.}

\RRChange{At least 3 teams also noted that they would anticipate VpH and a technique's overall effectiveness relative to other techniques would change if the technique were to continue to be performed over a longer period of time. For example, one team noted that for SMPT ``\emph{our guess is with time it will get even more difficult to come up with black box test cases manually thus giving lower efficiency eventually}''. Similarly, another team noted that for EMPT ``\emph{after [the analyst] tried every point [they] could imagine, there is less possib[ility of] detect[ing] vulnerabilities}''. In contrast, one team claimed that with DAST ``\textit{... if time and memory are not an issue you can run a local instance of the application and fuzz it for years}''.}

\subsubsection{Expertise}
\textit{\textbf{Summary:} Many types of expertise are needed to apply vulnerability detection techniques, particularly EMPT. Similar to findings from other works~\cite{itkonen2013role,itkonen2014test,votipka2018hackers}, students noted that EMPT in particular requires different types of expertise including technical expertise, security expertise, and expertise with the SUT. }

Overall, at least 8 teams commented on the role of expertise. Of the 8 teams who mentioned expertise, only 3 mentioned expertise in the context of tool-based techniques while 6 mentioned expertise when discussing EMPT, and one team mentioned expertise when discussing SMPT and manual techniques generally. Expertise was not clearly an advantage or disadvantage, with only 3 of the 8 teams who mentioned expertise suggesting that the expertise required to use a technique was a disadvantage, with the remaining 5 teams not clearly noting expertise as an advantage or disadvantage. Several specific types of expertise were discussed, three of which, technical, security, and SUT expertise, are similar to the types of expertise highlighted in related work~\cite{itkonen2013role,itkonen2014test,votipka2018hackers}. At least 1 team commented on the role of technical expertise in applying EMPT, 1 team commented on security expertise required for EMPT,  4 teams commented on the role of SUT expertise for EMPT, and 1 team commented on the role of SUT expertise for SMPT.

\bigskip
 \begin{tcolorbox}[colback=white,colframe=black,title=RQ3 - \rqQualitative ]
\textbf{Answer:} The three most frequently discussed factors other than Effectiveness and Efficiency were Effort, Time, and Expertise. Effort and Time were seen as a disadvantage of manual techniques. Perceptions of Effort and Time for automated techniques were more mixed, with some teams considering Effort and/or Time an advantage while others considered them a disadvantage. Expertise was associated with manual techniques, particularly EMPT, more than automated techniques. 
\end{tcolorbox}

\section{Limitations}
\label{sec:limitations}
We discuss the limitations to our approach in this Section. We group these limitations as threats to  Conclusion Validity, External Validity, Internal Validity, and Construct Validity~\cite{cook1979quasi,feldt2010validity,wohlin2012experimentation}. 

\subsection{Conclusion Validity}
\RRChange{Conclusion Validity is about whether conclusions are based on statistical evidence~\cite{cook1979quasi,wohlin2012experimentation}.} While we have empirical results for RQ1, a single case study is insufficient to draw statistically significant conclusions for effectiveness (RQ1).  The measures used to evaluate effectiveness are based on the number of vulnerabilities found by applying each technique thoroughly and systematically. Unlike the amount of time taken to apply each technique, the number of vulnerabilities found will be deterministic for deterministic techniques such as SAST. Measuring effectiveness with statistical significance would require the application of all four techniques to at least 10-20 additional applications~\cite{kirk2013experimentaldesign}. Applying all techniques to 10-20 similarly-sized SUT is impractical given the effort required to apply these techniques to a single application. To mitigate this threat to validity, we performed extensive review of the vulnerability counts, using the guidelines in Section~\ref{sec:rq1-guidelines}, and at least two individuals were involved in the review process for each technique to verify the accuracy of the results. For efficiency (RQ2) the measure used, VpH, was evaluated by having more individuals apply the technique to a subset of the application.

\subsection{Construct Validity}
 \RRChange{Construct Validity concerns the extent to which the treatments and outcome measures used in the study reflect the higher level constructs we wish to examine~\cite{cook1979quasi,wohlin2012experimentation, ralph2018construct}. In our study, the  cause construct of the vulnerability detection technique being used is reflected in our treatment of the four categories of vulnerability detection techniques. The two primary outcome constructs are effectiveness (RQ1) and Efficiency (RQ2) for which the proxy measures are the number and type of vulnerabilities (RQ1) and Vulnerabilities per Hour (RQ2). These treatments and outcome measures were the same treatments and measures used by Austin et al.~\cite{austin2011onetechniquenotenough,austin2013comparison}.} 

\RRChange{To measure effectiveness we examined the number, type, and severity of vulnerabilities found by each technique. The number and type of vulnerabilities found are commonly used measures of (in)security in academia and industry \cite{klees2018evaluating,delaitre2018sate,okun2013report,okun2011report,okun2010second,okun2009static}, including by the U.S. National Institute of Standards and Technology (NIST) Software Assurance Metrics and Tool Evaluation (SAMATE) program discussed in Section~\ref{sec:relwork}. Raw vulnerability counts do not fully capture the picture, since some vulnerabilities may be considered more important than others. We address this concern in several ways. First, as mentioned in Section \ref{sec:RQ1-1}, we excluded tool alerts which were marked as insignificant or inconsequential, assuming that these alerts would not be of interest to practitioners. Additionally, we use the OWASP Top Ten categorization to summarize our data, and indicate the severity of vulnerabilities found as described in Section~\ref{sec:Severity}. }

\RRChange{Efficiency, was measured in terms of vulnerabilities per hour. This allowed us to have a controlled experiment in which teams of students performed each technique on at least a subset of the application, through which we could examine whether the differences in efficiency were statistically significant. As discussed in Section~\ref{sec:results-rq3-time}, factors such as the length of time required to apply a technique may also be helpful in understanding efficiency. Applying SAST and DAST more comprehensively as part of RQ1 required over 20 hours per tool for both SAST tools as well as DAST-2. In a class where vulnerability detection was only part of the curriculum, it was not reasonable to expect students to spend 20 hours on such a small portion of their grade. Adding RQ3, which was not included in Austin et al.'s work~\cite{austin2011onetechniquenotenough,austin2013comparison}, allowed us to better understand how other factors such as time spent applying a technique were perceived by students, helping to mitigate this threat to validity.}

\subsection{Internal Validity}\label{sec:InternalValidity}
\RRChange{Internal Validity concerns whether the observed outcomes are due to the treatment applied, or whether other factors may have influenced the outcome~\cite{cook1979quasi,feldt2010validity}. For our study, the treatment applied is the type of vulnerability detection technique (SMPT, EMPT, DAST, or SAST). For RQ1 the observed outcome is the effectiveness as measured through number and type of vulnerabilities found, for RQ2 the outcome is effectiveness, measured in VpH. For RQ3 the outcome is the factors found through qualitative analysis.}

\RRChange{Running DAST based on more test cases may have found more vulnerabilities.  However, the resources available to our team were not significantly less than other small organizations, suggesting that resource limitations may be a factor to consider when using DAST.  Additionally, with OWASP ZAP we leveraged the tool's spider capability to expand on the 6 inputs, which helped mitigate this threat to validity by increasing system coverage. As noted in the description of DAST in Section \ref{sec:DAST-Desc}, other comparisons of vulnerability detection techniques have also used a spider, if a spider was available.}

Another threat to internal validity for this study is that the student data used for RQ2 is self-reported. The student data aligns with the experiences of the research team, but self-reported estimates of the length of time it took to complete a task are not necessarily representative of the actual time it takes to complete a task. However, perceived time it takes to complete a task must still be considered when making decisions on which vulnerability detection techniques are used. Additionally, as shown by the qualitative analysis, students' reported numeric efficiency was not necessarily indicative of the time and effort the students perceived was required for each technique.

\RRChange{Another limitation with RQ2 is posed by equipment constraints for the graduate level class. For SAST and DAST, the students were unable to scan some parts of the system due to insufficient memory. To mitigate the risk that memory-related processing issues would negatively impact student efficiency, the first author performed a scan of all modules in advance and directed students to modules which would not be impacted by equipment constraints. Furthermore, students were instructed to only report time spent reviewing results, not time spent trying to get the scan to run.}

The researchers performing qualitative analysis for RQ3 may have had biases which present threats to validity. Researcher biases may also have impacted the vulnerability review processes, and analysis of true and false positives from the results of vulnerability detection tools for RQ1. For this reason, two individuals jointly performed the qualitative analysis, and the vulnerability review was either performed by two independent individuals or performed by one individual and audited by a second individual for each technique.

\RRChange{Finally, although both DAST tools examined in RQ1 and RQ2 were the same, one of the two SAST tools examined differed between RQ1 and RQ2. Specifically, we did not use Sonarqube for RQ2, using a proprietary tool (SAST-3) that had been used in the course previously. Sonarqube may have been more or less efficient or effective as compared with SAST-3, which would influence our results. Student data was reported in aggregate and so we do not know how the efficiency of SAST-2 compares with the efficiency of SAST 3, we only have the average efficiency of SAST-2 and SAST-3. However, estimated researcher efficiency using Sonarqube was ~22 VpH while estimated researcher efficiency using SAST-2 was ~18 VpH, suggesting that tool differences may play less of a role as compared with other factors such as expertise. In RQ2 we control for expertise by comparing efficiency scores from the same group of individuals, i.e. we do not introduce researcher data or data from students whose teammates did not elect to participate in the study.}

\subsection{External Validity}\label{sec:limitations-external}
\RRChange{External Validity concerns the generalizability of our results~\cite{cook1979quasi,feldt2010validity,wohlin2012experimentation}. Our results may not generalize to software that is not similar to the SUT and the results may not generalize to other systems. For example, we know that a strongly-typed, memory-safe language such as Java, by design, is likely to have fewer memory-allocation-related vulnerabilities, such as buffer overflow, when compared with code in a non-memory-safe language such as C\cite{cowan2000buffer,nagarakatte2009softbound}. As discussed in Section \ref{sec:why-openmrs},OpenMRS is a large system built with commonly-used languages (e.g. Java, Javascript) and frameworks (e.g. Spring) that is comparable in terms of size and development practices to other systems in its domain. Additionally, many of our results are similar to other studies within the same domain. For example, in recent SATE comparisons\cite{delaitre2018sate}, the highest precision rates for SAST tools examined were 78-94\% in tests against Java applications, similar to those for the study and higher than we expected based on other prior work as we will discuss in Section~\ref{sec:disc-SAST-FP}.}
The tools used in this study may not be representative of DAST and SAST tools generally, posing a threat to external validity for the study.  We used two tools that are in prevalent use in industry when performing each technique to mitigate and understand possible biases introduced by tool selection. Our results as well as those of the SATE reports\cite{delaitre2018sate,okun2013report,okun2011report} suggest that the effectiveness of SAST tools may vary. As noted both in our own experience and by students, the two DAST tools were very different in terms of ease of use.

\RRChange{A related threat to external validity is that we are performing a scientific experiment in an academic setting, using industrial tools on a production system. We do not think the differences between our experiment and industry would impact our results and have worked to minimize differences. For example, the assignments were designed to mitigate the risk that differences from industry practice would impact the efficiency scores. When applying SAST tools in an industry project, once alerts are classified as true or false positive, practitioners are more concerned with resolving the true positive alerts than with handling false positive alerts, as supported by studies such as Imtiaz et al~\cite{imtiaz2019challenges}. However, true positive vulnerabilities require more analysis since the alert must be resolved, while false positives can ignored. In the SAST assignment\footnote{the full text of the assignment is available under Project Part 1 in Appendix~\ref{app:Assignments}}, students were required to analyze at least 10 alerts. To avoid incentivizing students to reduce their workload by classifying alerts as false positives, students were instructed ``If you have more than 5 false positives, keep choosing alerts until you have 5 true positives while still reporting the false positives''. Similarly, as noted in Section~\ref{RQ1-DC-Apply-SMPT} as part of ensuring that the vulnerability counts for RQ1 were not influenced by duplicate or erroneous test cases, researchers reviewed and de-duplicated the student-developed SMPT test cases, as well as writing additional test cases to increase coverage. In our experience\footnote{the first author has over 2 years of industry testing experience}, cursory review of test cases developed by less experienced testers, is necessary in some industry contexts to ensure the resulting test suite can be run efficiently and effectively. While our review was more extensive, reviewing the test cases for RQ1 was intended to ensure a more accurate test process and resulting vulnerability count. Since time was not considered in RQ1, the additional time spent on reviewing test cases for RQ1 would not impact the results. }

\section{Discussion}\label{sec:discussion}
\RRChange{In Section \ref{sec:objectives}, we provide examples of how our results may help inform practitioners' decisions based on their objectives, particularly for projects in a similar domain to OpenMRS. In Section~\ref{sec:resources} we discuss how the availability or limitation of resources, specifically Expertise, Time, and Equipment, may also impact which technique should be used. These two sections should be used together, and not independently. For example, as we note in Section~\ref{sec:objectives}, we found EMPT to be very effective at finding Injection vulnerabilities such as XSS. Hence if the objective is to find high-priority injection vulnerabilities, EMPT may be a good option. However, as we note in Section~\ref{sec:resources}, EMPT requires expertise. Hence if an organization does not have enough individuals with sufficient expertise to apply EMPT effectively, practitioner may need to look to other techniques. Additionally, there may be tradeoffs between techniques when practitioners focus on one objective over another.  A manager may want to avoid manual techniques in order to reduce the perceived effort for their team. However, in our context automated techniques were less effective in terms of the coverage of different vulnerability types and the severity of vulnerabilities found.}

\RRChange{In Section~\ref{sec:disc-researchers} we go over findings which have additional implications relevant to  research and other evaluations of vulnerability detection techniques. We would encourage any researcher or other individual comparing vulnerability detection techniques to also be aware of our findings in Sections~\ref{sec:objectives} and~\ref{sec:resources}.}

\subsection{\RRChange{Organizational Objectives}}\label{sec:objectives}
\RRChange{We provide four examples of how our results might inform practitioner decisions on which vulnerability detection techniques to use.}
\subsubsection{\RRChange{Specific Vulnerability Types (Effectiveness)}}
\RRChange{Practitioners may know that a certain class of vulnerabilities is likely prevalent in the system, or may be more concerned about a certain class of vulnerabilities than others. In OpenMRS, for example, there were many XSS and other injection vulnerabilities. EMPT was particularly effective at finding these vulnerabilities, as shown in Table~\ref{tab:OWASPTopTen2}, which may have contributed to EMPT's relatively high effectiveness overall. In other words, EMPT would be a good choice for someone working on an open-source Java-based medical application such as OpenMRS where Injection vulnerabilities are a problem. As noted by other comparisons of vulnerability detection tools\cite{delaitre2018sate,bau2012vulnerability}, which tool is most effective may vary across domains. Hence practitioners should look to vulnerability detection technique evaluations in their own domain. }
\begin{tcolorbox}[colback=black!10!white,colframe=black!10!white,halign=center,valign=center,boxrule=0.0mm, boxsep=0.1mm]
\RRChange{If an organization is trying to target a specific type of vulnerability, they should focus their vulnerability detection efforts on techniques that are effective at finding that type of vulnerability in systems from their domain. For example, a project similar to OpenMRS looking to find Injection (A03) vulnerabilities may benefit from EMPT as seen in Table~\ref{tab:OWASPTopTen2}.}
\end{tcolorbox}

\subsubsection{\RRChange{Coverage (Effectiveness)}}\label{disc-coverage}
\RRChange{While EMPT performed particularly well in our context, SMPT provided higher coverage across the OWASP Top Ten 2021 categories than other techniques, as shown in Table \ref{tab:OWASPTopTen2}. SMPT also provided more coverage of the OWASP Top Ten Categories in the current study as compared with the prior work by Austin et  al.\cite{austin2011onetechniquenotenough} as shown in Table~\ref{tab:PreviousWork}. As another example, if the goal of the practitioner is to thoroughly cover Logging and Monitoring concerns, a test suite based on Level 1 of the ASVS may not be preferable since the first level of the ASVS only contains two controls relating to logging. As seen in Table~\ref{tab:PreviousWork}, the previous test suite was much more effective at finding vulnerabilities associated with Security Logging and Monitoring failures.}
\begin{tcolorbox}[colback=black!10!white,colframe=black!10!white,halign=center,valign=center,boxrule=0.0mm, boxsep=0.1mm]
\RRChange{Our results suggest that if a practitioner needs a vulnerability detection technique that effectively covers important types of vulnerabilities, a more systematic technique, such as SMPT may be more effective than EMPT.}
\end{tcolorbox}

\subsubsection{\RRChange{Automation (Efficiency)}}
\RRChange{When considering automated tools, practitioners should note that automated techniques may not inherently be more efficient than manual techniques. An organization may choose to use automated tools for other reasons such as the need to integrate automated tools with continuous deployment pipelines\cite{rahman2015synthesizing}, and using automated tools is better than not performing vulnerability detection at all. However, our results indicate that using a limited range of techniques will also miss vulnerabilities that would be found by other techniques.}
\begin{tcolorbox}[colback=black!10!white,colframe=black!10!white,halign=center,valign=center,boxrule=0.0mm, boxsep=0.1mm]
\RRChange{Our results suggest that if an organization is considering whether to use an automated technique over a manual one, they should not assume that the automated technique will be more efficient. Our results suggest that manual techniques are comparable or better than automated techniques in terms of efficiency.}
\end{tcolorbox}

\subsubsection{\RRChange{Percieved Effort and Ease of Use(Other Factors)}}\label{disc-effort}
\RRChange{No one wants to do more hard work than necessary. Ease-of-use is associated with technology evaluation and adoption~\cite{davis1989perceived,kitchenham1996desmet}.Two of the three most-frequently-mentioned concepts in the students' free-form responses, \textit{Effort} and \textit{Expertise}, are associated with the broader concept of Perceived Ease of Use~\cite{davis1989perceived}.  As we discuss in Section~\ref{sec:results-qualitative}, \textit{effort} was predominantly seen as negative for manual techniques but views of effort were mixed for automated techniques. While views of expertise were less negative expertise was associated with manual techniques (SMPT and EMPT) more than automated techniques (DAST and SAST). In the same assignment where students reported spending more time to find fewer vulnerabilities with SAST and particularly DAST as compared with EMPT and to a lesser extent SMPT, students also claimed they considered Time and effort to be a disadvantage of manual techniques as discussed in our answer to RQ3.  Additionally, while Time may be an indicator of a technque's performance, the actual times recorded for manual techniques were, on average, no longer than the times recorded for automated techniques. Hence our findings suggest that time was \textit{perceived} as a disadvantage of manual techniques even if they actually required no more time than automated techniques. As noted by Gon{\c{c}}ales et al.\cite{gonccales2021measuring}, little empirical research has been done on the cognitive load of review-related tasks such as software testing. Pfahl et al's interviews of practitioners also found that EMPT was perceived as being less easy-to-use and requiring more skill. However, studies of SAST tools have also found Ease-of-Use concerns with SAST~\cite{smith2020SAST}.  Hence while we cannot make universal claims about all automated tools, practitioners looking for the ``easiest'' solution may wish to minimize their use of manual techniques.}
 
 \begin{tcolorbox}[colback=black!10!white,colframe=black!10!white,halign=center,valign=center,boxrule=0.0mm, boxsep=0.1mm]
\RRChange{Our results suggest that if an organization is looking for the technique that will be perceived as requiring the least amount of effort, they may want to avoid manual techniques (SMPT and EMPT).}
\end{tcolorbox}

\subsection{\RRChange{Resources to Consider}}\label{sec:resources}
\RRChange{The resources below represent factors that should be considered when selecting a vulnerability detection technique for a system such as OpenMRS. The availability of resources may be more important than an organization's objectives when selecting a vulnerability detection technique.  Two of the three resources were highlighted by student responses in RQ3, while the third resource provided a much more severe limitation on our experiment than anticipated.}

\subsubsection{\RRChange{Expertise}}
\RRChange{As noted by the students in RQ3 and supported by prior work \cite{itkonen2013role,itkonen2014test}, expertise plays a role in vulnerability detection, particularly EMPT. The effectiveness of the students both combined as shown in Table~\ref{tab:OWASPTopTen2}, and on average as discussed in Section~\ref{sec:effect-austin}, as well as their efficiency  shown in Figure~\ref{fig:EfficiencyBoxPlot} is promising. Students with an introductory knowledge of Security were efficient and effective with EMPT. Anecdotally based on our experience with RQ1 as well as in a student response to RQ3, experience may also impact the efficiency and effectiveness of automated techniques more than we expected. Particularly for SAST, the researchers were more efficient than the average student group.} 

\begin{tcolorbox}[colback=black!10!white,colframe=black!10!white,halign=center,valign=center,boxrule=0.0mm, boxsep=0.1mm]
\RRChange{Our findings support related work suggesting that availability of analysts with security expertise should be considered when selecting a technique. EMPT in particular is known to require analysts with some expertise. More research is needed to fully understand the role of expertise in applying automated techniques.}
\end{tcolorbox}

\subsubsection{\RRChange{Time}}\label{disc-time}
\RRChange{As noted by the students in RQ3, the amount of time an analyst can spend on a technique may influence the efficiency and effectiveness of a technique. While EMPT requires more expertise, little or no preparation is needed. SMPT, EMPT, and DAST take more time to setup. However, as noted by the students, as discussed in Section \ref{sec:results-rq3-time}, some techniques such as DAST may perform better if practitioners have an extended timeframe in which to apply the technique. As discussed in our comparison with Austin et al. in Section~\ref{sec:effect-austin}, a single individual performing EMPT for a longer period of time did not find more vulnerabilities than were found in a shorter timeframe.}

\begin{tcolorbox}[colback=black!10!white,colframe=black!10!white,halign=center,valign=center,boxrule=0.0mm, boxsep=0.1mm]
\RRChange{The amount of available time for using the technique should be considered when selecting a technique. Some techniques, such as DAST, may benefit from a longer timeframe. }
\end{tcolorbox}

\subsubsection{\RRChange{Equipment}}
\RRChange{Both SAST and DAST required significant equipment to run. As discuss in Section \ref{sec:InternalValidity}, students were only able to run SAST on smaller modules of OpenMRS using the base VMs provided by the school. Students also were only able to run DAST-2 against the login page of OpenMRS and equipment constraints played a role in determining how we could run DAST-2 systematically, and in a way that could be compared with other tools and techniques. Austin et al., as well as similar studies of industry scanners\cite{amankwah2020empirical,scandariato2013static,bau2012vulnerability} do not mention equipment constraints. Where the equipment used in the experiment is mentioned mentioned\cite{amankwah2020empirical}, it is implied that the tools were able to be run on machines similar to the VMs used by students of the graduate class. We found that a thorough evaluation of a ``large-scale'' system required more computing resources than we expected for both SAST and DAST tools. While OpenMRS is ``large'' for an evaluation of vulnerability detection techniques, OpenMRS is less than 4 million lines of code - relatively small compared to many realistic systems\cite{desjardins2017LoC,anderson2020LinuxLoC}.  Equipment constraints should be considered by practitioners when selecting a vulnerability detection technique, particularly when considering DAST and to a lesser extent SAST. Researchers evaluating vulnerability detection techniques should be aware of this potential constraint when setting up experiments. }

\begin{tcolorbox}[colback=black!10!white,colframe=black!10!white,halign=center,valign=center,boxrule=0.0mm, boxsep=0.1mm]
\RRChange{Equipment constraints may influence the effectiveness and efficiency of vulnerability detection techniques on realistic systems. Automated techniques require more computational power than manual techniques.}
\end{tcolorbox}

\subsection{\RRChange{Implications for Evaluating Vulnerability Detection Techniques}}\label{sec:disc-researchers}
\RRChange{The resource concerns highlighted in Section \ref{sec:resources} should be considered not only by practitioners but by researchers evaluating vulnerability detection techniques. Our results also have several implications specific to future evaluations of vulnerability detection techniques}

\subsubsection{True Positive Failure Count vs Vulnerability Count (Ratio)}
We start with an observation that is not discussed in much of the related work, but which may impact the results of any study comparing vulnerability detection tools or techniques. The ratio between the number of tool alerts or failing test cases, i.e. ``true positive failures'', and the number of vulnerabilities varies across tools and techniques. As can be seen in Table \ref{tab:VulnCounting}, particularly for DAST tools, the number of alerts was many times the number of vulnerabilities found. The high ratio of alerts to vulnerabilities is consistent with the finding from Klees et al.'s~\cite{klees2018evaluating} work with fuzzers that `` \textit{`unique crashes' massively overcount[s] the number of true bugs}''. For SMPT and SAST, on the other hand, the number of true positive alerts was slightly higher than the number of vulnerabilities but with a much lower ratio of True Positive Failures to Vulnerabilities when compared with DAST. 

More research is needed to fully understand the impact of having a higher or lower number of failures per vulnerability. If additional failures present more information about the vulnerability itself, having more alerts per vulnerability may be helpful for analysts and developers attempting to triage and fix the vulnerability. However, reviewing and analyzing these failures takes time, a potential disadvantage of having a higher number of alerts per vulnerability.  For developers who may use SAST tools built into their IDE to review code while it is being written, the actual vulnerability count and consequently the difference between SAST alert count and final vulnerability count may not have much impact. On the other hand, if a practitioner is using the overall alert count or vulnerability count to determine the cybersecurity risk of an application, such as for insurance estimates~\cite{dambra2020sok}, the difference between alert count and vulnerability count may have a larger impact. Researchers should be cautious when using alerts, failing test cases, or similar true positive failures as a proxy for the number of vulnerabilities found in a system.

\begin{tcolorbox}[colback=black!10!white,colframe=black!10!white,halign=center,valign=center,boxrule=0.0mm, boxsep=0.1mm]
 A consistent set of counting rules should be used when comparing the effectiveness of different tools or techniques. It cannot be assumed that tools or techniques use the same counting rules.
\end{tcolorbox}

\subsubsection{SAST tools had fewer False Positives than expected.}\label{sec:disc-SAST-FP}
High False Positive counts have historically been considered a drawback of SAST tools\cite{imtiaz2019challenges,johnson2013don,scandariato2013static,smith2015questions,hafiz2016game}. Our results suggest that, at least for our context, SAST actually has few false positives. The high precision of SAST tools for this study is similar to  results from recent SATE events~\cite{delaitre2018sate}. More research is needed to better understand the circumstances under which a lower false positive count may generalize and the relationship between the perception that SAST tools produce large numbers of false positives and the actual false positives produced by tools.
\begin{tcolorbox}[colback=black!10!white,colframe=black!10!white,halign=center,valign=center,boxrule=0.1mm, boxsep=0.1mm]
Our research supports the findings of recent SATE comparisons that some SAST tools have low false positive counts when applied to Java applications. The lower false positive rate opens up new questions about why the percentage of false positives is often perceived as a problem for SAST techniques, and whether false positive counts have improved in other contexts.
\end{tcolorbox}

\section{Conclusions and Future Work}\label{sec:conclusion}
The motivation for this paper came from practitioner questions about which vulnerability detection techniques they should use and whether the vulnerability detection could just be done automated. After ten years, with a changing vulnerability landscape, and many improvements in vulnerability detection techniques such as the more common use of symbolic execution and taint tracing in SAST tools\cite{Mallet2016SonarAnalyzer,Campbell2020TaintAnalysis} results from previous work by Austin et al. were no longer assured to hold true.  We replicated the previous work, this time examining at least two tools for each category of technique. The main finding of Austin et al. still holds - each approach to vulnerability detection found vulnerabilities NOT found by the other techniques. If the goal of an organization is to find ``all'' vulnerabilities in their system, they need to use as many techniques as their resources allow. 

We hope to leverage the lessons learned from this experience in future work. For an empirical comparison of vulnerability detection techniques in a large-scale application, we found that even simple measures, such as vulnerability count, are not entirely objective and require strict guidelines for the count to be interpretable and the results replicable. More research is needed to understand how vulnerability detection techniques compare in terms of other measures, such as exploitability, as well as how to apply those measures in the context of large-scale web applications. Additionally, an emerging class of automated vulnerability detection technique, sometimes referred to as ``hybrid'' techniques, combines static analysis with aspects of dynamic analysis\cite{chaim2018we,liu2019survey} and is considered ``promising''\cite{chaim2018we}. While out of scope for the replication study, we look forward to expanding our comparison of vulnerability detection techniques to include these and other tools and techniques.

An additional area of future work is further exploration of vulnerability severity and related measures such as exploitability. Although not included in the original study by Austin et al.~\cite{austin2011onetechniquenotenough,austin2013comparison}, finding and mitigating one high severity vulnerability may be more important than finding and mitigating multiple lower severity vulnerabilities. However, different perspectives on severity result in different prioritization. As can be seen from Table~\ref{tab:OWASPTopTen2}, analyses of severity or criticality do not always agree. Both DAST and SAST found vulnerabilities that the tools themselves classified as ``low severity'' but which were associated with ``Broken Access Control'', the \#1 most critical vulnerabilities according to the OWASP Top Ten. Similarly, vulnerabilities associated with information disclosure through error messages, associated with \#4 in the OWASP Top Ten - ``Insecure Design'' were not considered particularly critical in this context. There is more consensus between the severity measures when we look at what is \emph{not} important. Over half (382 out of 704) of the less severe vulnerabilities were associated with CWEs not mapped to the OWASP Top Ten. More research is needed to better understand which severity measures to use in a particular context.

\section{Acknowledgements}
We thank Jiaming Jiang for her support as teaching assistant for the security class. We are grateful to the I/T staff at the university for their assistance in ensuring that we had sufficient computing power running the course. We also thank the students in the software security class.  Finally, we thank all the members of the Realsearch research group for their valuable feedback through this project.

This material is based upon work supported by the National Science Foundation under Grant No. 1909516.  Any opinions, findings, and conclusions or recommendations expressed in this material are those of the author(s) and do not necessarily reflect the views of the National Science Foundation.

\bibliographystyle{spbasic}      % basic style, author-year citations
\bibliography{elder}   % name your BibTeX data base

\newpage
\appendix
\section{Appendix - Automated Technique CWEs}\label{app:AutomatedCWEs}
\renewcommand{\arraystretch}{1.3}
\begin{longtable}{ B{20pt} | B{185pt} |C{31pt} || C{21pt} | C{21pt} | C{21pt} | C{21pt} }

% \caption{CWEs of Medium or High Severity}\label{tab:long_CWE_new}\\
\caption{CWEs covered in the rules implemented by automated techniques}\label{tab:automated_cwe}\\
\centering
CWE ID & CWE Name & OWASP Top Ten & ZAP & DA-2  & Sonar & SA-2 \\
\endfirsthead
ID &  Name & Top Ten & ZAP & DA-2  &  Sonar & SA-2 \\
\endhead
\hline\hline
% ID&Name&Top Ten&ZAP&DAST-2&Sonar&SAST-2\\\hline
22&Path Traversal&A01&X&X&&X\\\hline
23&Relative Path Traversal&A01&&X&&\\\hline
200&Exposure of Sensitive Information to an Unauthorized Actor&A01&X&&&X\\\hline
201&Insertion of Sensitive Information Into Sent Data&A01&&&&X\\\hline
264&{Permissions, Privileges, and Access Controls}&A01&X&&&\\\hline
284&Improper Access Control&A01&X&&&X\\\hline
285&Improper Authorization&A01&&&&X\\\hline
352&Cross-Site Request Forgery (CSRF)&A01&X&X&X&X\\\hline
359&Exposure of Private Personal Information to an Unauthorized Actor&A01&&&&X\\\hline
425&Forced Browsing&A01&&&&X\\\hline
601&Open Redirect&A01&X&&&X\\\hline
668&Exposure of Resource to Wrong Sphere&A01&&&&X\\\hline
862&Missing Authorization&A01&&&&X\\\hline
863&Incorrect Authorization&A01&&&&X\\\hline
1275&Sensitive Cookie with Improper SameSite Attribute&A01&X&&&\\\hline
296&Improper Following of a Certificate's Chain of Trust&A02&&&&X\\\hline
319&Cleartext Transmission of Sensitive Information&A02&&&&X\\\hline
321&Use of Hard-coded Cryptographic Key&A02&&X&&X\\\hline
322&Key Exchange without Entity Authentication&A02&&X&&\\\hline
325&Missing Cryptographic Step&A02&&X&&\\\hline
326&Inadequate Encryption Strength&A02&X&X&X&\\\hline
327&Use of a Broken or Risky Cryptographic Algorithm&A02&&&X&\\\hline
328&Use of Weak Hash&A02&&&X&\\\hline
330&Use of Insufficiently Random Values&A02&&X&X&X\\\hline
336&Same Seed in Pseudo-Random Number Generator (PRNG)&A02&&&X&X\\\hline
337&Predictable Seed in Pseudo-Random Number Generator (PRNG)&A02&&&X&X\\\hline
523&Unprotected Transport of Credentials&A02&&X&&\\\hline
760&Use of a One-Way Hash with a Predictable Salt&A02&&&&X\\\hline
916&Use of Password Hash With Insufficient Computational Effort&A02&&&&X\\\hline
20&Improper Input Validation&A03&&&&X\\\hline
74&Injection&A03&&&&X\\\hline
78&OS Command Injection&A03&X&X&&X\\\hline
79&Cross-site Scripting&A03&X&X&&X\\\hline
83&Improper Neutralization of Script in Attributes in a Web Page&A03&&&&X\\\hline
88&Argument Injection&A03&&&&X\\\hline
89&SQL Injection&A03&X&X&&X\\\hline
90&LDAP Injection&A03&&X&&X\\\hline
91&XML Injection (aka Blind XPath Injection)&A03&&X&&\\\hline
93&CRLF Injection&A03&X&&&\\\hline
94&Code Injection&A03&X&&&X\\\hline
95&Eval Injection&A03&&&&X\\\hline
97&Improper Neutralization of Server-Side Includes (SSI) Within a Web Page&A03&X&&&\\\hline
98&PHP Remote File Inclusion&A03&X&&&\\\hline
99&Resource Injection&A03&&&&X\\\hline
113&HTTP Response Splitting&A03&&&&X\\\hline
184&Incomplete List of Disallowed Inputs&A03&&&&X\\\hline
470&Unsafe Reflection&A03&&&&X\\\hline
610&Externally Controlled Reference to a Resource in Another Sphere&A03&&&&X\\\hline
643&XPath Injection&A03&&&&X\\\hline
917&Expression Language Injection&A03&&&&X\\\hline
73&External Control of File Name or Path&A04&&&&X\\\hline
183&Permissive List of Allowed Inputs&A04&&&&X\\\hline
209&Generation of Error Message Containing Sensitive Information&A04&&X&&X\\\hline
311&Missing Encryption of Sensitive Data&A04&X&&&\\\hline
313&Cleartext Storage in a File or on Disk&A04&&&&X\\\hline
472&External Control of Assumed-Immutable Web Parameter&A04&X&&&\\\hline
501&Trust Boundary Violation&A04&&&&X\\\hline
522&Insufficiently Protected Credentials&A04&&&X&\\\hline
525&Use of Web Browser Cache Containing Sensitive Info. &A04&X&&&\\\hline
642&External Control of Critical State Data&A04&X&&&X\\\hline
646&Reliance on File Name or Extension of Externally-Supplied File&A04&&&&X\\\hline
650&Trusting HTTP Permission Methods on the Server Side&A04&&&&X\\\hline
770&Allocation of Resources Without Limits or Throttling&A04&&&&X\\\hline
807&Reliance on Untrusted Inputs in a Security Decision&A04&&&X&\\\hline
927&Use of Implicit Intent for Sensitive Communication&A04&&&&X\\\hline
1021&Improper Restriction of Rendered UI Layers or Frames&A04&X&&&\\\hline
7&J2EE Misconfiguration: Missing Custom Error Page&A05&&&&X\\\hline
% 11&ASP.NET Misconfiguration: Creating Debug Binary&A05&&&&X\\\hline
% 12&ASP.NET Misconfiguration: Missing Custom Error Page&A05&&&&X\\\hline
% 13&ASP.NET Misconfiguration: Password in Configuration File&A05&&&&X\\\hline
315&Cleartext Storage of Sensitive Information in a Cookie&A05&&&&X\\\hline
541&Inclusion of Sensitive Information in an Include File&A05&X&&&\\\hline
548&Exposure of Information Through Directory Listing&A05&X&&&\\\hline
611&Improper Restriction of XML External Entity Reference&A05&&&X&X\\\hline
614&Sensitive Cookie in HTTPS Session Without 'Secure' Attribute&A05&X&&&X\\\hline
776&XML Entity Expansion&A05&&&&X\\\hline
933&Security Misconfiguration&A05&X&&&\\\hline
942&Permissive Cross-domain Policy with Untrusted Domains&A05&&&&X\\\hline
1004&Sensitive Cookie Without 'HttpOnly' Flag&A05&X&&&X\\\hline
% 1032&Security Misconfiguration&A05&&&&X\\\hline
% 1035&Using Components with Known Vulnerabilities&A06&&&&X\\\hline
259&Use of Hard-coded Password&A07&&&&X\\\hline
263&Password Aging with Long Expiration&A07&&&&X\\\hline
287&Improper Authentication&A07&&X&&X\\\hline
288&Authentication Bypass Using an Alternate Path or Channel&A07&&&&X\\\hline
295&Improper Certificate Validation&A07&&X&X&X\\\hline
297&Improper Validation of Certificate with Host Mismatch&A07&&&&X\\\hline
300&Channel Accessible by Non-Endpoint&A07&&&&X\\\hline
307&Improper Restriction of Excessive Authentication Attempts&A07&&&&X\\\hline
346&Origin Validation Error&A07&&&&X\\\hline
384&Session Fixation&A07&&&&X\\\hline
521&Weak Password Requirements&A07&&&X&X\\\hline
613&Insufficient Session Expiration&A07&&&&X\\\hline
798&Use of Hard-coded Credentials&A07&&&&X\\\hline
345&Insufficient Verification of Data Authenticity&A08&X&&&X\\\hline
502&Deserialization of Untrusted Data&A08&&&X&X\\\hline
565&Reliance on Cookies without Validation and Integrity Checking&A08&X&&&X\\\hline
829&Inclusion of Functionality from Untrusted Control Sphere&A08&X&&&X\\\hline
915&Improperly Controlled Modification of Dynamically-Determined Object Attributes&A08&&&X&X\\\hline
532&Insertion of Sensitive Information into Log File&A09&&&&X\\\hline
778&Insufficient Logging&A09&&&&X\\\hline
4&J2EE Environment Issues (Deprecated)&NM&&&&X\\\hline
36&Absolute Path Traversal&NM&&X&&\\\hline
41&Improper Resolution of Path Equivalence&NM&&X&&\\\hline
67&Improper Handling of Windows Device Names&NM&&X&&\\\hline
102&Struts: Duplicate Validation Forms&NM&&&X&\\\hline
112&Missing XML Validation&NM&&X&&\\\hline
118&Range Error&NM&&X&&\\\hline
120&Buffer Overflow&NM&X&X&&\\\hline
124&Buffer Underflow&NM&&X&&\\\hline
134&Use of Externally-Controlled Format String&NM&X&X&&\\\hline
140&Improper Neutralization of Delimiters&NM&&X&&\\\hline
144&Improper Neutralization of Line Delimiters&NM&&X&&\\\hline
149&Improper Neutralization of Quoting Syntax&NM&&X&&\\\hline
150&{Improper Neutralization of Escape, Meta, or Control Sequences}&NM&&X&&\\\hline
154&Improper Neutralization of Variable Name Delimiters&NM&&X&&\\\hline
156&Improper Neutralization of Whitespace&NM&&X&&\\\hline
157&Failure to Sanitize Paired Delimiters&NM&&X&&\\\hline
158&Improper Neutralization of Null Byte or NUL Character&NM&&X&&\\\hline
166&Improper Handling of Missing Special Element&NM&&X&&\\\hline
172&Encoding Error&NM&&X&&\\\hline
174&Double Decoding of the Same Data&NM&&X&&\\\hline
175&Improper Handling of Mixed Encoding&NM&&X&&\\\hline
176&Improper Handling of Unicode Encoding&NM&&X&&\\\hline
177&Improper Handling of URL Encoding (Hex Encoding)&NM&&X&&\\\hline
185&Incorrect Regular Expression&NM&&X&&X\\\hline
189&Numeric Error&NM&&X&&\\\hline
190&Integer Overflow or Wraparound&NM&&&&X\\\hline
194&Unexpected Sign Extension&NM&&X&&\\\hline
215&Insertion of Sensitive Information Into Debugging Code&NM&&&&X\\\hline
242&Use of Inherently Dangerous Function&NM&&&&X\\\hline
252&Unchecked Return Value&NM&&&&X\\\hline
253&Incorrect Check of Function Return Value&NM&&&&X\\\hline
289&Authentication Bypass by Alternate Name&NM&&&&X\\\hline
299&Improper Check for Certificate Revocation&NM&&&&X\\\hline
314&Cleartext Storage in the Registry&NM&&&&X\\\hline
317&Cleartext Storage of Sensitive Info. in GUI&NM&&&&X\\\hline
332&Insufficient Entropy in PRNG&NM&&&X&\\\hline
366&Race Condition within a Thread&NM&&&&X\\\hline
369&Divide By Zero&NM&&&&X\\\hline
390&Detection of Error Condition Without Action&NM&&&&X\\\hline
% 391&Unchecked Error Condition&NM&&&&X\\\hline
398&Code Quality&NM&&&&X\\\hline
399&Resource Management Error&NM&&X&&\\\hline
400&Uncontrolled Resource Consumption&NM&&&&X\\\hline
403&File Descriptor Leak&NM&&&&X\\\hline
404&Improper Resource Shutdown or Release&NM&&&&X\\\hline
406&Insufficient Control of Network Message Volume (Network Amplification)&NM&&X&&\\\hline
427&Uncontrolled Search Path Element&NM&&&&X\\\hline
436&Interpretation Conflict&NM&X&&&\\\hline
476&NULL Pointer Dereference&NM&&&&X\\\hline
480&Use of Incorrect Operator&NM&&&&X\\\hline
483&Incorrect Block Delimitation&NM&&&&X\\\hline
484&Omitted Break Statement in Switch&NM&&&&X\\\hline
489&Active Debug Code&NM&&&X&\\\hline
493&Critical Public Variable Without Final Modifier&NM&&&X&\\\hline
500&Public Static Field Not Marked Final&NM&&&X&\\\hline
% 519&.NET Environment Issues (Deprecated)&NM&&&&X\\\hline
% 530&Exposure of Backup File to an Unauthorized Control Sphere&NM&&&&X\\\hline
543&Use of Singleton Pattern Without Synchronization in a Multithreaded Context&NM&&&&X\\\hline
561&Dead Code&NM&&&&X\\\hline
563&Assignment to Variable without Use&NM&&&&X\\\hline
567&Unsynchronized Access to Shared Data in a Multithreaded Context&NM&&&&X\\\hline
568&finalize() Method Without super.finalize()&NM&&&&X\\\hline
569&Expression Issue&NM&&&&X\\\hline
% 570&Expression is Always False&NM&&&&X\\\hline
573&Improper Following of Specification by Caller&NM&&&&X\\\hline
580&clone() Method Without super.clone()&NM&&&&X\\\hline
582&{Array Declared Public, Final, and Static}&NM&&&X&\\\hline
% 595&Comparison of Object References Instead of Object Contents&NM&&&&X\\\hline
599&Missing Validation of OpenSSL Certificate&NM&&&&X\\\hline
600&Uncaught Exception in Servlet&NM&&&X&\\\hline
607&Public Static Final Field References Mutable Object&NM&&&X&\\\hline
615&Inclusion of Sensitive Information in Source Code Comments&NM&&&&X\\\hline
% 617&Reachable Assertion&NM&&&&X\\\hline
628&Function Call with Incorrectly Specified Arguments&NM&&&&X\\\hline
661&Weaknesses in Software Written in PHP&NM&&&&X\\\hline
% 662&Improper Synchronization&NM&&&&X\\\hline
665&Improper Initialization&NM&&&&X\\\hline
% 667&Improper Locking&NM&&&&X\\\hline
670&Always-Incorrect Control Flow Implementation&NM&&&&X\\\hline
683&Function Call With Incorrect Order of Arguments&NM&&&&X\\\hline
688&Function Call With Incorrect Variable or Reference as Argument&NM&&&&X\\\hline
693&Protection Mechanism Failure&NM&X&&&\\\hline
704&Incorrect Type Conversion or Cast&NM&&&&X\\\hline
754&Improper Check for Unusual or Exceptional Conditions&NM&&&X&\\\hline
755&Improper Handling of Exceptional Conditions&NM&&&&X\\\hline
% 764&Multiple Locks of a Critical Resource&NM&&&&X\\\hline
777&Regular Expression without Anchors&NM&&&&X\\\hline
779&Logging of Excessive Data&NM&&&&X\\\hline
783&Operator Precedence Logic Error&NM&&&&X\\\hline
827&Improper Control of Document Type Definition&NM&&&X&X\\\hline
833&Deadlock&NM&&&&X\\\hline
835&Infinite Loop&NM&&&&X\\\hline
1022&Use of Web Link to Untrusted Target with window.opener Access&NM&&&X&X\\\hline
1023&Incomplete Comparison with Missing Factors&NM&&&&X\\\hline

\end{longtable}

\newpage
\section{Appendix - Student Experience Questionnaire}\label{app:ExperienceQuestions}
At the beginning of the course, students were asked to fill out a survey about their experience relevant to the course. The four questions asked to students were as follows:
\begin{enumerate}
    \item How much time have you spent working at a professional software organization -- including internships -- in terms of the \# of years and the \# of months?
    \item On a scale from 1 (none) to 5 (fully), how much of the time has your work at a professional software organization involved cybersecurity?
    \item Which of the follow classes have you already completed?  
    \item Which of the following classes are you currently taking? 
\end{enumerate}
Q1 was short answer. For Q2, students selected a single number between 1 and 5. For Q3, the students could check any number of checkboxes corresponding to a list of the security and privacy courses offered at the institution. For Q4, the students selected from the subset of classes from question 4 that were being offered the semester in which the survey was given.

 Fifty-nine of the sixty-three students who agreed to let their data be used for the study responded to the survey. Of these 59 responses, four students responses to Q1 provided a numeric value, e.g. ``3'', but did not specify whether the numeric value indicated years or months. We considered this invalid and summarize experience from the remaining 55 participants in Section~\ref{sec:studentData}
 
\newpage
\section{Appendix - Student Assignments}\label{app:Assignments}
  
The following are the verbatim assignments for the Course Project that guided the tasks performed by students. We have removed sections of the assignment that are not relevant to this project. Additionally, information that is specific to the tools used, such as UI locations, has also been removed. Text that has been removed is indicated by square brackets [~].
  
\subsection{\textbf{Project Part 1}}
Throughout the course of this semester, you will perform and document a technical security review of OpenMRS (\url{http://openmrs.org}).  This open-source systems provides electronic health care functionality for ``resource-constrained environments''.   While the system has not been designed for deployment within the United States, security and privacy concerns are still a paramount security concern for any patient.

\vspace*{8pt}
\noindent\textbf{\textit{Software:}}

\begin{sloppypar}
\noindent OpenMRS 2.9.0. There is no need to install OpenMRS.  You will use the VCL image CSC515\_SoftwareSecurity\_Ubuntu.
\end{sloppypar}

\vspace*{8pt}
\noindent\textbf{\textit{Deliverables:}}

\noindent Submit a PDF with all deliverables in Gradescope.  Only one submission should be performed per team.  Do not include your names/IDs/team name on the report to facilitate the peer evaluation of your assignment (see Part 3 of this assignment).  

\vspace*{12pt}
\noindent{\large 1.  Security test planning and execution (45 points) } 

\vspace*{4pt}
\noindent a.  Record how much total time (hours and minutes) your team spends to complete this activity (test planning and test execution). Compute a metric of how many true positive defects you found per hour of total effort.  

\vspace*{4pt}
\noindent b.  \textbf{Test planning.} Create 15 black box test cases to start a repeatable black box test plan for the OpenMSR (Version 2.9).   You may find the OWASP Testing Guide and OWASP Proactive Controls helpful references in addition to the references provided throughout the ASVS document.  

\vspace*{4pt}
\noindent For each test case, you must specify:
\begin{itemize}
    \item A unique test case id that maps to the ASVS, sticking to Level 1 and Level 2. Provide the name/description of the ASVS control. Only one unique identifier is needed (as opposed to the example in the lecture slides). The ASVS number should be part of the one unique identifier. 
    \item Detailed and repeatable (the same steps could be done by anyone who reads the instructions) instructions for how to execute the test case
    \item Expected results when running the test case.  A passing test case would indicate a secure system.
    \item Actual results of running the test case.  
    \item Indicate the CWE (number and name) for the vulnerability you are testing for.
\end{itemize}
\noindent\textbf{In choosing your test cases, we are looking for you to demonstrate your understanding of the vulnerability and what it would take to stress the system to see if the vulnerability exists. \underline{You may have only one test case per ASVS control.}}

\vspace*{4pt} 
\noindent c.  Extra credit (up to 5 points):  Create a black box test case that will reveal the vulnerability reported by the static analysis tool (Part 2 of this assignment) for up to 5 vulnerabilities (1 point per vulnerability).  Provide the tool output (screen shot of the alert) from each tool. 

\vspace*{12pt}
\noindent{\large 2.  Static analysis (45 points)}

\vspace*{4pt}
\noindent a.  Record how much total time (hours and minutes) your team spends to complete this activity (test planning and test execution).   Compute a metric of how many defects you found per hour of total effort.  

\vspace*{4pt}
\noindent b.  For each of the three tools (below), review the security reports.  Based upon these reports:
\begin{itemize}
    \item References: 
    \begin{itemize}
        \item Troubleshooting VCL
        \item Opening OpenMRS on VCL
    \end{itemize}
    \item Randomly choose 10 security alerts and provide a cross-reference back to the originating report(s) where the alert was documented. Explore the code to determine if the alert is a false positive or a true positive. The alerts analyzed MUST be security alerts even though the tools will report ``regular quality'' alerts -- you need to choose security alerts.
    \item If the alert is a false positive, explain why. If you have more than 5 false positives, keep choosing alerts until you have 5 true positives while still reporting the false positives (which may make you go above a total of 10).
    \item If the alert is a true positive, (1) explain how to fix the vulnerability; (2) map the vulnerability to a CWE; (3) map the vulnerability to the ASVS control.
    \item Find the instructions for getting [SAST-3] going on OpenMRS here[hyperlink removed].  [Tool-specific instructions] 
    \item Find the instructions for getting [SAST-2] going on OpenMRS here[hyperlink removed]. [Tool-specific instructions]
    % \item \textbf{(Deleted from Assignment) Find the instructions for getting [SAST-1] going on OpenMRS here[hyperlink removed] Take screenshots to document your work.}
\end{itemize}

\vspace*{4pt}
\noindent c.  Extra credit (up to 5 points):  Find 5 instances (1 point per instance) of a potential vulnerability being reported by multiple tools.  Provide the tool output (screen shot of the alert) from each tool. Explore the code to determine if the alert is a false positive or a true positive.   If the alert is a false positive, explain why.  If the alert is a true positive, explain how to fix the vulnerability.

\vspace*{12pt}
\noindent{\large 3.  Peer evaluation (10 points)}

\vspace*{4pt}
\noindent Perform a peer evaluation on another team.  Produce a complete report of feedback for the other team using this rubric [to be supplied].
\vspace*{4pt}
\noindent \textbf{Note: For any part of this course-long project, you may not directly copy materials from other sources.  You need to adapt and make unique to OpenMRS.  You should provide references to your sources.  Copying materials without attribution is plagiarism and will be treated as an academic integrity violation.}

\subsection{\textbf{Project Part 2}}

The fuzzing should be performed on the VCL Class Image (``CSC 515 Software Security Ubuntu''). 

\vspace*{12pt}
\noindent{\large  0.  Black Box Test Cases}

\vspace*{4pt}
\noindent Parts 1 (OWASP ZAP) and 2 ([DAST-2]) ask for you to write a black box test case.  We use the same format as was used in Project Part.  For each test case, you must specify:
\begin{itemize}
    \item A unique test case id that maps to the ASVS, sticking to Level 1 and Level 2. Provide the name/description of the ASVS control. Only one unique identifier is needed (as opposed to the example in the lecture slides).  The ASVS number should be part of the one unique identifier. 
    \item Detailed and repeatable (the same steps could be done by anyone who reads the instructions) instructions for how to execute the test case
    \item Expected results when running the test case.  A passing test case would indicate a secure system.
    \item Actual results of running the test case.  
    \item Indicate the CWE (number and name) for the vulnerability you are testing for.
\end{itemize}

\vspace*{12pt}
\noindent{\large  1.  OWASP ZAP (30 points, 3 points for each of the 5 test cases in the two parts) }

\vspace*{4pt}
\noindent \textbf{Client-side bypassing }
\begin{itemize}
    \item Record how much total time (hours and minutes) your team spends to complete this activity. Provide:
    \begin{itemize}
        \item Total time to plan and run the 5 black box test cases.
        \item Total number of vulnerabilities found.
    \end{itemize}
    \item Plan 5 black box test cases (using format provided in Part 0 above) in which you stop user input in OpenMRS with OWASP ZAP and change the input string to an attack. (Consider using the strings that can be found in the ZAP rulesets, such as jbrofuzz)  Use these instructions as a guide.  
    \item In your test case, be sure to document the page URL, the input field, the initial user input, and the malicious input.  Describe what ``filler'' information is used for the rest of the fields on the page (if necessary). 
    \item Run the test case and document the results.   
\end{itemize}

\vspace*{4pt}
\noindent \textbf{Fuzzing}
\begin{itemize}
    \item Record how much total time (hours and minutes) your team spends to complete this activity.  
    \begin{itemize}
        \item Do not include time to run ZAP  
        \item Provide:
        \begin{itemize}
            \item Total time to work with the ZAP output to identify the 5 vulnerabilities.
            \item Total time to plan and run the 5 black box test cases.
        \end{itemize}
    \end{itemize}
    \item Use the 5 client-side bypassing testcases (above) for this exercise.   
    \item Use the jbrofuzz rulesets to perform a fuzzing exercise on OpenMRS with the following vulnerability types: Injection, Buffer Overflow, XSS, and SQL Injection. 
    \item Take a screen shot of ZAP information on the five test cases.  
    \item Report the fuzzers you chose for each vulnerability type along with the results, and what you believe the team would need to do to fix any vulnerabilities you find. If you don't find any vulnerabilities, provide your reasoning as to why that was the case, and describe and what mitigations the team must have in place such that there are no vulnerabilities.
\end{itemize}

\vspace*{12pt}
\noindent{\large  2.   DAST-2 (25 points) }

\noindent [DAST-2] FAQ~[hyperlink removed] and [DAST-2] Troubleshooting~[hyperlink removed]
\begin{itemize}
    \item Record how much total time (hours and minutes) your team spends to complete this activity.   
    \begin{itemize}
        \item Do not include time to run [DAST-2].  
        \item Provide:
        \begin{itemize}
            \item Total time to work with the [DAST-2] output to identify the 5 vulnerabilities.
            \item Total time to plan and run the 5 black box test cases.
        \end{itemize}
    \end{itemize}
    \item Run [DAST-2] on OpenMRS.  Run any 5 of your test cases from Project Part 1 to seed the [DAST-2] run.  Run [DAST-2] long enough that you feel you have captured enough true positive vulnerabilities that you can complete five test case plans. Note: [DAST-2] will like run out of memory if you run all 5 together. It is best to run each one separately. Also, make sure you capture only the steps for your test cases, not other unnecessary steps.
    \item Export your results.
    \item Take a screen shot of [DAST-2] information on the five vulnerabilities you will explore further. Write five black box test plans (using format provided in Part 0 above) to expose five vulnerabilities detected by [DAST-2] (which may use a proxy).  Hint: Your expected results should be different from the actual results since these test cases should be failing test cases.
\end{itemize}

\vspace*{12pt}
\noindent{\large  3.  Vulnerable Dependencies (35 points) }

\vspace*{4pt}
[Assignment Section not Relevant]

\vspace*{12pt}
\noindent{\large 4.   Peer evaluation (10 points) }

\vspace*{4pt}
\noindent Perform a peer evaluation on another team.  Produce a complete report of feedback for the other team using this rubric (to be supplied).

\subsection{\textbf{Project Part 3}}

The project can be done on the VCL Class Image (``CSC 515 Software Security Ubuntu''). 

\vspace*{12pt}
\noindent{\large 0.  Black Box Test Cases }

\vspace*{4pt}
\noindent Parts 1 (Logging), 2 ([Interactive Testing]), and 3 (Test coverage) ask for you to write black box test cases.  We use the same format as was used in Project Part 1.  For each test case, you must specify:
\begin{itemize}
    \item A unique test case id that maps to the ASVS, sticking to Level 1 and Level 2.   Provide the name/description of the ASVS control. Only one unique identifier is needed (as opposed to the example in the lecture slides).  The ASVS number should be part of the one unique identifier. 
    \item Detailed and repeatable (the same steps could be done by anyone who reads the instructions) instructions for how to execute the test case
    \item Expected results when running the test case.  A passing test case would indicate a secure system.
    \item Actual results of running the test case.  
    \item Indicate the CWE (number and name) for the vulnerability you are testing for.
\end{itemize}

\vspace*{12pt}
\noindent{\large 1.  Logging (25 points) }

\noindent Where are the Log files? Check out the OpenMRS FAQ
\begin{itemize}
    \item Record how much total time (hours and minutes) your team spends to complete this activity (test planning and test execution).   Compute a metric of how many true positive defects you found per hour of total effort.  
    \item Write 10 black box test cases for ASVS V7 Levels 1 and 2.   You can have multiple test cases for the same control testing for logging in multiple areas of the application.  What should be logged to support non-repudiation/accountability should be in your expected results.
    \item Run the test.  Find and document the location of OpenMRS's transaction logs.   
    \item Write what is logged in the actual results column.   The test case should fail if non-repudiation/accountability is not supported (see the 6 Ws on page 3 of the lecture notes).  
    \item Comment on the adequacy of OpenMRS's logging overall based upon these 10 test cases.
 \end{itemize}
 
\vspace*{12pt}
\noindent{\large 2.  Interactive Application Security Testing (25 points) }

\vspace*{4pt}
[Assignment Section not Relevant]

\vspace*{12pt}
\noindent{\large 3.  Test Coverage (25 points)}

\vspace*{4pt}
\noindent \textbf{This test coverage relates to all work you have done in Project Parts 1, 2, and 3.}
\begin{enumerate}
    \item Compute your black box test coverage for each section of the ASVS (i.e. V1, V2, etc.) which includes the black box tests you write for Part 2 (Seeker) for Level 1 and Level 2 controls.  You get credit for a control (e.g. V1.1) if you have a test case for it. If you have more than one test case for a control, you do not get extra credit \textendash coverage is binary.  Coverage is computed as \# of test cases / \# of requirements.  
    \item (15 points, 3 points each) Write 5 more black box tests to increase your coverage of controls you did not have a test case for.
    \item (5 points) Recompute your test coverage.  Report as below.  Record how much total time (hours and minutes) your team spends to complete this activity (test planning and test execution).   Compute a metric of how many true positive defects you found per hour of total effort.  
    \item (5 points) Reflect on the controls you have lower coverage for.  Are these controls particularly hard to test, we didn't cover in class, you just didn't get to it, etc.
\end{enumerate}

\begin{center}\renewcommand{\arraystretch}{1.3}
\begin{tabular}{ | >{\raggedright\arraybackslash}p{130pt} | >{\raggedright\arraybackslash}p{50pt} |>{\raggedright\arraybackslash}p{50pt}  | >{\raggedright\arraybackslash}p{50pt} | }
\hline
{Control} & {\# of test cases} & {\# of L1 and L2 controls} & {Coverage}\\
\hline
{V1.1: Secure development lifecycle} & ? & 7 & ?/7\\
\hline
... & & & \\
\hline
Total & & & \\
\hline
\end{tabular}
\end{center}

\vspace*{12pt}
\noindent{\large  4.  Vulnerability Discovery Comparison (15 points)}
\begin{enumerate}
    \item (5 points) Compare the five vulnerability detection techniques you have used this semester by first completing the table below.
    \begin{itemize}
        \item A:  total number \# of true positives for this detection type for all activities (Project Parts 1-3)
        \item B:  total time spent on all for all activities (Project Parts 1-3)
        \item Efficiency is A/B
        \item Exploitability:  give a relative rating of the ability for this technique to find exploitable vulnerabilities
        \item Provide the CWE number for all the true positive vulnerabilities detected by this technique.  (This information will help you address the ``wide range of vulnerability types'' question below.)
    \end{itemize}

\begin{center}\renewcommand{\arraystretch}{1.3}
\begin{tabular}{ | >{\raggedright\arraybackslash}p{67pt} | >{\raggedright\arraybackslash}p{55pt} |>{\raggedright\arraybackslash}p{30pt}  | >{\raggedright\arraybackslash}p{60pt} | >{\raggedright\arraybackslash}p{65pt} | >{\raggedright\arraybackslash}p{32pt} | }
\hline
Technique & \# of true positive vulnerabilities discovered & Total time (hours) & {Efficiency:\newline \# vulnerabilities / total time} & Detecting Exploitable vulnerabilities? (High/Med/Low) & Unique CWE numbers\\
\hline
Manual black box & & & & & \\
\hline
Static analysis & & & & & \\
\hline
Dynamic analysis & & & & & \\
\hline
Interactive testing & & & & & \\
\hline
\end{tabular}
\end{center}

\vspace*{8pt}

    \item (10 points) Use this data to re-answer the question that was on the midterm (that people generally didn't do too well on).  Being able to understand the tradeoffs between the techniques is a major learning objective of the class.
    
\textit{As efficiently and effectively as possible, companies want to detect a wide range of exploitable vulnerabilities (both implementation bugs and design flaws).  Based upon your experience with these techniques, compare their ability to efficiently and effectively detect a wide range of types of exploitable vulnerabilities.}
\end{enumerate}

\vspace*{12pt}
\noindent{\large 5.   Peer evaluation (10 points) }

\vspace*{4pt}
\noindent Perform a peer evaluation on another team.  Produce a complete report of feedback for the other team using this rubric [to be supplied].

\subsection{\textbf{Project Part 4}}

\vspace*{4pt}
\noindent{\large 1. Protection Poker (20 points)}

\vspace*{4pt}
\noindent [Assignment Section not Relevant]

\vspace*{12pt}
\noindent{\large 2. Vulnerability Fixes (35 points)}

\vspace*{4pt}
\noindent [Assignment Section not Relevant]

\vspace*{12pt}
\noindent{\large 3.  Exploratory Penetration Testing (35 points)}

\vspace*{4pt}
\noindent Each team member is to perform 3 hours of exploratory penetration testing on OpenMRS.  This testing is to be done opportunistically, based upon your general knowledge of OpenMRS but without a test plan, as is done by professional penetration testers.  DO NOT USE YOUR OLD BLACK BOX TESTS FROM PRIOR MODULES. Use a screen video/voice screen recorder to record your penetration testing actions.  Speak aloud as you work to describe your actions, such as, ``I see the input field for logging in.  I'm going to see if 1=1 works for a password.'' or ``I see a parameter in the URL, I'm going to see what happens if I change the URL.''   \textbf{You should be speaking around once/minute to narrate what you are attempting.}  You don't have to do all 3 hours in one session, but you should have 3 hours of annotated video to document your penetration testing.  There's lots of screen recorders available -- if you know of a free one and can suggest it to your classmates, please post on Piazza.  

\vspace*{6pt}
\noindent Pause the recording every time you have a true positive vulnerability.  Note how long you have been working so a log of your work and the time between vulnerability discovery is created (For example, Vulnerability \#1 was found at 1 hour and 12 minutes, Vulnerability \#2 was found at  1 hour and 30 minutes, etc.) If you work in multiple sessions, the elapsed time will pick up where you left off the prior session -- like if you do one session for 1 hour 15 minutes, the second session begins at 1 hour 16 minutes.  Take a screen shot and number each true positive vulnerability .  Record your actions such that this vulnerability could be replicated by someone else via a black box test case.  Record the CWE for your true positive vulnerability.   Record your work as in the following table.  The reference info for video traceability is to aid a reviewer in watching you find the vulnerability.  If you have one video, the ``time'' should aid in finding the appropriate part of the video.  If you have multiple videos, please specify which video and what time on that video.  

\begin{center}\renewcommand{\arraystretch}{1.3}
\begin{tabular}{ | >{\raggedright\arraybackslash}p{60pt} | >{\raggedright\arraybackslash}p{35pt} |>{\raggedright\arraybackslash}p{70pt}  | >{\raggedright\arraybackslash}p{35pt} | >{\raggedright\arraybackslash}p{50pt} | }
\hline
Vulnerability \# & Elapsed Time & Ref Info for Video Traceability & CWE & Commentary \\
\hline
& & & & \\
\hline
\end{tabular}
\end{center}

\vspace*{6pt}
\noindent Replication instructions via a black box test and the screenshots for each true positive vulnerability should appear below the table, labeled with the vulnerability number.  Since you are not recording all your steps, the replication instructions may not work completely since you may change the state of the software somewhere along the line -- document what you can via a black box test and say the actual results don't match your screenshot.      

\vspace*{6pt}
\noindent After you are complete, compute an efficiency metric (true positive vulnerability/hour) metric for each student.  Submit a table:
\begin{center}\renewcommand{\arraystretch}{1.3}
\begin{tabular}{ | >{\raggedright\arraybackslash}p{50pt} | >{\raggedright\arraybackslash}p{50pt} |>{\raggedright\arraybackslash}p{50pt}  | >{\raggedright\arraybackslash}p{50pt} | }
\hline
& \# vuln & Time & Efficiency \\
\hline
Name 1 & & &  \\
\hline
Name 2 & & &  \\
\hline
Name 3 & & &  \\
\hline
Name 4 & & &  \\
\hline
Total & & &  \\
\hline
\end{tabular}
\end{center}

\vspace*{4pt}
\noindent Copy the efficiency table you turned in for Project Part 3 \#4.  Add an additional line for Penetration testing.  Compare and comment on this efficiency rate with the other vulnerability discovery techniques in the table you input in \#4 of Project Part 3.   
\begin{itemize}
    \item Each person on the team should submit one or more videos by uploading it/them to your own google drive and providing a link to the video(s), sharing the video with anyone who has the link and an NCSU login (which will allow peer evaluation and grading).  The video(s) should be approximately 3 hours in length.  
    \item A person who does not submit a video can not be awarded the points for this part of the project while the rest of the team can.
    \item It is possible to work for 3 hours and find 0 vulnerabilities -- real penetration tests constantly work more than 3 hours without finding anything.  That's part of the reason for documenting your work via video.  
    \item For those team members who do submit videos, the grade will be an overall team grade.
\end{itemize}

\vspace*{4pt}
\noindent Submission:  The team submits one file with the links to the team member's files.

\vspace*{12pt}
\noindent{\large 4. Peer Evaluation (10 points)}

\vspace*{4pt}
\noindent Perform a peer evaluation on another team.  Produce a complete report of feedback for the other team using this rubric [to be supplied].

\newpage
 
 \section{Appendix - Equipment Specifications}\label{app:Equipment}
 In this appendix we provide additional details of the equipment used in our case study. As noted in Section~\ref{sec:method-sut-equipment}, a key resource used in this project was the school's Virtual Computing Lab\footnote{https://vcl.apache.org/} (VCL), which provided virtual machine (VM) instances. Researchers used VCL when applying EMPT, SMPT, and DAST as part of data collection for RQ1. All student tasks were performed using VCL for RQ1 and RQ2. Researchers created a system image including the SUT (OpenMRS) as well as SAST and DAST tools. The base image was assigned 4-cores, 8G RAM, and 40G disk space. An instance of this image could be checked out by students and researchers and accessed remotely through a terminal using ssh or graphically using Remote Desktop Protocol (RDP). Researchers also used two expanded instances of the base image with 16 CPUs, 32GB RAM, and 80G disk space. For client-server tools, a server was setup in a separate VCL instance by researchers with assistance from the teaching staff of the course. The server UI was accessible from VCL instances of the base image, while the server instance itself was only accessible to researchers and teaching staff. The server instance had 4 cores, 8G RAM, and 60G disk space disk space, and contained the server software for SAST-1 used to answer RQ2. All VCL instances in this study used the Ubuntu operating system. 

The VCL alone was used for data collection for RQ2. However, the base VCL images were small, and the remote connection to VCL could lag. Researchers used two used additional resources as needed for RQ1 data collection. First, we created a VM in VirtualBox using the same operating system (Ubuntu 18.04 LTS) and OpenMRS version (Version 2.9) as the VCL images. This VM was used by researchers for SMPT and EMPT data collection, particularly when reviewing the output of each technique where instances of the SUT were needed on an ad hoc basis. The VM was assigned 2 CPUs, 4GB RAM, and 32G disk space and could be copied and shared amongst researchers to run locally. Researchers increased the size of the VM as needed, up to 8 CPUs and 16GB RAM when the host system could support the VM size. A second VM was created in VirtualBox with the same specifications and operating system, but with the server software for Sonarqube installed. We also used a desktop machine with 24 CPUs, 32G RAM, and 500G disk space. The desktop was running the Ubuntu operating system. This machine was accessible through the terminal via ssh and graphically using x2go\footnote{\url{https://wiki.x2go.org/doku.php}}. For RQ1 data collection we ran the SAST-1 server software directly on this machine. The desktop was also used to run VirtualBox VMs for resource-intensive activity such as running Sonarqube and DAST-2.

\newpage
\section{Appendix - All CWEs Table}\label{app:CWEs}
Table \ref{tab:long_CWE} shows the CWE for high and medium severity vulnerabilities found. Table \ref{tab:long_CWE_low} provides the same information for low severity vulnerabilities. The first column of the table indicates the CWE number. The CWEs are organized based on the OWASP Top Ten Categories. The second column of the table indicates which, if any, of the OWASP Top Ten the vulnerability maps to. Columns three and four of the table are the number of vulnerabilities found by the techniques SMPT and EMPT. Columns five through eight break down the vulnerabilities found by DAST and SAST by tool (ZAP, DA-2, Sonar, and SA-2). Column nine of Table \ref{tab:long_CWE} shows the total number of vulnerabilities found of each CWE type. The Total column is not the same as the sum of the previous six columns. Some vulnerabilities were found using more than one technique. Similarly, 20 Vulnerabilities were associated with more than one CWE; therefore the total vulnerabilities for each technique as shown in Table\ref{tab:VulnCounting} may be lower than the sum of each column in Table \ref{tab:long_CWE}.

\renewcommand{\arraystretch}{1.3}
\begin{longtable}{ B{130pt} | B{15pt} ||| C{25pt} || C{25pt} || C{21pt} | C{21pt} ||  C{21pt} | C{21pt} |||  C{17pt} }

\caption{CWEs associated with \textit{more severe} Vulnerabilities}\label{tab:long_CWE}\\
\centering
\multirow{2}{=}{CWE}  &\multirow{2}{*}{\parbox{22pt}{Top\newline Ten }}  & \multirow{2}{=}{SMPT} & \multirow{2}{=}{EMPT} & \multicolumn{2}{c||}{DAST} & \multicolumn{2}{c|||}{SAST} & \multirow{2}{=}{Total}\\
\cline{5-8}
& & & & ZAP & DA-2 & Sonar & SA-2 & \\
\endfirsthead
 CWE & {TT} & SMPT & EMPT & ZAP & DA-2 & Sonar & SA-2 & Total\\
\endhead
\hline\hline
\multicolumn{9}{H{344pt}}{A01 Broken Access Control}\\\hline
922 - Insecure Storage of Sensitive Information&A1&1&&&&&&1\\\hline
200 - Exposure of Sensitive Information to an Unauth. Actor&A1&&2&&&&&2\\\hline
601 - URL Redirection to Untrusted Site ('Open Redirect')&A1&&&&&&9&9\\\hline
% 22 - Improper Limitation of a Pathname to a Restricted Directory ('Path Traversal')&A1&&&&&&19&19\\\hline
285 - Improper Authorization&A1&1&13&&&&&13\\\hline
22 - Path Traversal&A1&&&&&&19&19\\\hline
% 352 - Cross-Site Request Forgery (CSRF)&A1&&&&&220&20&233\\\hline
\multicolumn{9}{H{344pt}}{A02 Cryptographic Failures}\\\hline
326 - Inadequate Encryption Strength &A1&&&1&&&&1\\\hline
319 - Cleartext Transmission of Sensitive Information &A2&1&1&&&&&1\\\hline
327 - Use of a Broken or Risky Cryptographic Algorithm &A2&&&&&2&&2\\\hline
\multicolumn{9}{H{344pt}}{A03 Injection}\\\hline
643 - XPath Injection&A3&&&&&&1&1\\\hline
89 - SQL Injection &A3&&&&&&4&4\\\hline
20 - Improper Input Validation &A3&3&19&&2&&&21\\\hline
79 - Cross-site Scripting &A3&2&100&3&7&&19&124\\\hline
\multicolumn{9}{H{344pt}}{A04 Insecure Design}\\\hline
269 - Improper Privilege Management&A4&&1&&&&&1\\\hline
313 - Cleartext Storage in a File or on Disk&A4&&&&&&1&1\\\hline
770 - Allocation of Resources Without Limits or Throttling&A4&1&1&&&&&1\\\hline
% 602 - Client-Side Enforcement of Server-Side Security&A4&&2&&&&&2\\\hline
419 - Unprotected Primary Channel&A4&2&1&&&&&2\\\hline
807 - Reliance on Untrusted Inputs in a Security Decision&A4&&&&&3&&3\\\hline
73 - External Control of File Name or Path&A4&&&&&&4&4\\\hline
% 419 - Unprotected Primary Channel&A4&3&3&&&&&5\\\hline
598 - Use of GET Request Method With Sensitive Query Strings&A4&2&5&&1&&&6\\\hline
\multicolumn{9}{H{344pt}}{A05 Security Misconfiguration}\\\hline
548 - Exposure of Information Through Directory Listing &A5&&&1&&&&1\\\hline
614 - Sensitive Cookie in HTTPS Session Without `Secure' Attribute&A5&1&1&&&&&1\\\hline
16 - Configuration&A5&1&1&1&&&&3\\\hline
611 - Improper Restriction of XML External Entity Reference &A5&&&&&13&1&14\\\hline
\multicolumn{9}{H{344pt}}{A06 Vulnerable and Outdated Components}\\\hline
\multicolumn{9}{H{344pt}}{A07 Identification and Authentication Failures}\\\hline
308 - Use of Single-factor Authentication &A7&&1&&&&&1\\\hline
384 - Session Fixation &A7&1&1&&&&&1\\\hline
620 - Unverified Password Change &A7&1&&&&&&1\\\hline
346 - Origin Validation Error &A7&&&&&&2&2\\\hline
613 - Insufficient Session Expiration &A5&1&1&&1&&&2\\\hline
521 - Weak Password Requirements&A7&10&7&&&&&10\\\hline
\multicolumn{9}{H{344pt}}{A08 Software and Data Integrity Failures}\\\hline
829 - Inclusion of Functionality from Untrusted Control Sphere&A8&1&&&&&&1\\\hline
502 - Deserialization of Untrusted Data&A8&&&&&&10&10\\\hline
\multicolumn{9}{H{344pt}}{A09 Security Logging and Monitoring Failures}\\\hline
532 - Insertion of Sensitive Information into Log File&A9&1&1&&&&&1\\\hline
778 - Insufficient Logging&A9&2&9&&&&&11\\\hline
\multicolumn{9}{H{344pt}}{A10 Server-Side Request Forgery (SSRF)}\\\hline
918 - Server-Side Request Forgery (SSRF)&A10&1&&&&&&1\\\hline
\multicolumn{9}{H{344pt}}{Not Mapped to OWASP Top Ten}\\\hline
509 - Replicating Malicious Code (Virus or Worm)&NA&1&1&&&&&1\\\hline
1022 - Use of Web Link to Untrusted Target with window.opener Access&NA&&&&&1&&1\\\hline
674 - Uncontrolled Recursion&NA&&&&&&2&2\\\hline
567 - Unsynchronized Access to Shared Data in a Multithreaded Context&NA&&&&&&4&4\\\hline
543 - Use of Singleton Pattern Without Synchronization in a Multithreaded Context&NA&&&&&&8&8\\\hline
827 - Improper Control of Document Type Definition&NA&&&&&13&1&14\\\hline
404 - Improper Resource Shutdown or Release&NA&&&&&&39&39\\%\hline
\end{longtable}

\renewcommand{\arraystretch}{1.3}
\begin{longtable}{ B{130pt} | B{15pt} ||| C{25pt} || C{25pt} || C{21pt} | C{21pt} ||  C{21pt} | C{21pt} |||  C{17pt} }
% \caption{CWEs of Medium or High Severity}\label{tab:long_CWE_low}\\
\caption{Low Severity Vulnerability CWEs}\label{tab:long_CWE_low}\\
\centering
\multirow{2}{=}{CWE}  &\multirow{2}{*}{\parbox{22pt}{Top\newline Ten }}  & \multirow{2}{=}{SMPT} & \multirow{2}{=}{EMPT} & \multicolumn{2}{c||}{DAST} & \multicolumn{2}{c|||}{SAST} & \multirow{2}{=}{Total}\\
\cline{5-8}
& & & & ZAP & DA-2 & Sonar & SA-2 & \\
\endfirsthead
 CWE & {TT} & SMPT & EMPT & ZAP & DA-2 & Sonar & SA-2 & Total\\
\endhead
\hline\hline
\multicolumn{9}{H{344pt}}{A01 Broken Access Control}\\\hline
% 922 - Insecure Storage of Sensitive Information&A1&1&&&&&&1\\\hline
% 200 - Exposure of Sensitive Information to an Unauth. Actor&A1&&3&&&&&3\\\hline
% 601 - URL Redirection to Untrusted Site ('Open Redirect')&A1&&2&&&&9&11\\\hline
% % 22 - Improper Limitation of a Pathname to a Restricted Directory ('Path Traversal')&A1&&&&&&19&19\\\hline
% 22 - Path Traversal&A1&&&&&&19&19\\\hline
% 285 - Improper Authorization&A1&1&24&&&&&24\\\hline
% % 352 - Cross-Site Request Forgery (CSRF)&A1&&&&&220&20&233\\\hline
352 - Cross-Site Request Forgery&A01&&&1&&220&20&234\\\hline
\multicolumn{9}{H{344pt}}{A02 Cryptographic Failures}\\\hline
% 352 - Cross-Site Request Forgery&A1&&&&&220&20&233\\\hline
% \multicolumn{9}{H{344pt}}{A02 Cryptographic Failures}\\\hline
\hline
760 - Use of a One-Way Hash with a Predictable Salt &A02&&&&&&2&2\\\hline
% 327 - Use of a Broken or Risky Cryptographic Algorithm &A2&&&&&2&&2\\\hline
% 319 - Cleartext Transmission of Sensitive Information &A2&3&5&&&&&6\\\hline
% % 328 - Use of Weak Hash &A2&&&&&&6&6\\\hline
\multicolumn{9}{H{344pt}}{A03 Injection}\\\hline
470 - Use of Externally-Controlled Input to Select Classes or Code ('Unsafe Reflection')&A03&&&&&&34&34\\\hline
\multicolumn{9}{H{344pt}}{A04 Insecure Design}\\\hline
209 - Generation of Error Message Containing Sensitive Information &A04&2&18&&1&&&18\\\hline
501 - Trust Boundary Violation&A04&&&&&&28&28\\\hline
\multicolumn{9}{H{344pt}}{A05 Security Misconfiguration}\\\hline
7 - Missing Custom Error Page&A05&1&1&1&1&&1&1\\\hline
933 - Security Misconfiguration&A05&&&1&&&&1\\\hline
16 - Configuration&A05&2&1&2&&&&2\\\hline
\multicolumn{9}{H{344pt}}{A06 Vulnerable and Outdated Components}\\\hline
\multicolumn{9}{H{344pt}}{A07 Identification and Authentication Failures}\\\hline
\multicolumn{9}{H{344pt}}{A08 Software and Data Integrity Failures}\\\hline
345 - Insufficient Verification of Data Authenticity&A08&&&1&&&&1\\\hline
502 - Deserialization of Untrusted Data&A08&&&&&1&&1\\\hline
\multicolumn{9}{H{344pt}}{A09 Security Logging and Monitoring Failures}\\\hline
\multicolumn{9}{H{344pt}}{A10 Server-Side Request Forgery (SSRF)}\\\hline
% 918 - Server-Side Request Forgery (SSRF)&A10&1&&&&&&1\\\hline
\multicolumn{9}{H{344pt}}{Not Mapped to OWASP Top Ten}\\\hline

% % 400 - Uncontrolled Resource Consumption&N/A&&&&&&&1\\\hline % The only vuln assigned to 400 also maps to 770. 770 maps into the OWASP Top 10, probably "good enough" for both
% 1022 - Use of Web Link to Untrusted Target with window.opener Access&NA&&&&&1&&1\\\hline
242 - Use of Inherently Dangerous Function&NA&&&&&&2&2\\\hline % Low
615 - Inclusion of Sensitive Information in Source Code Comments&NA&&&&&&5&5\\\hline % Low
404 - Improper Resource Shutdown or Release&NA&&&&&&17&17\\\hline % Low
489 - Active Debug Code&NA&&&&&26&&26\\\hline % Low
582 - Array Declared Public, Final, and Static&NA&&&&&31&&31\\\hline % Low
754 - Improper Check for Unusual or Exceptional Conditions&NA&&&&&31&&31\\\hline % Low
600 - Uncaught Exception in Servlet&NA&&&&&60&3&60\\\hline
493 - Critical Public Variable Without Final Modifier&NA&&&&&210&&210\\\hline% Low
% 544 - Missing Standardized Error Handling Mechanism&NA&&5&&&&&5\\\hline % Low % Updated to be CWE 7, and CWE 209 (standardization

% 674 - Uncontrolled Recursion&NA&&&&&&2&2\\\hline
% 567 - Unsynchronized Access to Shared Data in a Multithreaded Context&NA&&&&&&4&4\\\hline
% % % 615 - Inclusion of Sensitive Information in Source Code Comments&NA&&&&&&5&5\\\hline % Low
% 543 - Use of Singleton Pattern Without Synchronization in a Multithreaded Context&NA&&&&&&8&8\\\hline
% 827 - Improper Control of Document Type Definition&NA&&&&&13&1&14\\\hline
% 404 - Improper Resource Shutdown or Release&NA&&&&&&39&39\\%\hline
\end{longtable}

\end{document}